\documentclass[%
 a4paper,
 reprint,
 floatfix,
 superscriptaddress,
 nofootinbib,
 amsmath,amssymb,
 aps,
 prd,
]{revtex4-2}

\usepackage{graphicx}
\usepackage{dcolumn}
\usepackage{bm}

\usepackage[separate-uncertainty=true]{siunitx}
\DeclareSIUnit\parsec{pc}
\DeclareSIUnit\year{yr}
\DeclareSIUnit\erg{erg}
\DeclareSIUnit\gauss{G}

\usepackage{orcidlink}
\usepackage{xspace}
\usepackage{commath}
\usepackage{hyperref}
\definecolor{ntnu}{RGB}{0, 80, 158}
\hypersetup{
   colorlinks = true,
   citecolor  = ntnu,
   urlcolor   = ntnu,
   linkcolor  = ntnu,
}

\usepackage[capitalise]{cleveref}
\crefname{figure}{Fig.}{Figs.}
\crefname{section}{Sec.}{Secs.}
\crefname{equation}{Eq.}{Eqs.}
\crefname{align}{Eq.}{Eqs.}

\usepackage{relsize}
\usepackage{tikz}
\usetikzlibrary{shapes.geometric, arrows, matrix, fit}
\tikzstyle{box} = [rectangle, minimum width=2cm, minimum height=1cm, text centered, draw=black, fill=none]
\tikzstyle{rbox} = [rectangle, rounded corners, minimum width=2cm, minimum height=1.5cm, text centered, draw=black, fill=none]
\tikzstyle{arrow} = [thick,->,>=stealth]

\graphicspath {{images/}}

\newcommand{\xmax}{X_\text{max}}
\newcommand{\mxmax}{\langle X_\text{max}\rangle}
\newcommand{\sxmax}{\sigma(X_\text{max})}
\newcommand{\rmax}{\ensuremath{R_\text{max}}\xspace}
\newcommand{\src}{\text{src}}
\newcommand{\pp}{\text{pop}}
\newcommand{\bp}{\ensuremath{{\beta_\pp}}\xspace}
\newcommand{\gs}{\gamma_\src}
\newcommand{\gp}{\gamma_\pp}
\newcommand{\prmax}{\ensuremath{p(\rmax)}\xspace}
\newcommand{\prmaxvar}[1]{\ensuremath{p(\rmax|#1)}\xspace}
\newcommand{\lc}{{\lambda_\text{cut}}\xspace}
\newcommand{\hesi}{\text{hs}}
\newcommand{\expo}{\text{exp}}
\newcommand{\bexp}{\text{b-exp}}
\newcommand{\sexp}{\text{s-exp}}
\newcommand{\rOne}{\beta_1}
\newcommand{\rTwo}{\beta_2}
\newcommand{\dila}{\xi}
\newcommand{\rll}{\text{LL}}
\newcommand{\rul}{\text{UL}}

\begin{document}

\title{Curious case of the maximum rigidity distribution of cosmic-ray accelerators}

\author{D.~Ehlert \orcidlink{0000-0002-4322-6400}}
    \email{domenik.ehlert@ntnu.no}
    \affiliation{Institutt for fysikk, Norwegian University of Science and Technology (NTNU), 7491 Trondheim, Norway}
\author{F.~Oikonomou \orcidlink{0000-0002-0525-3758}}
    \affiliation{Institutt for fysikk, Norwegian University of Science and Technology (NTNU), 7491 Trondheim, Norway}
\author{M.~Unger \orcidlink{0000-0002-7651-0272}}
    \affiliation{Institute for Astroparticle Physics, Karlsruhe Institute of Technology (KIT), 3640 Karlsruhe, Germany}
    \affiliation{Institutt for fysikk, Norwegian University of Science and Technology (NTNU), 7491 Trondheim, Norway}

\date{\today}

\begin{abstract}
In many models, the sources of ultra-high-energy cosmic rays (UHECRs) are assumed to accelerate particles to the same maximum energy.
Motivated by the fact that candidate astrophysical accelerators exhibit a vast diversity in terms of their relevant properties such as luminosity, Lorentz factor, and magnetic field strength, we study the compatibility of a population of sources with non-identical maximum cosmic-ray energies with the observed energy spectrum and composition of UHECRs at Earth.
For this purpose, we compute the UHECR spectrum emerging from a population of sources with a power-law, or broken-power-law, distribution of maximum energies, applicable to a broad range of astrophysical scenarios.
We find that the allowed source-to-source variance of the maximum energy must be small to describe the data if a power-law distribution is considered. Even in the most extreme scenario, with a very sharp cutoff of individual source spectra and negative redshift evolution of the accelerators, the maximum energies of 90\% of sources must be identical within a factor of three -- in contrast to the variance expected for astrophysical sources. Substantial variance of the maximum energy in the source population is only possible if the maximum energies follow a broken power-law distribution with a very steep spectrum above the break. However, in this scenario, the individual source energy spectra are required to be unusually hard with increasing energy output as a function of energy. 

\end{abstract}

\maketitle

\section{Introduction}
Ultra-high-energy cosmic rays (UHECRs) are charged particles that
reach Earth with energies of up to several
$10^{20}$~\si{\electronvolt}. The identification of the astrophysical sources capable of accelerating particles to these energies is one of
the unsolved mysteries of high energy astrophysics (see e.g.~\cite{Anchordoqui:2018qom,AlvesBatista:2019tlv} for recent reviews).
A correlation between astrophysical objects and the measured arrival directions of cosmic rays has not yet been established at high
significance~\cite{TelescopeArray:2021gxg}, but the source properties are constrained by measurements of the diffuse particle
flux and composition at Earth; see e.g.~\cite{Allard:2005cx,Hooper:2006tn,Allard:2007gx,Allard:2008gj,Globus:2014fka,Globus+15,Unger:2015laa,PierreAuger:2016use,Fang+17,Kachelriess:2017tvs,Boncioli:2018lrv,Muzio:2019leu,Heinze:2020zqb,Muzio:2021zud,PierreAuger:2021mmt,Bergman:2021djm,PierreAuger:2022atd}.

Most of these studies assume an acceleration mechanism that is universal in rigidity\,\footnote{The rigidity of a particle with charge $Ze$ and momentum $p$ is $R=p\,c/(Ze) \simeq E/Z$ (using natural units and with energy $E$).} up to a maximum rigidity of $\rmax$, leading to consecutive flux suppressions of the elemental spectra  at energies of $E_\text{max} = Z\rmax$, where $Z$ denotes the cosmic-ray charge. Assuming such a ``Peters cycle''~\cite{Peters:1961,Gaisser:2016uoy} at the sources gives a good description of the flux and composition measured at Earth at ultra-high energies, see e.g.\ \cite{PierreAuger:2016use}. Depending on the source environment, the maximum energy can follow a different functional form~\cite{Ptitsyna:2008zs}. Here, we consider primarily the canonical scenario where the maximum energy scales in proportion to the nuclear charge and briefly discuss the effect of alternate scalings on our results.

A major caveat of many studies is that the sources are typically assumed to be identical -- a description that is unlikely to hold for realistic sources. The most probable astrophysical candidates for the sources of UHECRs, e.g.\ active galactic nuclei (AGN) and gamma-ray bursts (GRBs), are generally not very similar -- even within a single source class -- but exhibit an enormous diversity in terms of key parameters like luminosity, size, magnetic field and jet power.

Only a few studies have relaxed the assumption of identical sources in the past, by focusing on a low number of discrete local sources~\cite{Ahlers:2012az,Eichmann:2017iyr,Eichmann:2022ias} or by considering the superposition of a few ($\leq$ 3) source classes, e.g.~\cite{Muzio:2019leu,Rodrigues:2020pli,PierreAuger:2021mmt,Das:2020nvx,Mollerach:2020mhr,PierreAuger:2022atd}. The time variation of $\rmax$ in AGN jets was studied in Ref.~\cite{Matthews:2021nik}, and the effective spectrum produced by sources with non-identical spectral shapes and spectral indices has been discussed in the context of gamma-ray spectra~\cite{Lipari:2020szc} and Galactic cosmic-ray sources~\cite{Yuan:2011ys}. Populations of sources with non-identical maximum cosmic-ray rigidities were considered previously, for a pure proton UHECR composition~\cite{Kachelriess:2005xh}, for Galactic cosmic rays~\cite{Shibata:2010zza}, and for gamma-ray bursts~\cite{Heinze:2020zqb}.

Here we present, for the first time, a rigorous exploration of the population variance of $\rmax$ compatible with current observations of the spectrum and composition of UHECRs.
This is achieved by convolving the distribution of source properties, parameterised by the maximum rigidity, with the individual source spectra to obtain an analytical description of the total population spectrum, as detailed in \cref{sec:theory}. We simulate the propagation of UHECRs to Earth through the extragalactic photon
fields and find the best source parameters by
comparing the model predictions to UHECR data in \cref{sec:methods}. From this, we derive lower limits on the source variance allowed by the data, as described in \cref{sec:results}. We conclude in \cref{sec:conclusion} that only a limited amount of population variance is permitted, and UHECR sources are required to be nearly identical in terms of maximum rigidity under realistic choices of the model parameters if a power-law distribution of maximum rigidities is assumed. The variance can be large for broken power-law distributions, provided the source spectra are sufficiently hard. However, we find the distributions predicted for most candidate source classes -- blazars, gamma-ray bursts, and tidal disruption events -- to be incompatible with the limits obtained from the UHECR fit. Only sources with luminosity distribution similar to the one of Seyfert galaxies can potentially satisfy the constraints.

\section{Population Spectrum of non-identical Sources}\label{sec:theory}
In this study, we assume that the rigidity spectra of individual sources
of UHECRs are well described by the aforementioned Peters Cycle, and thus we assume a power law with a high-rigidity cutoff,
\begin{equation}
    \phi_\src = \frac{\dif^{\,2}N}{\dif R \dif t} = \sum_{i}\phi_0(Z_i)\,R^{-\gs}\, f(R,\rmax)\,,
    \label{eq:source}
\end{equation}
where the sum runs over all accelerated chemical elements with charge $Z_i$ and the spectral index is assumed to be universal.\footnote{$\gs\approx2$ for diffusive shock acceleration~\cite{Bell:1978zc,Bell:1978fj}, but in this study the value of $\gs$ is a free parameter.} The term $f(R,\rmax)$ describes the high-rigidity cutoff at maximum rigidity \rmax.
We refer to the sum of the spectra of all sources within a certain volume as the {\itshape population spectrum} $\phi_\pp$.

In the limit of identical sources, the population spectrum will necessarily have the same shape as the spectra of individual sources. A source-by-source variation of the normalisation factors $\phi_0(Z_i)$ does not lead to a qualitatively different population spectrum since it is equivalent to identical sources with the
source-averaged normalisations $\phi_0$, and thus we will not consider it in the following. It should, however, be kept in mind that the elemental fractions obtained from the fits presented later in this paper should be understood as source-averaged fractions. A phenomenologically more interesting source property is the maximum rigidity. If the probability for an individual accelerator to reach a certain maximum rigidity is distributed as $\prmax\equiv\dif p/\dif\rmax$ and the source spectra follow $\phi_\src(R,\rmax)$, then the combined spectrum of the entire population is given by the convolution
\begin{equation}
    \phi_\pp(R) = \int_{0}^{\infty} \phi_\src(R, \rmax) \,  \prmax \,  \dif\rmax.
\end{equation}
Here and in subsequent occurrences of $\phi_\src$, the sum over all chemical elements of charge $Z_i$ (cf.\ \cref{eq:source}) is assumed implicitly.
In the following, we specify the functional forms of individual source spectra and probability distribution of \rmax that will be studied in this work.

\subsection{Source Spectra}\label{sec:intro_source_spectrum}
    In general, the rigidity cutoff of astrophysical accelerators depends on the acceleration mechanism and the source environment, in particular on the dominant energy-loss process, see e.g.~\cite{Protheroe:1998pj,Protheroe:2004rt}. The simplest description of the shape is given by a sharp termination when particles exceed the maximum rigidity
    \begin{equation}
        \phi_\src^\hesi =  \phi_0\,R^{-\gs}\,\theta(\rmax - R),
    \end{equation}
    where $\theta(x)$ denotes the Heaviside step function.
    The population spectrum corresponding to this function has been studied previously in Ref.~\cite{Kachelriess:2005xh}. Since $\phi_\src^\hesi$ describes the sharpest-possible rigidity cutoff, it provides a useful extreme case that will allow for a maximum variation of maximum rigidity among the sources in the population. 

    A more commonly used choice, expected under certain astrophysical conditions, see e.g.~\cite{Protheroe:2004rt,Zirakashvili:2006pv}, is given by an exponential cutoff
    \begin{equation}
        \phi_\src^\expo = \phi_0\,R^{-\gs}\exp{\left(-\frac{R}{\rmax}\right)}\,.
    \end{equation}
    However, this function has the disadvantage that the effect of the cutoff already starts to become noticeable well below the maximum rigidity and the interpretation of the spectral index $\gs$ is complicated. For this reason, some phenomenological studies assume a broken-exponential source spectrum, e.g.~\cite{PierreAuger:2016use}
    \begin{equation}
        \phi_\src^\bexp = \phi_0 \, R^{-\gs}\,
        \begin{cases}
            1 &R < \rmax \\
            \exp\left(1-\frac{R}{\rmax}\right)&\text{otherwise},
        \end{cases}
    \end{equation}
    which alleviates the issue but lacks physical motivation.

    Finally, we consider spectra with an exponential cutoff raised to the power of $\lc$,
    \begin{equation}\label{eq:source_super_expon}
        \phi_\text{src}^\sexp = \phi_0 \, R^{-\gs}\,\exp\left(-\frac{R}{\rmax}\right)^\lc,\quad\lc>0\,.
    \end{equation}
    We refer to this description as a ``super-exponential" cutoff. The function can be used to interpolate continuously between classical, exponential cutoffs ($\lc=1$) and sharp, Heaviside-like terminations ($\lc=\infty$). Additionally, for $\lc<1$ the cutoff shape becomes sub-exponential up to no cutoff when $\lc\rightarrow0$. Super-exponential cutoff profiles were obtained e.g.\ in Ref.~\cite{Zirakashvili:2006pv} with $\lc=2$ for synchrotron losses during acceleration.

\begin{figure*}
		\centering
		\includegraphics[width=0.48\linewidth]{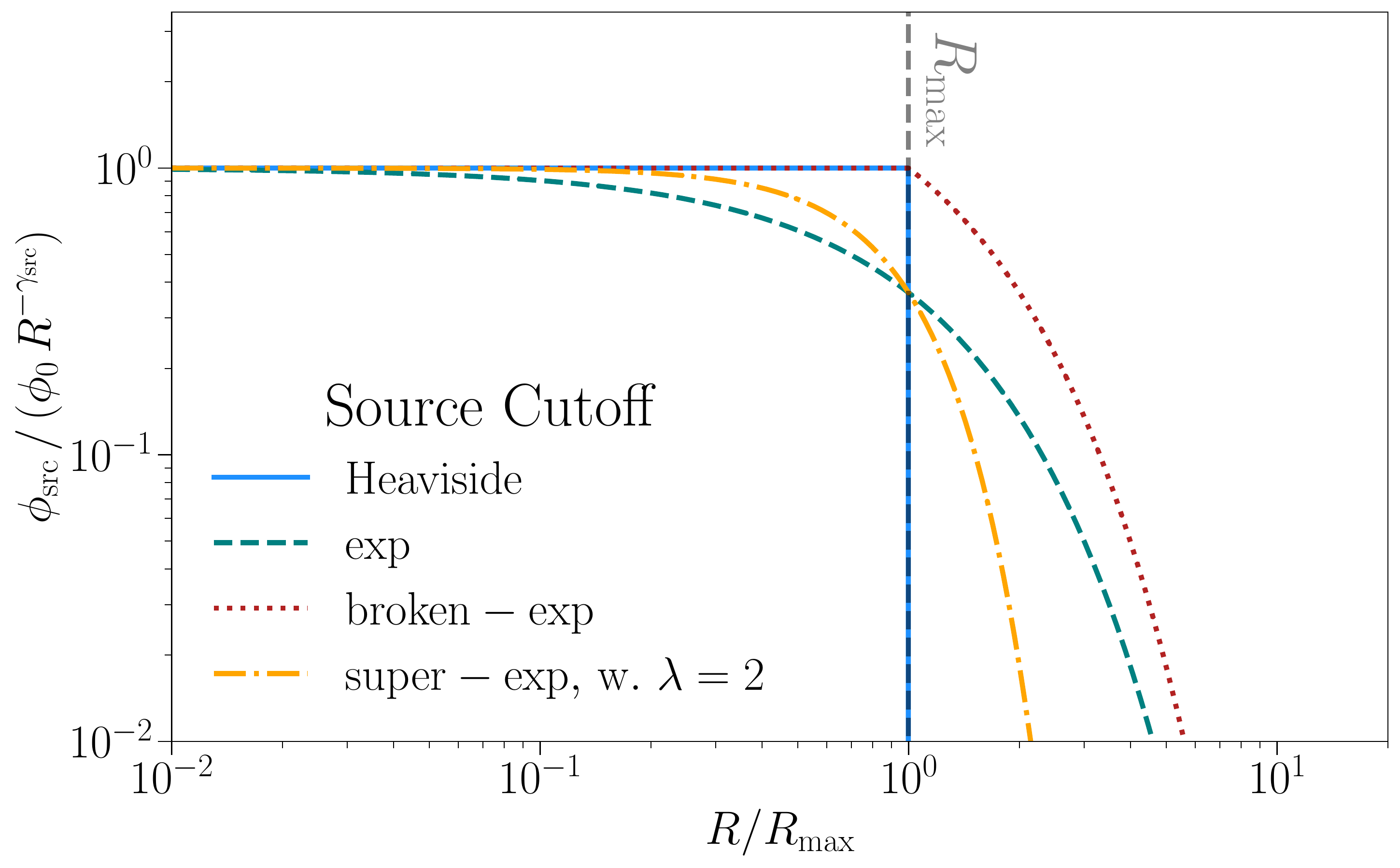}
		\includegraphics[width=0.48\linewidth]{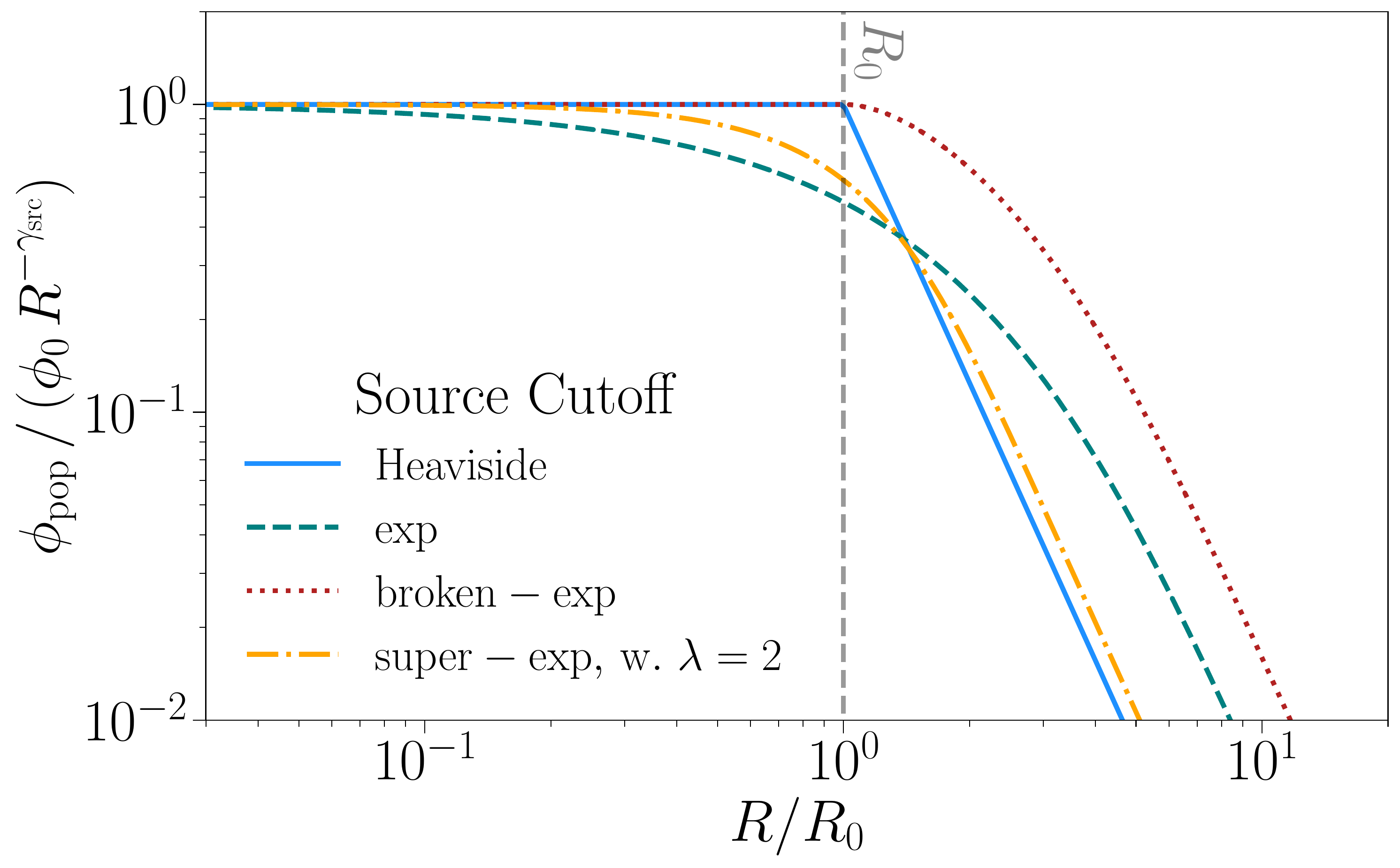}
		\caption{{\bf Left:} Cosmic-ray source spectra for the different rigidity cutoff functions. \rmax denotes the maximum rigidity and the $y$-axis is scaled to show the ratio to an unmodified power law with spectral index $\gs$. {\bf Right:} Population spectra resulting from the convolution of a power-law distribution of maximum rigidities above rigidity $R_0$ and the source spectra displayed in the left panel ($\bp=4$).}\label{fig:sample_spectra}
	\end{figure*}

    An illustration of the source spectrum for different choices of the cutoff is shown in the left panel of \cref{fig:sample_spectra}.

\subsection{Distribution of Maximum Rigidities}
\label{subsec:Rmax_distribution}
\paragraph{Power Law}
    In this paper, we mainly consider a population of sources with a distribution of maximum rigidities that follows a power law (PL) with spectral index \bp above a minimum allowed maximum rigidity $R_0$
    \begin{equation}\label{eq:dpdRmax_spl}
        \prmax =
        \begin{cases}
            0 &\rmax < R_0 \\
            \frac{\bp-1}{R_0}\,\left( \frac{\rmax}{R_0}\right)^{-\bp}&\text{otherwise,}
        \end{cases}
    \end{equation}
    which is also known as a Pareto distribution. This \rmax distribution was
    previously considered in Ref.~\cite{Kachelriess:2005xh}.
    Because of the asymmetric nature of the power-law distribution, the standard deviation
    is of limited use to characterize the source variance, and instead we will report the one-sided $90\%$ quantile $\rmax^{0.90}$ defined as
    \begin{equation}\label{eq:Rq90_def}
        \int_{R_0}^{\rmax^{0.90}}\dif \rmax\,p(\rmax) = 0.90\,.
    \end{equation}
    For a power-law distribution of maximum rigidities, the quantile $q$ is given by the relation
    \begin{equation}\label{eq:Rq_spl}
        \rmax^q / R_0 = [ 1 - q]^{1/(1-\bp)}.
    \end{equation}
    For illustration, population diversity of more than a decade, i.e.\ $\rmax^{0.90}/R_0\geq10$, is obtained if $\bp<2$.

\paragraph{Broken Power Law}
    Alternatively, the distribution of maximum rigidities can be modelled as a broken power law (BPL) which can be written as
    \begin{equation}\label{eq:dpdrmax_bpl_main}
        \prmax = \frac{ R_0^{-1} }{ C }
        \times\begin{cases}
            \left(\frac{\rmax}{R_0}\right)^{-\rOne}& \rmax \leq R_0    \\
            \left(\frac{\rmax}{R_0}\right)^{-\rTwo}&\rmax > R_0,
        \end{cases}
    \end{equation}
    with $C$ a normalisation constant. A detailed discussion of the BPL scenario is provided in \cref{apx:broken_power-law_distribution}.
    
\subsection{Population Spectrum}\label{sec:pop}
\paragraph{Power Law}
Assuming a power-law distribution of $\rmax$, it is possible to derive an analytical description of the population spectrum for all source spectra presented in \cref{sec:intro_source_spectrum}. In the case of sources with a Heaviside termination in rigidity, the population spectrum is given by,
\begin{equation}
\phi_\text{pop}^\hesi = \phi_0\,R^{-\gs}
    \begin{cases}
         1  & R < R_0 \\
        \left(\frac{R}{R_0}\right)^{-\bp+1} & \text{otherwise}.
    \end{cases}
    \label{eq:heaviside_pop}
\end{equation}
For source spectra with a broken exponential cutoff, the population spectrum becomes,
\begin{align}
  \phi_\text{pop}^\bexp = \phi_0\,R^{-\gs}
\begin{cases}
  1  & R < R_0 \\
  \left(\frac{R}{R_0}\right)^{-\bp+1} f(\frac{R}{R_0}, \bp) & \text{otherwise}
\end{cases}
\end{align}
with
\begin{align*}
  f\left(\textstyle\frac{R}{R_0}, \bp\right) = \,&1+e\,(\bp-1)\\&\times\left[\gamma\left(\bp-1,\frac{R}{R_0}\right)  -\gamma\big(\bp-1,1\big)\right].
\end{align*}
Finally, for sources with a (super-)exponential cutoff, the population spectrum is given by,
\begin{align}\label{eq:pop_super_expon}
  \phi_\text{pop}^\sexp = \phi_0\,R^{-\gs}\,&\left(\textstyle\frac{R}{R_0}\right)^{-\bp+1}\,\textstyle\frac{\bp-1}{\lc} \nonumber\\
  &\times\,\gamma\left(\textstyle\frac{\bp-1}{\lc},\left(\frac{R}{R_0}\right)^{\lc}\right).
\end{align}
For a standard exponential distribution, i.e.\ $\lc=1$ this simplifies to
\begin{align}\label{eq:pop_expon}
  \phi_\text{pop}^\expo = \phi_0\,R^{-\gs}&\left(\textstyle\frac{R}{R_0}\right)^{-\bp+1}(\bp-1) \nonumber   \\
  &\times\gamma\left(\textstyle\bp-1,\frac{R}{R_0}\right),
\end{align}
 whereas for $\lc\to\infty$, \cref{eq:heaviside_pop}  is recovered.
Here $\gamma$ denotes the lower incomplete gamma function, not to be confused with the source spectral index $\gs$.

Population spectra for a particular choice of parameters are shown in \cref{fig:sample_spectra}, right. The main impact of the cutoff parameter $\lc$ is a later onset and faster turnover of the spectral break around $R_0$.
The limiting behaviour of the population spectrum is independent of the source cutoff function. The asymptotic rigidity dependencies of the population spectra are
\begin{equation}\label{eq:pp_Rlim0}
  \lim\limits_{R\to0} \phi_\pp(R) \propto R^{-\gs}
\end{equation}
and
\begin{equation}\label{eq:pp_RlimInfty}
  \lim\limits_{R\to\infty} \phi_\pp(R) \propto R^{-\gs-\bp+1}.
\end{equation}
\paragraph{Broken Power-Law}\label{sec:pop_bpl}
In the broken power-law scenario, the resulting population spectrum for sources with super-exponential cutoff is
\begin{align}\label{eq:pp_BPL_sexp_main}
    \phi_\pp^\sexp &= \frac{ \phi_0 }{ C\cdot\lc }\,R^{-\gs}\times\Big[L + H\Big],   
\end{align}
    with
\begin{align} 
    &L = \left(\frac{R}{R_0}\right)^{-\rOne+1}\cdot\Gamma\left(\frac{\rOne-1}{\lc},\left(\frac{R}{R_0}\right)^{\lc}\right) \nonumber    \\
    &H = \left(\frac{R}{R_0}\right)^{-\rTwo+1}\cdot\gamma\left(\frac{\rTwo-1}{\lc},\left(\frac{R}{R_0}\right)^{\lc}\right).
\end{align}
Here ``$C$'' is a normalisation factor (see \cref{apx:broken_power-law_distribution}), ``$L$'' reflects the contribution from sources with $\rmax\leq R_0$ and ``$H$'' represents sources above the break. $\Gamma$ and $\gamma$ are the upper and lower incomplete gamma function respectively. The single power-law scenario is retrieved as a limiting case of the more general broken power-law case for $\rOne\to-\infty$, i.e.\ when the contribution ``$L$'' of low-$\rmax$ sources vanishes. We discuss the broken power-law scenario in more detail in \cref{apx:broken_power-law_distribution}. Here we only mention that the same expression for the population spectrum as in the power-law picture holds approximately after defining the effective spectral index $\gp=\gs+\beta_1-1$ and $\rmax$-distribution index $\bp=\beta_2-\beta_1+1$.

\subsection{Relation to Astrophysical Quantities}
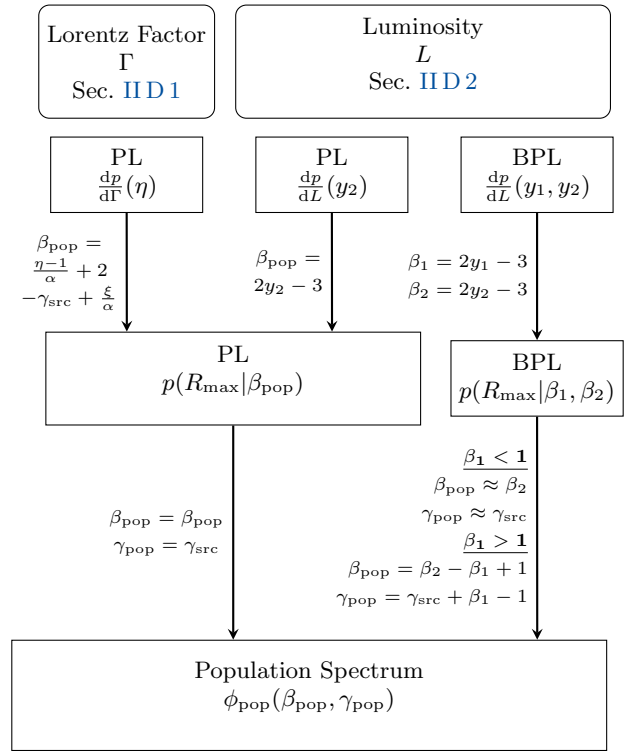
\begin{figure}
    \begin{tikzpicture}[node distance=1.7cm, bend angle=90]
        \node[box, align=center] (lorentz_SPL) {PL\\$\frac{\dif p}{\dif\Gamma}(\eta)$};
        \node[box, align=center, right of=lorentz_SPL, xshift=1cm] (lumi_SPL) {PL\\$\frac{\dif p}{\dif L}(y_2)$};
        \node[box, align=center, right of=lumi_SPL, xshift=1cm] (lumi_BPL) {BPL\\$\frac{\dif p}{\dif L}(y_1,y_2)$};
        \node[rbox, align=center, above of=lorentz_SPL, yshift=-0.2cm] (lorentz) {Lorentz Factor\\$\Gamma$\\\cref{sec:intro_lorentz_factor}};
        \node[rbox, align=center, fit=(lumi_SPL)(lumi_BPL), above of=lumi_SPL, xshift=1.2cm, yshift=-0.2cm] (lumi) {Luminosity\\$L$\\\cref{sec:intro_luminosity}};
        \node[box, fit=(lorentz_SPL)(lumi_SPL), align=center, minimum width=3cm, below right of=lorentz_SPL, yshift=-1.5cm, xshift=0.2cm] (rmax_SPL) {PL\\\prmaxvar{\bp}};
        \node[box, align=center, below of=lumi_BPL, yshift=-1cm] (rmax_BPL) {BPL\\\prmaxvar{\rOne,\rTwo}};

        \draw[arrow, align=center] (lorentz_SPL.south) -- node[anchor=east] {\smaller[1.5]$\bp=$\\\smaller[1.5]$\frac{\eta-1}{\alpha}+2$\\\smaller[1.5]$-\gs+\frac{\dila}{\alpha}$} ([xshift=-1.4cm]rmax_SPL.north);
        \draw[arrow, align=center] (lumi_SPL.south) -- node[anchor=east] {\smaller[1.5]$\bp=$\\\smaller[1.5]$2y_2-3$} ([xshift=1.3cm]rmax_SPL.north);
        \draw[arrow, align=center] (lumi_BPL.south) -- node[anchor=east] {\smaller[1.5]$\rOne=2y_1-3$\\\smaller[1.5]$\rTwo=2y_2-3$} (rmax_BPL.north);
        \node[box, fit=(rmax_SPL)(rmax_BPL), align=center, minimum width=3cm, below of=rmax_SPL, xshift=1cm, yshift=-2.5cm] (phi_pop) {Population Spectrum\\$\phi_\pp(\bp,\gp)$};
        \draw[arrow, align=center] (rmax_SPL.south) -- node[anchor=east] {\smaller[1.5]$\bp=\bp$\\\smaller[1.5]$\gp=\gs$} ([xshift=-1cm]phi_pop.north);
        \draw[arrow, align=right] (rmax_BPL.south) -- node[anchor=east] {\smaller[1.5]$\bf\underline{\rOne<1}$\\\smaller[1.5]$\bp\approx \rTwo$\\\smaller[1.5]$\gp\approx\gs$\\\smaller[1.5]$\bf\underline{\rOne>1}$\\\smaller[1.5]$\bp=\rTwo-\rOne+1$\\\smaller[1.5]$\gp=\gs+\rOne-1$} ([xshift=3.0cm]phi_pop.north);
    \end{tikzpicture}

    \caption{Schematic description of how the parameters of the considered astrophysical scenarios ($\eta$, $\alpha$, $\
    \dila$, $y_1$, $y_2$, $\rOne$, $\rTwo$, $\gs$) are connected to the parameters $(\bp,\gp)$ that describe the effective population spectrum $\phi_\pp$. In the first step, the single power-law (PL) or broken power-law (BPL) distributions of physical parameters are converted to a PL or BPL distribution of maximum rigidities. This conversion is exact, assuming that \cref{eq:rmax_lorentz,eq:rmax_lumi} are valid. If $\rmax$ follows a PL distribution, with slope $\rmax^{-\bp}$ above the threshold at $R_0$, we obtain our default case for the population spectrum. If $\rmax$ is distributed according to a broken power law, with break at $R_T$ and slope $\rmax^{-\rOne} / \rmax^{-\rTwo}$ before/after the break, the same parameterisation of $\phi_\pp$ is possible but only after a re-interpretation of the parameters. This approach is exact except around the break at $R_0$.}\label{fig:beta_flowchart}
    \end{figure}

    The population spectra derived in the previous section provide simple analytic expressions that are well suited for fits to UHECR observations at Earth, with which the key parameters $\bp$ and $\gs$ can be derived. The connection of these parameters to the properties of UHECR sources is discussed in the following for a few examples. We will show that the assumed power-law distribution in maximum rigidity can be attributed to different acceleration scenarios. The relevant parameters of these scenarios are summarised in \cref{tab:parameter_key} and the re-interpretation of the fitted parameters $\bp,\,\gs$ in terms of proposed underlying physical properties are listed in \cref{tab:beta_key}. Through the re-interpretation of the fitted parameters, all considered physical scenarios can be reduced to the same population spectra that are obtained for a power-law distribution of $\rmax.$ The process is illustrated in \cref{fig:beta_flowchart}.

	\begin{table*}
		\caption{\label{tab:parameter_key}
		Summary of  parameters used for the three
		illustrative scenarios. See sections given
		in the first column for further details.}
		\renewcommand{\arraystretch}{1.3}
		\begin{ruledtabular}
		\begin{tabular}{l|c|ll}
		    Scenario    & Parameter & Description   & Equation  \\ \toprule
            Power law    & $\gs$  & True spectral index of the sources   & $\phi_\src\propto R^{-\gs}$ \\
		    (\cref{subsec:Rmax_distribution})        & $\gp$     & Effective spectral index of the sources (same as the true spectral  & \\
		            &           & index for a power-law distribution of \prmax)  & $\phi_\pp\propto R^{-\gp}$  \\
		            & $\bp$     & (Effective) spectral index of the $\prmax$ distribution   & $\prmax\propto \rmax^{-\bp}$    \\
		            & $\rOne,\,\rTwo$     & Spectral index of the distribution of maximum rigidities below (above) & \\
		            &           & the break for a broken power law in $\prmax$   & $\prmax\propto \rmax^{-\rOne}\,(\rmax^{-\rTwo})$\\ \hline
		   Lorentz  & $\eta$    & Spectral index of the power-law distribution of Lorentz-factors     & $\dif p/\dif\Gamma\propto\Gamma^{-\eta}$\\
		   factor   & $\alpha$  & Energy boosting by the relativistic motion of the jet    & $E=E'\,\Gamma^\alpha$ \\
		  (\cref{sec:intro_lorentz_factor})          & $\dila$       & Time dilation caused by the relativistic motion of the source region  & $t=t'\,\Gamma^\dila$  \\  \hline
		  Luminosity     & $y_1,\,y_2$    & Spectral index of the broken power-law luminosity function of sources & \\
		  (\cref{sec:intro_luminosity})          &           & below (above) the break  & $\dif p/\dif L\propto
		  L^{-y_1}\,(L^{-y_2})$
		\end{tabular}
		\end{ruledtabular}
	\end{table*}

    \subsubsection{Jet Lorentz Factor}\label{sec:intro_lorentz_factor}
        In some scenarios, for sources with relativistic jets, the maximum rigidity is directly related to the bulk Lorentz factor of the motion, $\Gamma_{\rm jet}$. For instance, the Hillas criterion~\cite{Hillas:1984ijl} for relativistic sources gives $\rmax = R_0\Gamma_{\rm jet}$, with $R_0\propto l B$ where $l$ is the size of the source and $B$ the magnetic field.

        It is also possible that UHECRs are galactic cosmic rays that receive a ``one-shot" boost of a factor of $\sim\Gamma_{\rm jet}^2$ in the jet of their host galaxies, in which case $\rmax \sim R_0\,\Gamma_{\rm jet}^2$, where $R_0$ is the maximum energy of the cosmic rays before re-acceleration. This is referred to as the \emph{Espresso} mechanism~\cite{Caprioli:2015zka,Mbarek:2019glq,Mbarek:2021bay}. Cosmic rays that do not enter the most relativistic parts of the jet are only partially boosted with $\rmax\propto\Gamma_{\rm jet}^\alpha,\,\alpha<2$.
        Thus, here we investigate the general case of
        \begin{equation}\label{eq:rmax_lorentz}
            \rmax = R_0\,\Gamma_{\rm jet}^\alpha
        \end{equation}
        where the aforementioned cases are described by $\alpha=1$ (Hillas) and $\alpha\leq2$ (Espresso).

        Assuming, for example, a power-law distribution of the Lorentz factors, as found consistent with observations of jetted AGN in Refs.\ \cite{Lister_1997,2019ApJ...874...43L},
        \begin{equation}
            \frac{\dif p}{\dif\Gamma_{\rm jet}} = (\eta-1) \,\Gamma_{\rm jet}^{-\eta},\label{eq:lister}
        \end{equation}
        the distribution of maximum rigidities can be calculated as
        \begin{align}\label{eq:listerrmax}
            \prmax &= \frac{\dif p}{\dif\Gamma_{\rm jet}}\left|\dod{\Gamma_{\rm jet}}{\rmax}\right| \\
            &= \frac{\eta-1}{\alpha}\,R_0^{-1}\,\left(\frac{\rmax}{R_0}\right)^{\frac{1-\eta}{\alpha}-1}\theta(\rmax-R_0). \nonumber
        \end{align}
        Of course, $R_0$ is also expected to vary from source to source, and therefore the distribution of $\rmax$ should be {\itshape broader} and the above equation can be understood as a lower limit on the
       source-to-source variation of \rmax.

        The boosting of particle energies also affects the expected flux emitted by individual sources. This introduces additional terms into \cref{eq:listerrmax} but we show in \cref{apx:flux_boosting} that the convolution of the source spectra and the \rmax distribution lead to the same functional forms as derived in the last section. However, the parameter $\bp$ can now be related to physical properties of the source population, namely the spectral index $\gs$, the Lorentz-boosting factor $\rmax\propto\Gamma_\text{jet}^\alpha$, and the distribution of Lorentz factors $p(\Gamma_\text{jet})\propto\Gamma_\text{jet}^{-\eta}$ via
        \begin{equation}
            \bp = \frac{\eta-1}{\alpha} + 2 - \gs + \dila/\alpha\,,
        \end{equation}
        where the time dilation factor $\dila=1$ for an acceleration region co-moving with
        the jet and $\dila=0$ for espresso-type re-acceleration.
    \subsubsection{Luminosity}\label{sec:intro_luminosity}
        Another plausible distribution of the maximum rigidity can be derived from the minimum luminosity requirement for particle acceleration in expanding flows. Here, the minimum luminosity, $L_{\rm 0}$, needed to accelerate CRs to maximum rigidity $R_0$ is given by ~\cite{Lovelace:1976,Waxman:1995vg,Waxman:2001tk,Blandford_2000,Lemoine:2009pw,Kachelriess:2022phl,Rieger:2022qhs},\footnote{The normalisation value varies slightly between different papers.}
        \begin{equation}
            L_{\rm 0} \approx 10^{45.5}~\frac{1}{\beta} \left( \frac{R_{0}}{10^{20}\,\si{\volt}} \right)^2~{\rm erg~s^{-1}},
        \end{equation}
        where $\beta$ is the speed of the flow in units of $c$.
        In this scenario, we can relate $\rmax$ and observed luminosity $L$ of a source via
        \begin{equation}\label{eq:rmax_lumi}
            \rmax \sim R_0\,\beta^{1/2} \,\left(\frac{L}{L_0}\right)^{1/2}.
        \end{equation}
        The impact of the variance of $\beta$ in a population can be approximately neglected for highly-, as well as mildly- and non-relativistic sources.\footnote{If the acceleration region is highly relativistic, then $\beta\sim1$ and there is no additional variance introduced by the non-identical outflow speeds. Even in non- or mildly-relativistic source environments, the impact is expected to be small. For example, the authors of \cite{Santoro_2020} found that the relation between luminosity and outflow speed in a sample of AGN outflows is $L\sim v_\text{out}^{4.6}$. The additional contribution in \cref{eq:rmax_lumi} is thus $\beta^{1/2}\sim L^{1/9}$, which constitutes only a subdominant effect compared to the original $\rmax\propto L^{1/2}$.}
        
        In this scenario, where we relate the maximum rigidity to the source luminosity,
        \begin{equation}
          \prmaxvar{z} = \od{p}{L}(z)\left|\dod{L}{\rmax}\right|
        \end{equation}
        where $\dif p(z)/\dif L$ is the luminosity function of the sources. For a single power-law distribution of luminosities, which can adequately describe many proposed source classes, and without taking into account the redshift evolution,\footnote{For our purposes, luminosity and density redshift evolution of sources are indistinguishable in terms of the total contribution to the observed UHECR energy flux. However, as the maximum rigidity is related to the source luminosity (\cref{eq:rmax_lumi}), an evolution of maximum rigidities, $L(z)\to\rmax(z)$, would be introduced. The impact of such an evolution is studied in \cref{sec:results_dRdz}, where we find that the cosmic-ray fit is not very sensitive to this behaviour but negative evolutions, i.e.\ $\rmax\propto z^{\zeta}\,,\zeta<0$, are preferred at moderate significance.} we can write
        \begin{equation}
            \frac{\dif p}{\dif L} = \frac{y_2 - 1}{L_0}\,\left(\frac{L}{L_0}\right)^{-y_2}\,.
        \end{equation}
        We assume that the emitted flux of a single source scales with the luminosity as $\phi_\src\propto L/L_0$. Noting \cref{eq:rmax_lumi}, this introduces an additional dependency of the source flux on the maximum rigidity, which can be absorbed into the $\prmax$ distribution by adjusting the definition of the effective slope $\bp$. The distribution of maximum rigidities is then
        \begin{equation}
            \prmax = \frac{2\,(y_2 - 1)}{R_0}\,\left(\frac{\rmax}{R_0}\right)^{-2y_2 + 3},
        \end{equation}
        which, except for an additional normalisation constant $\kappa$, reduces to the PL expression in \cref{eq:dpdRmax_spl} after defining
        \begin{equation}
            \bp=2y_2-3\quad\text{and}\quad \kappa = \frac{\bp+1}{\bp-1}\,.
        \end{equation}
        The situation is more complex if sources follow a broken power-law luminosity function $\dif p/\dif L(y_1,y_2)$. By defining $\rOne=2y_1-3,\,\text{and}\,\rTwo=2y_2-3$, it is possible to express the maximum rigidity distribution as a broken power law.
        
        The expected values of $\bp$ (labelled $\beta_{\rm pop, max}$) are listed in the last column of \cref{tab:beta_key} for a selection of possible source candidates. These can be compared directly to the fitted values of $\bp$ discussed below. For the investigated sources $\beta_{\rm pop, max}$ is in general low, meaning that we would expect to observe the effect of the variance of the population in the UHECR data. It should be kept in mind that the estimates given in this section are only a lower limit on the source variance (upper limit on $\bp$) as we focused only on the variation of a few key parameters and treated others as a constant (e.g.\ $R_0$) and therefore the real source variance will be larger. In addition, for sources observing a broken power-law distribution the variance predicted from \cref{eq:Rq90_def} with the tabled $\bp$ will underestimate the true population variance since only sources above the break are considered if the approximation as a single power law is made (case III.2 of \cref{tab:beta_key}). Depending on the distribution of sources below the break, the true variance can be much larger if $\beta_1$ is not small. The discrepancy is smaller if the sub-break distribution is inverted ($\beta_1 < 0$) and approaches zero for $\beta_1\to-\infty$, in which case the expression reduces to a single power law (\cref{eq:dpdRmax_spl}).

        \begin{table*}
    	\caption{\label{tab:beta_key}
            Effective fit parameters, $\bp$, the spectral index of the maximum rigidity distribution of the UHECR source population, $\gs$, the assumed spectral index of the UHECR spectrum of individual sources, and their interpretation in terms of source properties for various scenarios considered in this work.
            The scenarios are: (I) a distribution of $\prmax$ that follows an \textit{ad hoc} single power law (PL) or broken power law (BPL); (II) maximum rigidity that scales as $\rmax\propto\Gamma^\alpha$ with $\Gamma$ the PL-distributed bulk Lorentz factor of the acceleration region (see \cref{sec:intro_lorentz_factor}) with $\dif p/\dif\Gamma\propto\Gamma^{-\eta}$; and (III) $\rmax$ as a function of source luminosity, $\rmax\propto\sqrt{L}$, with PL or BPL distribution of $\dif p/\dif L$ (see \cref{sec:intro_luminosity}). For power-law distributions the parameter in brackets denotes the slope, e.g.\ $\dif p/\dif\Gamma(\eta)\propto\Gamma^{-\eta}$, while for broken power-law distributions the parameters give the slope before and after the break, respectively. Scenario I represents our baseline model that we use to compute the population spectra for different source spectral cutoff functions. Cases II and III can be reduced to the former after re-interpretation of the source and population parameters $(\eta,\alpha,y_1,y_2,\rOne,\rTwo,$ and $\gs)$ in terms of the parameters $\beta_{\rm pop, max}$ and $\gp$. For scenario III we quote the slopes for $\dif p/\dif L$, as opposed to $\dif p/\dif\log(L)$ which is often used in the literature. This introduces a factor of $L^{-1}$, i.e.\ $y_{1/2}\to y_{1/2}+1$.}
    		\renewcommand{\arraystretch}{1.3}
    		\begin{ruledtabular}
    		\begin{tabular}{l|ll|l|l|l|l}
    		    ID      & Param.    & Distribution      & $\bp$     & $\gp$     & Sources  & $\beta_{\rm pop, max}$  \\ \toprule
    			I.1     & $\rmax$   & PL, $\prmaxvar{\bp}$       & $\bp$     & $\gs$     &  &          \\
    			I.2     & $\rmax$   & BPL, $\prmaxvar{\rOne,\rTwo}$   &          &           &   &             \\
    			        &           & $\rOne<1$           & $\approx\rTwo$   & $\approx\gs$  &     &   \\
    		            &           & $\rOne>1$           & $\rTwo-\rOne+1$     & $\gs+\rOne-1$  &     &  \\ \hline
    			II      & $\rmax\propto\Gamma^\alpha$  & PL, $\dif p/\dif\Gamma(\eta)$     & $(\eta-1)/\alpha+2$    & $\gs$    & Blazars~\cite{2019ApJ...874...43L}\,\footnotemark[1]: $\eta=1.4$\tiny{$\pm{0.2}$} & \\
    			        &           &       &  $-\gs+\xi/\alpha$ &           & + Hillas: $\quad\,\alpha=1,~\xi=1$    & $3.4$\tiny{$\pm0.2$}\small{$-\gs$}  \\
    			        &           &       &           &           & + Espresso: $\alpha=2,~\xi=0$     & $2.2$\tiny{$\pm0.1$}\small{$-\gs$}   \\ \hline
    		    III.1   & $\rmax\propto\sqrt{L}$       & PL, $\dif p/\dif L(y_2)$   & $2\,y_2-3$  & $\gs$     & BL\,Lacs~\cite{Ajello:2009ip}\,\footnotemark[2]: $\quad\, y_2=2.61$\tiny{$\pm0.37$}   & $2.22$\tiny{$\pm0.74$}   \\
    		    &           &                   &                     &                 & FSRQs~\cite{BASS:2022gdj}\,\footnotemark[2]: \quad\, $y_2=2.36$\tiny{$\pm0.10$}  & $1.72$\tiny{$\pm0.20$}   \\
    		    &           &                   &                     &                 & Blazars~\cite{BASS:2022gdj}\,\footnotemark[2]: $\quad\,~ y_2=2.32$\tiny{$\pm0.08$}  & $1.64$\tiny{$\pm0.16$}   \\
    		    &           &                   &                     &                 & TDEs~\cite{vanVelzen:2017qum,Lin:2022jvw}: $\quad~ y_2=2.30$\tiny{$\pm0.20$}  & $1.60$\tiny{$\pm0.40$}   \\
    		    III.2  & $\rmax\propto\sqrt{L}$       & BPL, $\dif p/\dif L(y_1,y_2)$ &         &           &     &  \\
    		        	&           & $y_1<2$           & $\approx2\,y_2-3$   & $\approx\gs$ 
    		            & GRBs~\cite{Wanderman:2009es}: ~~~$y_1=1.2$\tiny{$^{+0.2}_{-0.1}$}\small{$,\,y_2=2.4$}\tiny{$^{+0.3}_{-0.6}$}  & $1.8$\tiny{$^{+0.6}_{-1.2}$}   \\
    		            &           &                   &                     &                 & FSRQs~\cite{BASS:2022gdj}\,\footnotemark[2]: $y_1=0$\tiny{$\pm2.07$}\small{$,\,y_2=2.67$}\tiny{$\pm0.17$}  & $2.34$\tiny{$\pm0.34$}   \\
    		            &           &                   &                     &                 &
    		            Blazars~\cite{BASS:2022gdj}\,\footnotemark[2]: $y_1=0.49$\tiny{$\pm1.15$}\small{$,\,y_2=2.79$}\tiny{$\pm0.19$}  & $2.58$\tiny{$\pm0.38$}   \\
    		            &           &                   &                     &                 & Seyferts~\cite{Ueda:2014tma}: ~$y_1 = 1.96$\tiny{$\pm0.04$}\small{$,\,y_2=3.71$}\tiny{$\pm0.09$}  & $4.42$\tiny{$\pm0.18$}   \\
    		\end{tabular}
    		\end{ruledtabular}
    		\footnotetext[1]{A steeper distribution of $\eta=2.1\pm0.4$ was found in~\cite{Saikia:2016blk} when fitting only blazars with $\Gamma=1-40$, resulting in reduced population variance of $\bp=4.1\pm0.4$ for the Hillas and $\bp=2.6\pm0.2$ for the Espresso scenario.}
    		\footnotetext[2]{Assuming the pure-luminosity-evolution (s/m)PLE model.}
	    \end{table*}

\section{Methods}\label{sec:methods}
\subsection{UHECR Data}
    We use the latest publicly available data from the Pierre Auger Observatory for comparison with our numerical simulations. These are the energy spectrum of UHECRs from Ref.~\cite{PierreAuger:2020qqz}, and the
    mean and standard deviation of the maximum depth of air showers~\cite{PierreAuger:2014sui, Yushkov:2020nhr} that are sensitive to the composition
    of cosmic rays, see e.g.~\cite{Kampert:2012mx}.

\subsection{UHECR Propagation}
    UHECR injection and propagation are simulated with the numerical Monte Carlo framework {\scshape CRPropa3}~\cite{Batista:2016yrx}, including the production of cosmogenic neutrinos and gamma rays. Upper limits and measurements of the latter are qualitatively taken into account in what follows. These are the Fermi-LAT isotropic diffuse gamma-ray background~\cite{Fermi-LAT:2014ryh}, the observed IceCube high-energy starting-event neutrino flux~\cite{Kopper:2017zzm}, and the IceCube $90\%$ upper limits above \SI{5e6}{\giga\electronvolt}~\cite{IceCube:2018fhm}.  The UHECR sources are simulated in the continuous-source approximation out to maximum redshift $z_\text{max} = 4$.
    For UHECRs in the energy range that we fit, the effective horizon is much closer at no more than $z\lesssim1$, e.g.~\cite{AlvesBatista:2018zui}, but sources at larger distances can have a strong impact on the predicted flux of cosmogenic neutrinos.

    All relevant interactions are taken into account during propagation~\cite{Batista:2016yrx,PierreAuger:2016use,AlvesBatista:2015jem}; these are (i) redshift energy loss, (ii) photo-pion production and (iii) electron-positron pair production on the cosmic microwave background (CMB) and the infrared background (IRB~\cite{Gilmore:2012}), and for heavier cosmic rays also (iv) photo-disintegration on CMB \& IRB and (v) nuclear decay.

    We assume that UHECRs propagate in the ballistic regime and neglect the effects of extragalactic magnetic fields on the trajectories of UHECRs. Based on the results of previous studies~\cite{Aloisio:2004jda,Gonzalez:2021ajv,Mollerach:2013dza,Mollerach:2020mhr,Globus:2007bi,Wittkowski:2017ZK}, these propagation effects are mainly important at low  rigidities, where they can lead e.g.\ to an apparent hardening of the UHECR flux at Earth, but are not expected to alter our conclusions about the source variance of maximum rigidities.

\subsection{Model Fit}
    We compare the model-predicted UHECR spectrum and composition after propagation to observations by the Pierre Auger Observatory. For that purpose we convert the model composition into the air shower observables -- the mean depth of the shower maximum $\mxmax$ and its standard deviation $\sxmax$ -- following Ref.~\cite{PierreAuger:2013xim}.

    The agreement between simulation and observations is evaluated with a standard $\chi^2$ estimator plus additional penalty terms
    \begin{align}\label{eq:chi2}
        \chi^2 = \sum_{E_i\geq E_\text{min}}\left(\frac{d_i-m(E_i,{\bf p})}{\sigma_\text{stat}(d_i)}\right)^2
                    + \chi^2_\text{UL} + \chi^2_\text{zero} + \chi^2_\text{shifts}\,,
    \end{align}
    and minimized adjusting the model parameters ${\bf p}$.
    The sum runs over all Auger data points at energies $E_i$ above the threshold energy. Here, $d_i$ denotes the three measured quantities, i.e.\ the energy spectrum, average $\xmax$ and standard deviation of $\xmax$. We select a high value of $E_\text{min}=10^{18.8}\,\si{\electronvolt}$ as the  minimum fitted UHECR energy to reduce the impact of a possible low-energy cosmic-ray component, different from the one responsible for the highest energies (Hillas'  ``component B''~\cite{Hillas:2005cs}). We have verified that the results are consistent within uncertainties under small changes of $E_\text{min}$.
    
    The smallest set of free fit parameters $\bf{p}$ are the minimum rigidity $R_0$ and slope $\bp$ of the single power-law distribution of maximum rigidities, source spectral index $\gs$, total population emissivity $L_0$, and elemental injection fractions $f_A^R$ which are defined as relative flux ratios at the same rigidity\,\footnote{Fractions are sometimes defined at the same energy instead of rigidity. A transformation between the two parametrisations can be achieved via $f_A^R = f_A^E\cdot Z(A)^{-\gs+1}$.}. A combination of five injection elements - $^1$H, $^4$He, $^{14}$N, $^{28}$Si and $^{56}$Fe - is used as an effective approximation of mass groups in the cosmic-ray composition.
    
    For alternative ``interaction limited'' scenarios, e.g.~\cite{Ptitsyna:2008zs}, a scaling of the maximum energy with the CR mass is typically predicted. Yet, for most stable injection elements, except hydrogen, there is an approximately constant relation between CR mass and charge, with $A/Z\approx2$, and the transition from $E_\text{max}\propto Z$ to $E_\text{max}\propto A$ would thus only result in a re-scaling of the reference rigidity $R_0$ by a constant factor. For proton injection, where $A/Z\neq2$, the maximum energy will not follow the same linear scaling but will be offset by a relative factor of $(A/Z)_\text{p}/(A/Z)_{\text{elem}\neq\text{p}}$. However, since in our fits this component consistently sits below the ankle, the impact on the results is expected to be negligible.

    We treat spectral data points below the threshold as one-sided $\chi^2$ penalty terms that only contribute to the overall goodness-of-fit if the predicted flux exceeds the observations. This component is denoted as $\chi^2_\text{UL}$ in \cref{eq:chi2} and is analogous to the first term but only evaluated if the model exceeds the data.

    No cosmic rays were observed in the two highest-energy bins at $E\geq\,10^{20.2}\,\si{\electronvolt}$ and only $90\%$ upper limits are given in \cite{PierreAuger:2020qqz}. The $\chi^2$ penalty derived for this type of zero-event data points follows from the asymptotic $\chi^2$ term assuming a Poissonian distribution of events~\cite{Baker:1983tu}, and is estimated as
    \begin{equation}
        \chi^2_\text{zero} =\sum_{i=1}^{\rm ULs} 2n_i^\text{model},
    \end{equation}
    where $n_i^\text{model}$ is the number of particles predicted by the simulation at the energy $E(\text{UL}_i)$ after taking into account the detector exposure~\cite{PierreAuger:2020qqz}.

    Finally, we consider the systematic uncertainties in the absolute scale of the energy, $\mxmax$ and $\sxmax$. They are included in the fit as nuisance parameters with
    \begin{equation}
        \chi^2_\text{shifts} = \sum_{k\in\{E,\mxmax,\sxmax\}} \left(\frac{\delta_k}{\sigma_k}\right)^2,
    \end{equation}
    where the energy uncertainty is assumed as $\sigma_E\leq14\%$~\cite{Dawson:2020bkp}, and the shower depth uncertainties are taken directly from the data set~\cite{Yushkov:2020nhr,PierreAuger:2014sui}. The scale shifts can be fitted freely in the range $\delta_k\in[-\sigma_k,\sigma_k]$, but unless indicated otherwise, we fix the values of the systematic shifts $\delta_k$ to fiducial values, as detailed below.

\section{Results and Discussion}\label{sec:results}
\subsection{Fiducial Model}\label{sec:results_fiducial}
    The observed variance of $\xmax$ consists of two separate contributions: (i) shower-to-shower variations, and (ii) intrinsic shower variability. To allow for stronger source diversity, it is necessary that $\sxmax$ is not already dominated by the latter. Air showers originating from light primary cosmic rays have a larger variability \cite{PierreAuger:2013xim}, and intrinsic shower variations are minimised by shifting the observational data to the heaviest composition that is allowed within systematic uncertainties -- corresponding to an adjustment of $\mxmax$ by about \SI{-8.5}{\gram\per\centi\meter\squared} at the ankle and \SI{-7}{\gram\per\centi\meter\squared} at the highest energies \cite{Yushkov:2020nhr,PierreAuger:2014sui}. In addition, we shift the observed variance of the shower maximum up by the systematic uncertainty to allow for the largest reasonable source variance.  We adopt these shifts as our fiducial model to allow for maximum intrinsic source diversity. An adjustment of the energy scale is also possible, but we found our conclusions to be invariant under such a shift and neglect it for simplicity.

    We furthermore select {\scshape Sibyll2.3}c~\cite{Riehn:2017mfm} as our default hadronic interaction model to convert the predicted UHECR composition to mean shower depth and variance. This is motivated by the fact that interpreting the $\xmax$ data of Auger with {\scshape Sibyll} yields  a heavier composition than when using {\scshape Epos-LHC}~\cite{Pierog:2013ria}. As default, we assume a flat redshift evolution of the source density and an exponential source cutoff. Alternative redshift evolutions are explored in \cref{sec:results_dndz} and different cutoff functions in \cref{sec:results_other}.

    In line with our outlined considerations, we find that the fiducial scale shifts allow for increased population variance compared to the un-shifted observations, as shown in \cref{tab:results_baseline}. This is true independent of the choice of hadronic interaction model.

    \begin{figure}
		\centering
		\includegraphics[width=\linewidth]{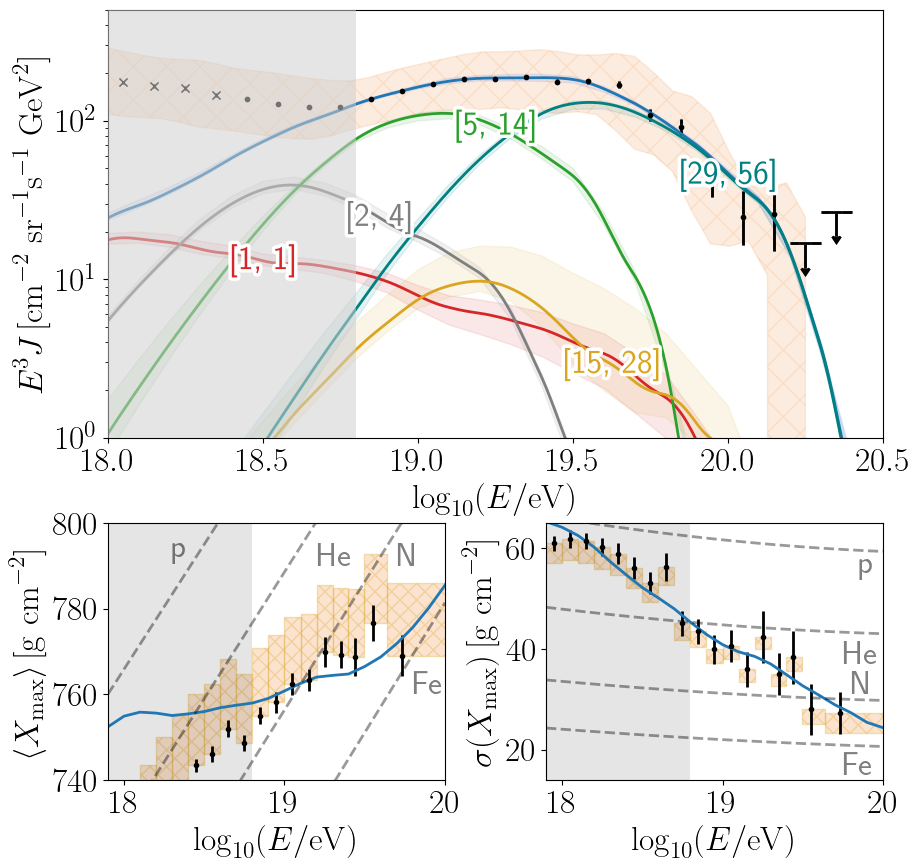}
		\caption{Predicted spectrum and composition at Earth for the best-fit scenario of the fiducial model ({\scshape Sibyll2.3}c, $\mxmax-\sigma_\text{syst}$, $\sxmax+\sigma_\text{syst}$). The coloured bands indicate the contributions of the separate mass groups with $[A_\text{min},A_\text{max}]$, including the $68\%$ uncertainties (1 d.o.f.). Hatched areas indicate systematic uncertainties of the data. Data points at $E<10^{18.4}\,\si{\electronvolt}$ (crosses) are taken from~\cite{Verzi:2019AO} and are only shown for visual guidance. Only points above $10^{18.8}\,\si{\electronvolt}$ are used in the fit.}\label{fig:fiducial_spectrum}
	\end{figure}
	\begin{figure}
		\centering
		\includegraphics[width=\linewidth]{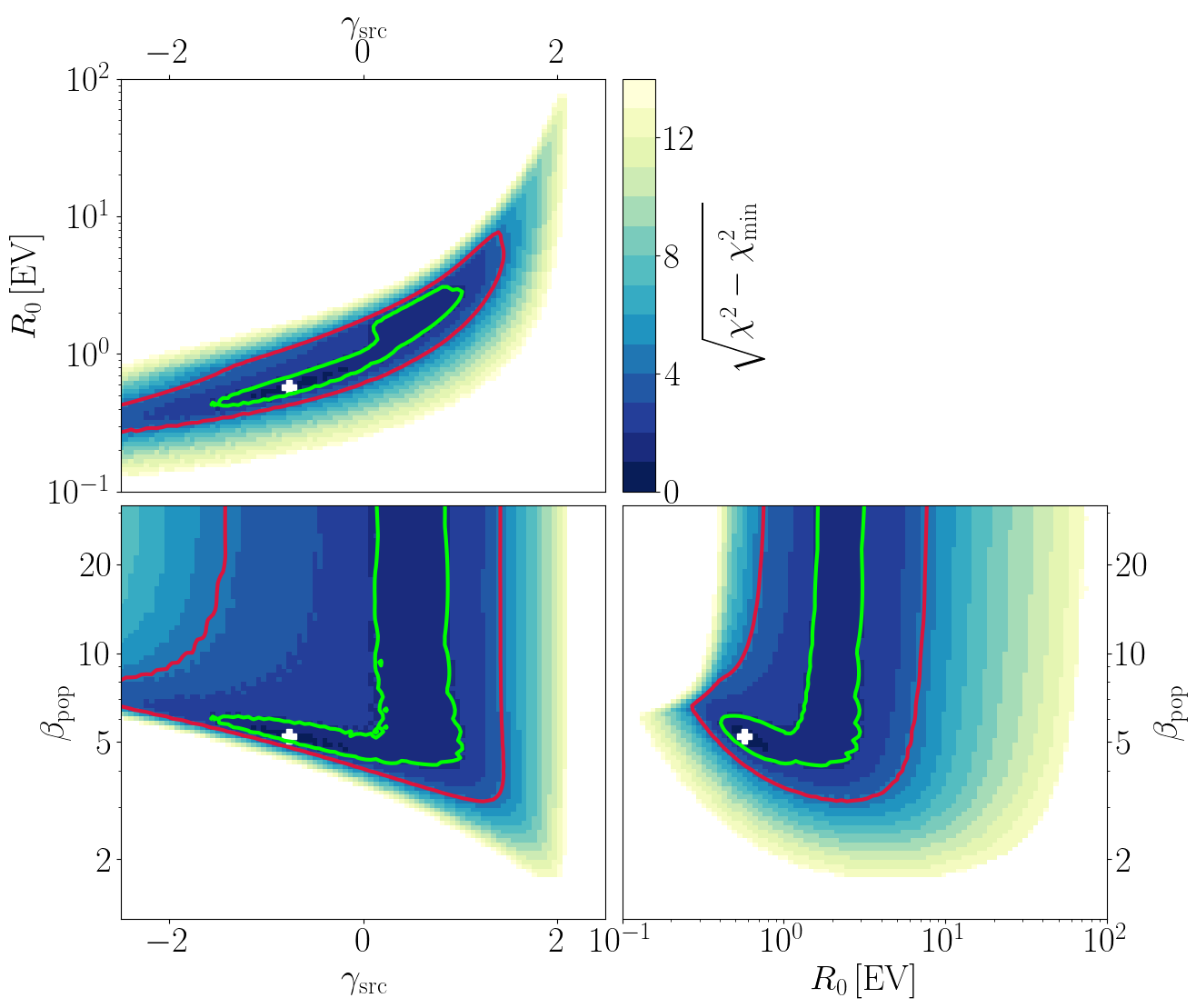}
		\caption{Results of the source parameter scan for the fiducial model marginalised along all but two axes respectively. The surface plot shows the agreement between prediction and Auger observations in terms of the $\chi^2$ estimator and the contour lines indicate the one (green) and three (red) sigma confidence interval for two degrees of freedom. The best fit is marked with a white cross.}\label{fig:fiducial_grid}
	\end{figure}
    The predicted spectrum and composition at Earth for the best-fit source parameters of the fiducial model are shown in \cref{fig:fiducial_spectrum}. The viable range of $\bp$ is sharply bounded from below (\cref{fig:fiducial_grid}), approximately as
    \begin{equation}\label{eq:index_condition}
        \bp\gtrsim-\gs+4\,
    \end{equation}
    and appreciable source diversity is only possible for soft source spectra $\gs\gtrsim1$. Yet, when assuming acceleration following a Peters cycle, soft source spectra imply a significant amount of mixing between the different mass groups, which leads to an increase in shower variance, a prediction that is in tension to the low values of $\sxmax$ measured by the Pierre Auger Collaboration, and the resulting agreement between simulation and observations is poor. This problem is exacerbated for the soft spectra expected from diffusive shock acceleration, and values above $\gs\approx2$ are excluded.

    Realisations of the source model where \cref{eq:index_condition} is violated are characterised by a population spectrum with an extreme UHE tail; a consequence of $\lim_{R\to\infty}(\phi_\pp)\propto R^{-\bp-\gs+1}$. Such extremely-UHE cosmic rays experience strong interactions during propagation, producing a large flux of light secondary cosmic rays with energies up to the Greisen-Zatsepin-Kuzmin (GZK)~\cite{Greisen:1966jv,Zatsepin:1966jv} limit. In combination with the remaining non-disintegrated component of heavy primaries, the predicted flux at Earth exhibits a large amount of mixing between the mass groups, which leads to strong intrinsic shower variance in excess of observations and an overall bad fit to the observed spectral shape.

    The parameter space where a good fit to the measured UHECR spectrum and composition is achieved can be divided into two different regimes; one that runs approximately parallel to the boundary with $\bp+\gs\approx4-6$ in the range $\gs\in[-1,0.5]$, and a second that is effectively degenerate in the population variance $\bp\gtrsim5$ with $\gs\in[0,1]$. The former is associated with a sub-EV maximum rigidity threshold $R_0$ and a heavy composition dominated by nitrogen-like nuclei with little contribution from lighter elements. The second regime allows for a lighter composition of up to $\sim50\%$ protons/helium with $R_0\sim \SI{2}{\exa\volt}$. Only the second region is present in the scenario without fiducial scale shifts applied. In both regimes, sources are effectively identical and population variance of half a decade or more, i.e.\ $\rmax^{0.90}/R_0 > 5$, and $ \bp\lesssim2.4$, can be excluded at a confidence level above $6\sigma$\,\footnote{A penalty factor $S$ that corrects for the quality of the global best-fit point $\chi^2_\text{min}$ is included in this estimate. We adopt the form $S^{-1}=\sqrt{\chi^2_\text{min}/N_\text{d.o.f.}}$ proposed in~\cite{Rosenfeld:1975fy}. In essence, the penalty factor reduces the rejection strength of sub-optimal fit points if the overall best fit is poor.}. With {\scshape Epos-LHC}, we obtain increased source variance at the best fit $\Delta\bp\approx-0.8$
    but at the cost of a significantly reduced fit quality.

    A statistically significant lower limit on the population variance cannot be established, and identical sources cannot be rejected.

    \begin{table}
        \caption{\label{tab:results_baseline}
        Best-fit parameters for several variations of the source model. From left to right: the base scenario with {\scshape Sibyll2.3}c as air shower model and no shifts of the energy- and $X_\text{max}$-scales of the data; our fiducial model, {\scshape Sibyll2.3}c and the $\langle X_\text{max}\rangle$ / $\sigma(X_\text{max})$ data points shifted by $-1/+1\,\sigma_\text{syst}$ respectively; the same scale shifts but with {\scshape Epos-LHC} as air shower model. The injection fractions are given in descending order for p, He, N, Si and Fe. An asterisk indicates that the confidence interval extends to the edge of the scan range and the parameter is not properly constrained in that direction.}
        \renewcommand{\arraystretch}{1.3}
        \begin{ruledtabular}
        \begin{tabular}{l|rrr}
			Model	                & {\scshape Sibyll2.3}c     & {\scshape Sibyll2.3}c     & {\scshape Epos-LHC}    \\
			                        & (no shifts)               & (fid. shifts)             & (fid. shifts) \\
			\toprule
			$R_0$ [EV]	            & $1.73^{+0.20}_{-0.18}$    & $0.57^{+1.88}_{-0.11}$	& $1.6^{+0.6}_{-0.4}$\\
			$\bp$		            & $29.9^{+1.7*}_{-18.1}$    & $5.2^{+26.4*}_{-0.5}$     & $4.4^{+0.5}_{-0.5}$\\
			$\gs$		            & $-0.23^{+0.18}_{-0.26}$   & $-0.8^{+1.4}_{-0.5}$      & $0.1^{+0.4}_{-0.5}$\\  \hline
			$f_A^R [\%]$			& $0^{+0}_{-0}$             & $0^{+36.4}_{-0}$          & $0^{+0}_{-0}$          \\
			                        & $58.1^{+0.4}_{-1.9}$      & $0^{+51.3}_{-0}$          & $36.9^{+7.4}_{-22.8}$ \\
			                        & $35.0^{+1.6}_{-0.2}$      & $93.7^{+0.5}_{-53.5}$     & $50.3^{+16.3}_{-5.4}$ \\
			                        & $5.7^{+0.5}_{-0.6}$       & $0.3^{+7.7}_{-0.3}$       & $11.3^{+6.6}_{-3.8}$  \\
			                        & $1.16^{+0.12}_{-0.11}$    & $6.0^{+0.2}_{-3.8}$       & $1.41^{+0.27}_{-0.04}$  \\ \hline
			$\rmax^{0.90}$ $[R_0]$  & $1.083^{+0.155}_{-0.005}$ & $1.72^{+0.13}_{-0.64}$    & $1.97^{+0.22}_{-0.17}$  \\
			$\chi^2 / \text{d.o.f.}$	& $45.0 / 26$               & $40.4 / 26$               & $56.3 / 26$           \\
        \end{tabular}
        \end{ruledtabular}
    \end{table}

\subsection{Broken Power-Law Distribution of the Maximum Rigidity}
\begin{figure}
    	\centering
    	\includegraphics[width=\linewidth]{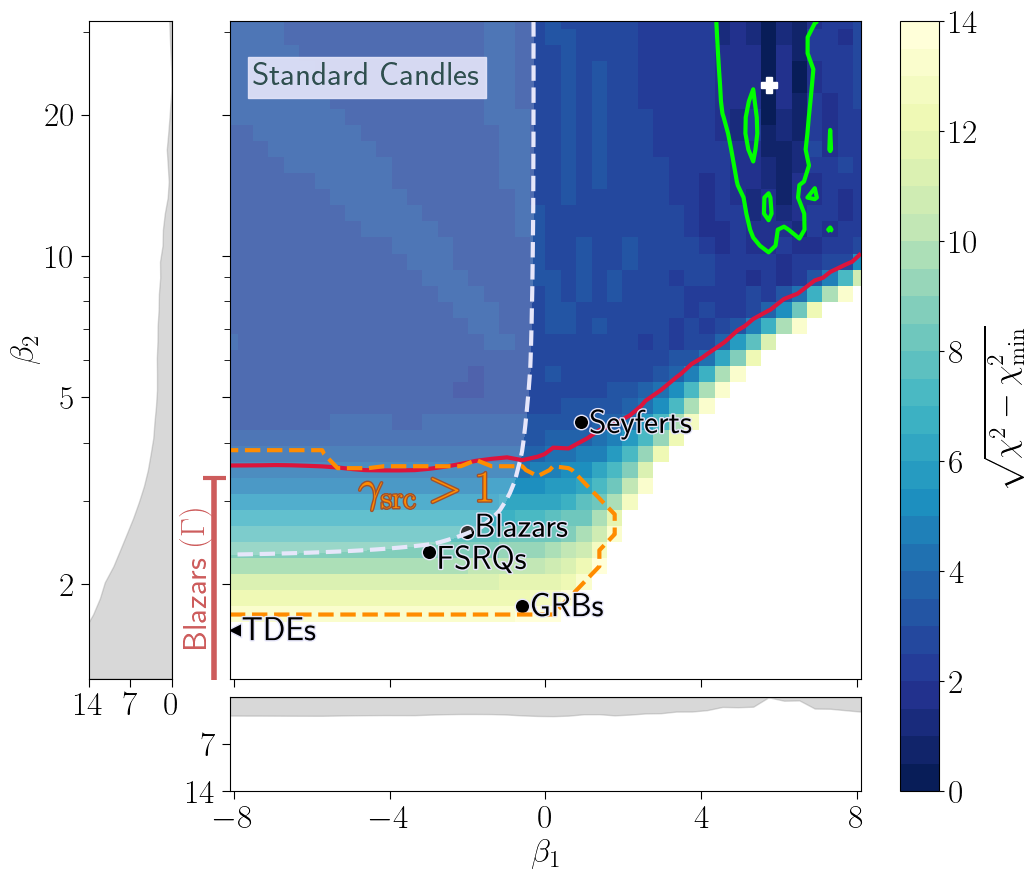}
    	\caption{Fit quality for the population model with broken power-law distribution of maximum rigidities, marginalised onto $\beta_1\times\beta_2$ space (the spectral index of the population spectrum below and beyond the break). Contours indicate the one (green) and three (red) sigma confidence intervals (2 d.o.f.). The white-shaded region denotes the parameter space where $\rmax^{0.9}/R_0<10$, i.e.\ where the spread in maximum rigidity is less than a decade for $90\%$ of sources. The best fit is marked with a white cross. Parameters of potential source classes predicted based on their luminosity functions (\cref{eq:rmax_lumi}) are shown as black points. The allowed values of $\beta_2$ for blazars in the $\rmax(\Gamma)$ scenario (\cref{eq:rmax_lorentz}) for the Hillas-constrained case and under the assumption that $\gs\geq2$ are indicated on the left side.}\label{fig:results_BPL}
	\end{figure}

\begin{figure}
    	\raggedleft
    	\includegraphics[width=0.9\linewidth]{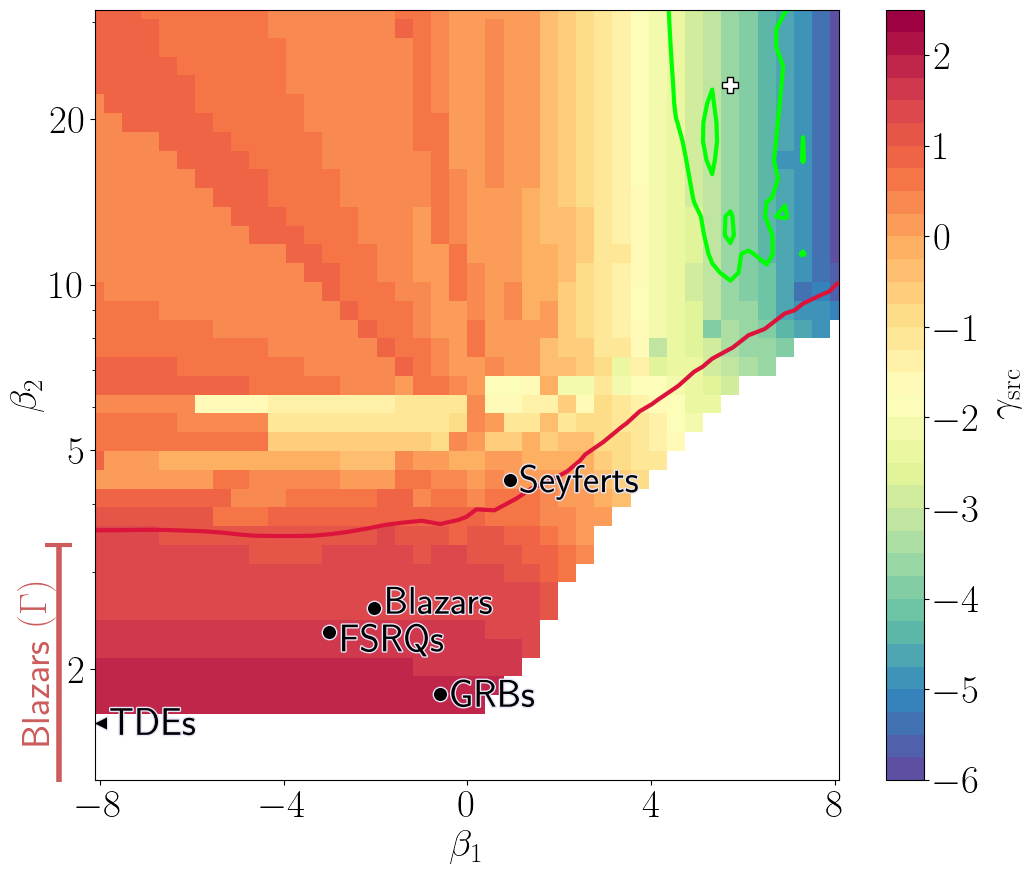}
    	\includegraphics[width=0.9\linewidth]{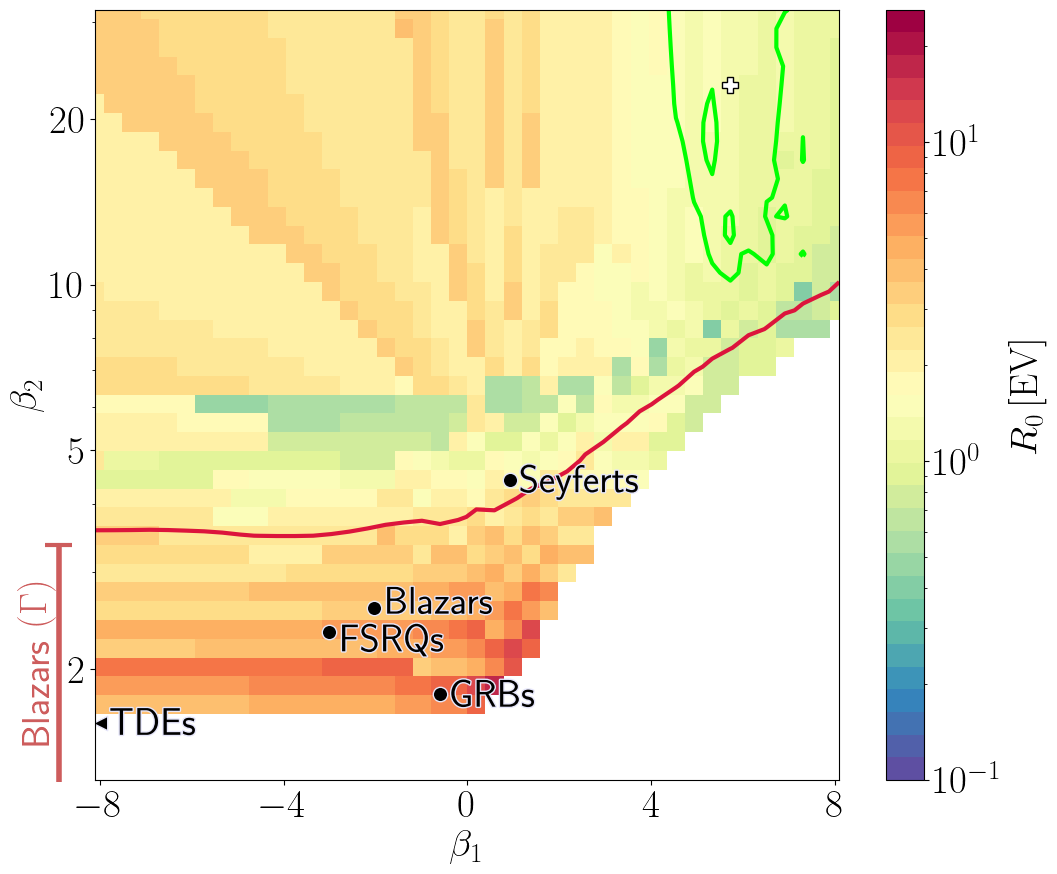}
    	\caption{Best-fit source spectral index $\gs$ (\textbf{top}) and break rigidity $R_0$ (\textbf{bottom}) as a function of $\beta_1$ and $\beta_2$ for the population model with a broken power-law distribution of maximum rigidities. Best-fit confidence contours and source candidates are indicated as in \cref{fig:results_BPL}. The structures in the upper-left of the plots are artefacts from the limited resolution of our sampling grid and some degeneracy in the choice of $(R_0,\gs)$ for a particular combination of $(\beta_1,\beta_2)$.}\label{fig:results_BPL_b1xb2_parameters}
	\end{figure}

    We find that a broken power-law distribution in $\rmax$ can fit the observed UHECR flux in the regime of highly non-identical sources $(\beta_1\gtrsim1)$ if
    \begin{enumerate}
    \item[I.a] The $\rmax$ distribution satisfies $$\beta_2 \gtrsim \beta_1 + 3$$ as can be seen in \cref{fig:results_BPL} and, 
    \item[I.b] The effective index of the rigidity spectra of the sources satisfies
    $$\gp=1.22^{+0}_{-0.04},$$
    as illustrated in \cref{fig:results_BPL_b1xb2_parameters}. 
    \end{enumerate}
    This suggests that the effective population spectra $R^{-\gp}$ must be moderately soft and the break must steepen the $\rmax$ distribution by at least $\times \rmax^{-3}$. Assuming reasonable source spectra, i.e.\ $\gs>0\,(>1)\,[>2$], we can place an upper limit on the steepness of the sub-break $\rmax$ distribution of $\beta_1 \lesssim 2.2\,(1.2)\,[0.2]$.
    
    If sources are in the regime of limited population variance $(\beta_1\lesssim1)$ we find that they must satisfy 
    \begin{enumerate} 

    \item[II.a] \hspace{2.5cm} $\beta_2\gtrsim4.5$,\\
    and 
    \item[II.b] \hspace{2cm} $\gs=-0.6^{+1.2}_{-1.0},$\\
    \end{enumerate} 
    consistent with the results for standard-candle sources. In summary, the key constraints that must be observed by potential source classes for $68\%$ of realisations are
    \begin{align}\label{eq:bpl_criteria}
        \gs &=
        \begin{cases}
                    2.22^{+0}_{-0.04} &\text{for }\beta_1\geq1,  \\
                    -0.6^{+1.2}_{-1.0} &\text{otherwise,}
        \end{cases}\\
        \beta_2 &\geq
        \begin{cases}
                    \beta_1 + 3 &\text{for }\beta_1\geq1,  \\
                    4.5 &\text{otherwise.}
        \end{cases}
    \end{align}
    
    Our results clearly reflect the correlation of $\gs$ and $\beta_1$ expected from \cref{eq:index_redefinition_ys}. Large population variance (large $\beta_1$) is possible if individual source spectra are hard ($\gs\lesssim1$). Pareto distributions ($\beta_2\ll1$) with sufficient flatness to produce substantial source variance ($\beta_2\lesssim3$) are disfavoured at $4.7\sigma$. All studied astrophysical source classes (\cref{tab:beta_key}) are located in a region of the parameter space where ``normal'' (non-inverted) spectral indices are preferred (\cref{fig:results_BPL_b1xb2_parameters}) based on the values of $\beta_1$ and $\beta_2$ that we can infer from the studied luminosity functions. However, we found the predicted maximum rigidity distributions to be generally incompatible with the constraints of the UHECR fit. Only for Seyfert-like galaxies is the predicted $\rmax$ distribution above the break approximately compatible with the fit to the UHECR data.
    
    The horizontal band at $\beta_2\approx6$ in \cref{fig:results_BPL_b1xb2_parameters} corresponds to a secondary family of viable solutions separate from the global minimum, with sub-EV break rigidity and harder injection spectra.
    
    The best-fit parameters of the BPL and fiducial PL fit are compared in \cref{tab:results_other}. The more-general, broken power-law approach provides an improvement of $\Delta\chi^2\approx5.7$ -- which corresponds to a weak preference with respect to a single power-law description at $2\sigma$.
    

\subsection{Redshift Evolution of the Source Density}\label{sec:results_dndz}
    For simplicity, we have so far assumed the distribution of sources to be flat in redshift. However, the most probable source classes of UHECRs do not exhibit this behaviour but have densities evolving as a function of redshift. A common parametrisation of the evolution is given by
    \begin{equation}\label{eq:dnzd}
        n(z) =
        \begin{cases}
			(1+z)^m&\text{for }m\leq0, \\
			(1+z)^m&\text{for }m>0\text{ and }z<z_0,\\
			z_0^m&\text{for }m>0\text{ and }  z_0 < z < z_\text{max},\\
			0 & \text{otherwise}
		\end{cases}
    \end{equation}
    which captures the general trends for the expected source classes: a power-law increase or decrease in density up to some break point and an approximate flattening above that. Positive redshift evolution ($m>0$) is observed e.g.\ for active galactic nuclei~\cite{Hasinger:2005sb} and gamma-ray bursts \cite{Sun:2015bda, Wanderman:2009es}, and negative evolution for some BL\,Lac sub classes~\cite{Ajello:2013lka} and tidal disruption events~\cite{Sun:2015bda,Guepin:2017abw,Kochanek:2016zzg}. Source densities following the star formation rate are approximately reproduced for $m=3.4$~\cite{Hopkins:2006bw}.

    We study the effect of source density evolution on the allowed level of population variance by evaluating the fiducial model also for redshift evolutions of $m=-3,3,6$ with $z_0=1.5$ and $z_\text{max}=4$. Results are shown in \cref{tab:results_dndz}.

    Best agreement with observations is found for predominantly local sources ($m=-3$), and a continuous decrease in fit quality is identified for stronger density evolutions. The improved fit for small $m$ is driven primarily by a better match of the observed composition, in particular $\mxmax$, but the difference in $\chi^2$ only becomes large once $m>3$. Redshift evolutions stronger than $m=6$ could be excluded at more than $3\sigma$ based on their cosmogenic neutrino signature by future neutrino detectors such as GRAND-200k~\cite{GRAND:2018iaj} or IceCube Gen2~\cite{IceCube:2019pna}.

    We find a clear anti-correlation between source density evolution $m$ and spectral index $\gs$, in agreement with previous studies, e.g.\ \cite{Unger:2015laa,PierreAuger:2016use,AlvesBatista:2018zui,Heinze:2019jou}. This is caused by the, on average, larger source distance for stronger density evolutions and consequently increased interactions during propagation. Since interactions soften the spectrum, a harder injection spectrum is required at the sources. The same argument applies to the progressively heavier source composition at the best fit. For strong evolution, the viability of the $\bp$-degenerate regime of the fiducial model is reduced and the $\bp+\gs\approx4$ regime is preferred more strongly but shifted to harder source spectra $\gs<0$. Extremely identical sources are disfavoured in this case because they would lead to a worse description of the observed spectral shape and an underestimation of the shower variance.

    As established previously for the fiducial model, there exists an approximate boundary of $\bp>-\gs+4$, which dictates that larger population variance requires softer source spectra. Local source distributions allow for softer spectra, and the source density redshift evolution and population variance are therefore positively correlated in the sense that smaller values of $m$ allow for smaller values of $\bp$.

    To summarise, negative redshift evolutions of the source density provide many attractive benefits: (i) a quantitatively better fit to the observed UHECR spectrum and composition, (ii) lighter required injection composition, (iii) more natural spectral indices $\gs>0$, and (iv) a potentially higher, but still not very large, population variance. This makes classes of astrophysical objects with a negative redshift evolution, such as tidal disruption events~\cite{Sun:2015bda,Kochanek:2016zzg} and high-spectral-peak BL\,Lacs~\cite{Ajello:2013lka}, appealing as sources of ultra-high-energy cosmic rays.

    \begin{table}
		\caption{\label{tab:results_dndz}
		Best-fit parameters for the fiducial model but with different source density redshift evolutions, where $n(z)\sim(1+z)^m,\,z<1.5$ (see \cref{eq:dnzd}). The second column ($m=0$) is identical to the fiducial scenario presented in \cref{tab:results_baseline}. The injection fractions are given in descending order for p, He, N, Si and Fe. An asterisk indicates that the confidence interval extends to the edge of the scan range and the parameter is not properly constrained in that direction.}
		\renewcommand{\arraystretch}{1.3}
		\begin{ruledtabular}
		\begin{tabular}{l|rrrr}
			Redshift        &                           &                           &                           &   \\
			evolution $m$   & -3                        & 0                         & 3                         & 6 \\
			\toprule
			$R_0$ [EV]	    & $0.80^{+1.88}_{-0.16}$	& $0.57^{+1.88}_{-0.11}$    & $0.46^{+0.05}_{-0.09}$    & $0.52^{+0.06}_{-0.05}$\\
			$\bp$		    & $4.4^{+23.9}_{-0.5}$      & $5.2^{+26.4*}_{-0.5}$     & $6.46^{+0.36}_{-0.34}$    & $6.46^{+0.36}_{-0.34}$\\
			$\gs$		    & $0.2^{+0.8}_{-0.4}$       & $-0.8^{+1.4}_{-0.5}$	    & $-2.0^{+0.4}_{-0.5}$	    & $-2.24^{+0.35}_{-0.18}$\\ \hline
			$f_A^R [\%]$    & $3.5^{+46.8}_{-3.5}$      & $0^{+36.4}_{-0}$          & $0^{+0.01}_{-0}$          & $0^{+0}_{-0}$  \\
			                & $8.7^{+49.8}_{-8.7}$      & $0^{+51.3}_{-0}$          & $2.6^{+17.0}_{-2.6}$      & $0^{+0}_{-0}$  \\
			                & $81.3^{+11.5}_{-46.7}$    & $93.7^{+0.5}_{-53.5}$     & $90.5^{+2.0}_{-16.2}$     & $38.5^{+1.8}_{-15.7}$ \\
			                & $1.7^{+3.7}_{-0.8}$	    & $0.3^{+7.7}_{-0.3}$       & $0^{+0.9}_{-0}$           & $53.0^{+16.2}_{-3.9}$ \\
			                & $4.8^{+0.8}_{-2.8}$	    & $6.0^{+0.2}_{-3.8}$       & $6.8^{+0.5}_{-1.3}$       & $8.5^{+2.1}_{-0.5}$ \\ \hline
			$\rmax^{0.90}$ $[R_0]$  &  $1.97^{+0.22}_{-0.88}$   & $1.72^{+0.13}_{-0.64}$   & $1.53^{+0.04}_{-0.04}$  & $1.53^{+0.04}_{-0.04}$ \\
			$\chi^2 / \text{d.o.f.}$	&  $37.3 / 26$              & $40.6 / 26$           & $42.5 / 26$           & $68.9 / 26$ \\
		\end{tabular}
		\end{ruledtabular}
	\end{table}

\subsection{Redshift Evolution of the Maximum Rigidity}\label{sec:results_dRdz}
     In addition to the interactions with ambient photon fields, cosmic rays lose energy due to the adiabatic expansion of the Universe, with $E_\text{obs}=E_\text{inj}/(1+z)$. For a population of sources, this will lead to different effective maximum rigidities for sources at different distances and result in a naturally broadened population spectrum at Earth, even in the limit of identical sources.

    We have previously assumed that the distribution of maximum rigidities $\prmax$ does not depend on distance. However, most classes of astrophysical objects exhibit larger luminosities at higher redshifts~\cite{Hasinger:2005sb,Sun:2015bda,Wanderman:2009es}. If the maximum rigidity and luminosity of a cosmic-ray source are connected, as outlined in \cref{sec:intro_luminosity}, then $\rmax$ should also evolve as a function of redshift. We study this scenario by evolving the starting point of the $\rmax$ distribution with redshift,
    \begin{equation}
        R_0(z) = R_0\,(1+z)^q,~q\in{\rm I\!R}\,.
    \end{equation}
    In the limit of $q=0$, we obtain the default no-redshift-scaling case while for $q=1$ adiabatic losses are exactly compensated, and sources would have the same effective maximum rigidity at all redshifts. Overcompensation ($q>1$) and even enhancement of local sources ($q<0$) are also possible.

    \begin{figure}
		\centering
		\includegraphics[width=\linewidth]{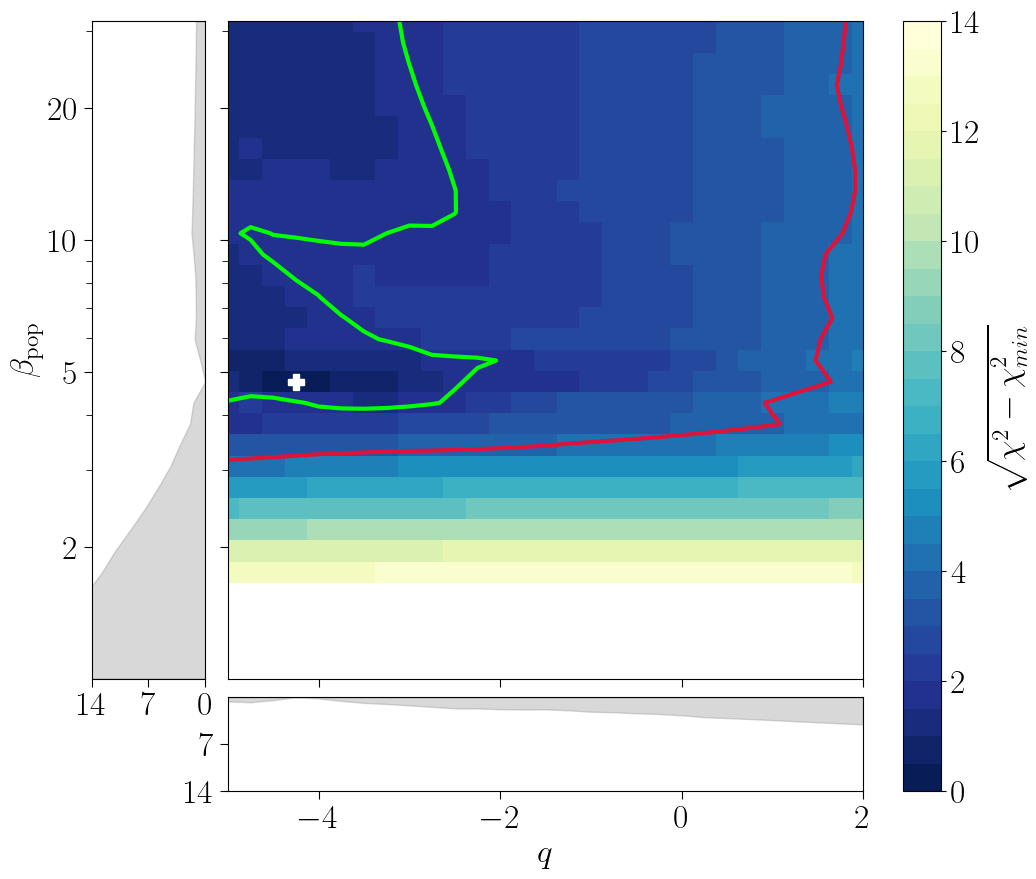}
		\caption{Results of the source parameter scan for the population model with redshift evolution of the distribution of maximum rigidities marginalised onto $\bp\times\gs$ space. The agreement between prediction and Auger observations in terms of the $\chi^2$ estimator is displayed with different colour levels, and the contour lines indicate the one (green) and three (red) sigma confidence intervals for two degrees of freedom. The best fit is marked with a white cross.}\label{fig:results_drmaxdz_slice}
	\end{figure}

    We find that the cosmic-ray fit has only moderate sensitivity to the value of $q$, and no appreciable correlation with $R_0$, $\bp$ or $\gs$ is observed. Nevertheless, negative evolutions are preferred, with the best fit at $q=-4.3^{+1.0}_{-0.8*}$, and positive values of $q\geq1$ excluded at $3\sigma$ confidence level. The difference is explained by a marginally better fit of the mean shower depth and a better description of the observed spectral shape, which is related to the stacking of contributions from different redshift shells.

    Intuitively, the largest possible population rigidity variance should be allowed for a redshift scaling of $\rmax(z)=\rmax^{z=0}\cdot(1+z)$ as this would compensate the intrinsic broadening of the maximum rigidity termination via adiabatic energy losses. This expectation is not reflected in the results (\cref{fig:results_drmaxdz_slice}), and we find a lower limit of $\bp^\text{LL}\approx4$ almost independent of $q$. A weak positive correlation of $\bp^\text{LL}$ and $q$ can be observed, especially for the $3\sigma$ contour. This trend is related to the reduced fragmentation of heavy cosmic rays during propagation if the highest energies are reached only at the most local sources. As a consequence, negative evolutions of $\rmax(z)$ allow for more extrinsic mixing of the mass groups due to non-identical sources.

\begin{figure}
	\centering
	\includegraphics[width=\linewidth]{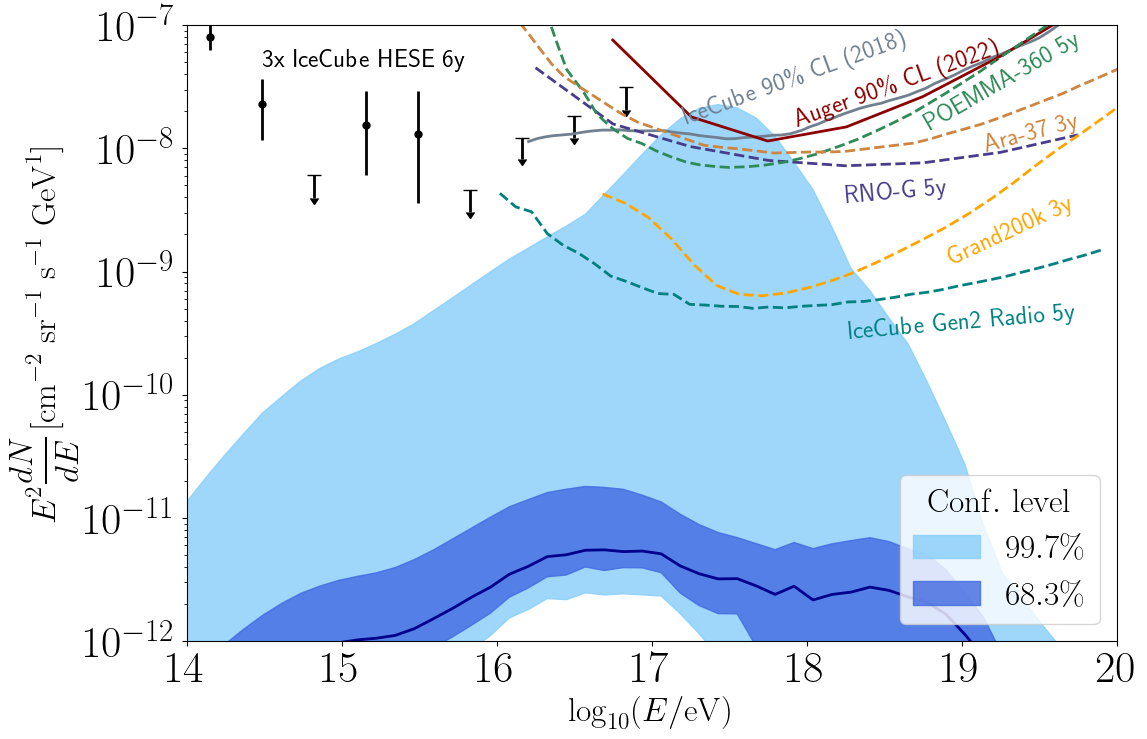}
        \caption{Predicted cosmogenic neutrino flux associated with the best fit (solid line) of the population model with redshift evolution of $\prmax$. The shaded one and three sigma uncertainty bands correspond to the contours in \cref{fig:results_drmaxdz_slice}. We show the observed IceCube HESE flux~\cite{Kopper:2017zzm}, upper limits from IceCube~\cite{IceCube:2018fhm} and Auger~\cite{Pedreira:2021gcl}, and predicted sensitivities of planned detectors \cite{IceCube:2019pna,GRAND:2018iaj,IceCube:2019pna,ARA:2015wxq,Cummings:2020ycz} as a reference.}\label{fig:results_drmaxdz_neutrinos}
    \end{figure}

    Perhaps the most tantalising result of this scan is not the precise best fit but rather the realisation that strongly positive rigidity-redshift scalings lead to an important multimessenger signature in the form of a large flux of cosmogenic high-energy and ultra-high-energy neutrinos (\cref{fig:results_drmaxdz_neutrinos}). If the evolution is strong, then the more distant, high-$\rmax$ sources are screened from our view in cosmic rays because of the interactions experienced during propagation. At the same time, these interactions produce a large flux of cosmogenic neutrinos that can reach us even from high redshifts. Around the peak at $E_\nu\approx10^{17.4}\,\si{\electronvolt}$ more than $95\%$ of the predicted neutrino flux is produced by UHECRs from sources at redshift $z>1$. Based on this prediction, existing UHE limits of IceCube~\cite{IceCube:2018fhm} and Auger~\cite{Pedreira:2021gcl} are able to constrain the redshift evolution of maximum rigidities to $q\lesssim2$. However, we stress that this high neutrino flux is only obtained as the most extreme scenario within $3\sigma$ and only for a low number of specific realisations. With the increased sensitivity of future detectors, such as IceCube Gen2 and GRAND-200k, this upper limit can be reduced to $q\lesssim1$ assuming a non-detection of UHE neutrinos.
    
    The expected neutrino spectra differ substantially from predictions that are obtained without redshift scaling of the maximum rigidity both in shape and magnitude, especially for non-identical sources and for strongly positive evolutions of $\rmax$. The $99.7\%$ upper limit (2 d.o.f.) in \cref{fig:results_drmaxdz_neutrinos}, with a peak in the neutrino flux between \SIrange{0.1}{1}{\exa\electronvolt} of up to $E^2\,\dif N/\dif E\gtrsim10^{-8}\,\si{\giga\electronvolt\per\centi\meter\squared\per\steradian\per\second}$, is obtained for approximately $R_0=\SI{1.8}{\exa\volt},\,\bp=14.5,\,\gs=0.3,\,q=2$, and injection fractions $f_A^R:0,0.020,0.875,0.055,0.050$ ($^1$H,$^4$He,$^{14}$N,$^{28}$Si,$^{56}$Fe). The large value of $\bp$ means that population variance is not an important ingredient for the large predicted neutrino flux in this scenario. The positive redshift evolution of $\rmax$ decreases the quality of the UHECR fit. For the same source parameters but no evolution ($q=0$), and matching injection fractions, the fit improves by $\Delta\chi^2=-7.9$.
    
    We have verified that the predicted neutrino flux is in qualitative agreement with Refs.\ \cite{AlvesBatista:2018zui,Heinze:2019jou} if no redshift evolution of $\rmax$ and the same maximum source distance are assumed. Our upper limit of the cosmogenic neutrino flux is similar in magnitude to the flux predicted from interactions in the source environment in Ref.~\cite{Muzio:2021zud} but offset in energy by a factor of $\sim10$.
    
    Also visible in \cref{fig:results_drmaxdz_neutrinos} is a new ``shoulder'' feature at ultra-high neutrino energies beyond the classical CMB-induced peak. This feature is produced by cosmic rays with trans-GZK energies from the highest-energy tail of the population spectrum and is also present in the other scenarios we have studied. The extent and magnitude are linked to the strength of the UHECR tail and consequently to the amount of population variance.

    The neutrino spectra in \cref{fig:results_drmaxdz_neutrinos} are derived under the assumption of a source density that does not evolve as a function of distance. For sources that are more abundant in the local Universe the predicted flux is reduced while it is enhanced for most other realistic source classes (AGN~\cite{Hasinger:2005sb}, GRB~\cite{Sun:2015bda,Wanderman:2009es}).

\subsection{Other Variations of the Source Model}\label{sec:results_other}
    We have considered additional variations of the source model to study the impact on the allowed level of population diversity in maximum rigidity (\cref{tab:results_other}). They are described briefly in the following. In all scenarios, a small but generally non-zero level of population variance on the order of $\bp\sim4-6$ is recovered at the best fit to Auger observations. The largest amount of source diversity -- the most extreme case -- is obtained for negative redshift evolution of the source density and Heaviside cutoff of the source spectra, which yields $\bp\sim3-3.5$ depending on the choice of hadronic interaction model.
    \paragraph{Minimum Source Distance}
        Cosmic rays are attenuated during propagation depending on their source distance. UHECRs with energies around the cutoff and with the observed heavy composition are expected to reach us only from relatively local sources. To avoid artifacts in the simulations we have chosen $z_\text{min}=10^{-3}$, or about \SI{4.3}{\mega\parsec}, as minimum source distance. Setting a larger minimum distance, i.e.\ \SI{43}{\mega\parsec}, we observe a decrease in fit quality for our fiducial model. This is because nearby sources contribute primarily at the highest energies. If these are removed, sources from larger distances must compensate for the loss in flux. However, for a Peters cycle progression of maximum rigidities with a preference for low maximum rigidities $\mathcal{O}(\SI{1}{\exa\electronvolt})$ this compensation must come predominantly from heavier elements since only they can reach the required energies. Because of increased interaction due to the larger source distance, this also leads to the production of a substantial flux of lighter secondary cosmic rays and a stronger mixing of the mass groups -- in contrast with observations. The tension can be partially mitigated when sources are essentially identical and the high-rigidity tail is very small. This shift to larger values of $\bp$ for the best fit is observed in our simulations; however, the viable range is not affected strongly, and the lower limit remains approximately the same.
    \paragraph{Source Cutoff Function}
        Sources with an exponential UHE cutoff already include an intrinsic dispersion in the maximum rigidity of the produced cosmic rays even for a single source. This contribution is reduced for sharper-than-exponential cutoffs and becomes zero for a sudden, Heaviside-like limit. An increased level of population variance should therefore be expected for sources with a steeper cutoff function.

        As proposed in \cref{eq:source_super_expon,eq:pop_super_expon}, a super-exponential cutoff can be assumed with adjustable exponent $\lc$. Exactly Heaviside-like sources are obtained only for $\lc\to\infty$ but effectively the population spectra become very similar already for $\lc\gtrsim2$. Beyond that point, the difference manifests mainly in the increasing sharpness of the break at $R_0$.

\begin{figure}
    		\centering
    		\includegraphics[width=\linewidth]{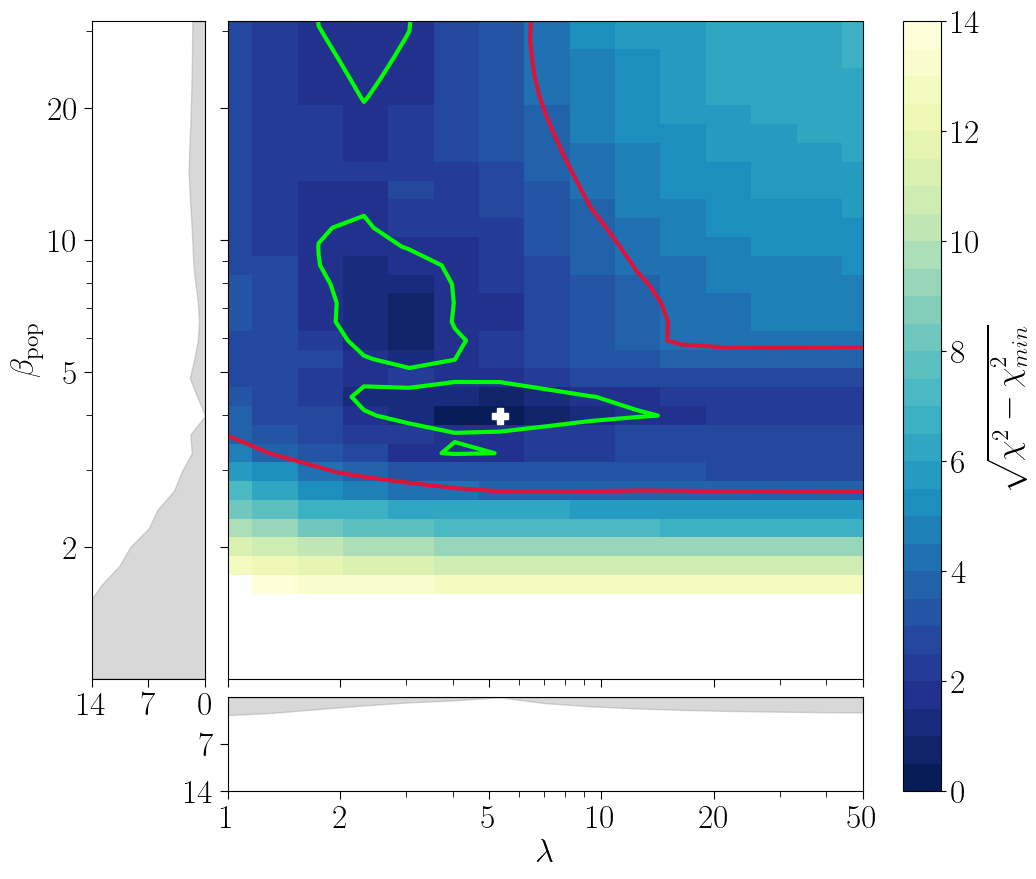}
    		\caption{Results of the source parameter scan for the population model with super-exponential source cutoff function, with exponent $\lc$, marginalised onto $\bp\times\lc$ space. The surface plot shows the agreement between prediction and Auger observations in terms of the $\chi^2$ estimator and the contour lines indicate the one (green) and three (red) sigma confidence interval for two degrees of freedom. The best fit is marked with a white cross.}\label{fig:results_super_exp_slice}
	    \end{figure}

        Re-simulating the fiducial model for a range of exponents, $\lc\in[1,50]$, we find the global best fit at $\lc=5.4^{+1.7}_{-2.3}$, which indicates effectively Heaviside-like sources. Overall the fit exhibits only moderate sensitivity to the precise shape of the cutoff except when it is close to an ordinary exponential (\cref{fig:results_super_exp_slice}). The latter is disfavoured at a level of $2.3\sigma$. As expected from previous results, the population variance is poorly constrained for sources with exponential cutoff function, but even for close-to-Heaviside source cutoffs an upper limit cannot be placed -- only at $\lc\gtrsim8$ does a significant preference of intermediate diversity $\bp\sim4$ emerge. This reveals substantial dependence of the population variance on the precise shape of the spectrum at the break. We note that for approximately Heaviside-like sources the best-fit shifts to softer source spectra and hard spectra of $\gs\lesssim1$ are disfavoured. This is in better agreement with the expectation from diffusive shock acceleration.
        The correlation between the sharpness of the cutoff and rigidity variance is not as strong as expected, and even sources with effectively instantaneous limit do not allow for diversities much greater than $\bp\sim3.5$.

    \paragraph{Fixed Injection Composition}
        We investigate the scenario with injection fractions reflecting the composition observed for Galactic cosmic rays (GCRs). It is peculiar to note that an adequate fit of the UHECR data can be achieved with (near-)GCR-like composition if fractions are specified at the same energy~\cite{Unger:2015laa,Rodrigues:2020pli,Kimura:2017ubz}, however, this is not the case if fractions at the same rigidity are used. In the latter case, the abundance of heavy elements is insufficient to power the UHECR flux at the highest energies, or in other words, the predicted composition at UHE is much lighter than indicated by observations.
        
        We argue that rigidity fractions are the correct choice since the most relevant interactions for cosmic rays, such as electromagnetic re-acceleration, magnetic horizon effects, and approximately also disintegration, are universal in rigidity and thus preserve the rigidity fractions but not the energy fractions. The fit is possible with fractions at the same energy since then the effective composition can be modified by changing the spectral index, due to $f_A^R=f_A^E\, Z^{-\gs+1}$.
        
        We select the GCR composition at a fixed rigidity ($f_\text{gal}^\text{R}:\,0.794,0.190,0.013,0.002,0.001$)~\cite{ParticleDataGroup:2020ssz} as our baseline scenario but allow for a free-floating rescaling of the metallicity, i.e.\ the abundance of elements heavier than helium, by a factor of $\mu\in[1,100]$. With this model, the best agreement with observations is obtained for hard source spectra $\gs\approx0.7$ and population variance of $\bp\approx4.9$, which is similar to the source variance allowed in our fiducial scenario. At the best fit, a rescaling of the injection metallicity by a factor of approx.\ 60 is required, although values as low as $\mu\approx20$ are allowed within the $3\sigma$ confidence interval.

    \begin{table}
		\caption{\label{tab:results_other}
		Best-fit source spectral index $\gamma_\text{src}$ and maximum rigidity variance $\beta_\text{pop}$ plus corresponding $\chi^2$ for different variations of the source model. These are (fd) the fiducial model, (bp) broken power-law distribution of $\prmax$, (zr) redshift evolution of $\prmax$, (zn) redshift evolution of the source density, (zm) larger minimum source distance, (sc) super-exponential source cutoff function, (fg) relative injection fractions similar to the composition observed for Galactic cosmic rays, and (ex) the extreme scenario that yields the largest amount of population variance with negative redshift evolution of the source density ($m=-3$) and Heaviside rigidity cutoff at the source. The best-fit values of the additional free parameters are, $\beta_1=5.7^{+0.8}_{-0.4}\,\&\,\beta_2=23.1^{+8.5*}_{-10.8}$ for (bp), $q=-4.3^{+1.0}_{-0.8}$ for (zr), $\lambda=5.4^{+1.7}_{-2.3}$ for (sc), and $\mu=64.0^{+10.3}_{-8.8}$ for (fg). Confidence intervals that reach a limit of the scan range are marked with an asterisk.}
		\renewcommand{\arraystretch}{1.3}
		\begin{ruledtabular}
		\begin{tabular}{ll|rrr}
		    Model   & Parameter & $\beta_\text{pop}$                     & $\gamma_\text{src}$                     & $\chi^2$  \\ \toprule
		    fd  &           & $5.2^{+26.4*}_{-0.5}$     &   $-0.8^{+1.4}_{-0.5}$    & $40.4$    \\
		    bp  & $\beta_1,\beta_2$ & $18.4^{+8.5}_{-11.2}$ & $-3.5^{+0.2}_{-0.8}$    & $34.7$    \\ \hline
			zr  & $q\in[-5,2]$ & $4.8^{+26.9*}_{-0.5}$  &   $-0.19^{+0.89}_{-0.18}$    & $33.7$    \\
			zn  & $m=-3$    & $4.4^{+23.9}_{-0.5}$      &   $0.2^{+0.8}_{-0.4}$     & $37.3$  \\
			    & $m=3$     & $6.46^{+0.36}_{-0.34}$    &   $-2.0^{+0.4}_{-0.5*}$   & $42.5$  \\
			    & $m=6$     & $6.46^{+0.36}_{-0.34}$    &   $-2.24^{+0.35}_{-0.18}$    & $68.9$  \\
			zm  & $z_\text{min}=0.01$   & $29.9^{+1.7*}_{-25.5}$    &   $0.38^{+0.18}_{-1.22}$     & $46.2$   \\   \hline
			sc  & $\lambda\in[1,50]$    & $4.0^{+3.2}_{-0.4}$   &   $1.43^{+0.16}_{-0.16}$ & $33.6$  \\
			fg  & $f_A^R$   & $4.9^{+0.5}_{-0.5}$       &   $0.73^{+0.16}_{-0.16}$     & $45.5$  \\
            ex  & {\scshape Epos-LHC}   & $3.17^{+0.18}_{-0.17}$    &   $1.43^{+0.09}_{-0.09}$     & $40.6$  \\
                & {\scshape Sibyll2.3}c & $3.5^{+0.6}_{-0.5}$       &   $1.69^{+0.09}_{-0.09}$     & $34.7$  \\
		\end{tabular}
		\end{ruledtabular}
	\end{table}

\section{Conclusion}\label{sec:conclusion}
We have performed the first systematic investigation of the allowed population variance in maximum UHECR rigidity. To this end, we have derived analytical expressions for the population spectrum of an ensemble of non-identical UHECR sources, assuming a (broken-)power-law distribution of maximum rigidities $\prmax\propto\rmax^{-\bp}$ and different choices of the spectral high-energy cutoff at the sources. For the first time, we have integrated this approach into a fit of the energy spectrum and composition data to quantify the constraints on source similarity from existing observations by the Pierre Auger Observatory.

If maximum rigidities are distributed according to a power law with a sudden start (Pareto distribution) -- which appears as a suitable choice for several source candidates (e.g.\ AGN, blazars, TDEs) based on the observed luminosity functions or under the assumption of power-law distributed Lorentz factors -- our results show that sources are required to be effectively identical if only Auger data at the
nominal energy and composition scale are considered. After adjusting the measured mean shower depth and variance within systematic uncertainties to the favourable directions that result in the heaviest composition interpretation, we find that large yet finite values of $\bp\sim5$ are preferred, corresponding to a dispersion in maximum rigidity of sources by a factor of approximately two.

Increased levels of population variance up to $\bp\sim3-4$ are possible for sources with sharp UHE cutoff and for source densities evolving negatively with redshift. Even then, maximum rigidities do not differ between sources by more than a factor of a few. In contrast, if sources are more abundant at larger redshifts they are required to be more identical because the preferred source spectrum becomes harder with redshift due to increased interactions during propagation. Since the population spectrum behaves as $\lim\limits_{R\to\infty}(\phi_\pp)\propto R^{-\gs-\bp+1}$ a smaller source diversity (larger $\bp$) is required to limit the strength of the spectral UHE tail.

For some source classes (e.g.\ GRBs, blazars, Seyferts), the luminosity function motivates a broken power-law distribution of maximum rigidities. In this scenario, the population variance can be large, driven by sources below the break rigidity $R_0$, provided the break is sharp and the spectral index of individual sources is sufficiently hard to counteract the variance introduced by the non-identical sources. This requires hard spectral indices of $\gs\lesssim1.2$ if the rigidity distribution below the break is softer than $\beta_1 \sim 1$. In addition, for any value of $\beta_1$, the $\rmax$ distribution must steepen at the break by at least $\rmax^{-3}$. For $\beta_1\to-\infty$ we obtain the power-law scenario as an asymptotic limit, with $\gs\lesssim1$ and $\beta_2\gtrsim4$.

We have derived the UHECR population spectra of plausible astrophysical source classes by connecting luminosity and maximum rigidity via the Lovelace--Blandford--Waxman relation~\cite{Lovelace:1976,Waxman:1995vg,Waxman:2001tk,Blandford_2000,Lemoine:2009pw,Kachelriess:2022phl}, \cref{eq:rmax_lumi}. For all proposed source classes, the preferred spectral index corresponding to their respective distribution $p(\rmax|\beta_1,\beta_2)$ is in the physically plausible range of $\gs=0-2$ if they produce UHECRs with a maximum rigidity distribution which follows the one we have derived using their observed luminosity functions and \cref{eq:rmax_lumi}. 
The predicted UHECR population spectra produced by blazars, tidal disruption events, and GRBs are inconsistent with the UHECR data within our formalism. 
Seyfert-like galaxies are the only investigated population (see \cref{tab:beta_key}) with sufficiently steep post-break slope to explain the required small variance at ultra-high energies. However, a hard spectral index of $\gs\approx0-0.5$ is necessary. The variance in maximum rigidity obtained using \cref{eq:rmax_lumi} represents a lower limit. Additional variance of the maximum rigidity is expected as a consequence of the distribution of other relevant source properties that we have not considered here, which will likely reduce this compatibility.

In summary, we have found that the maximum rigidity distribution of UHECR sources is remarkably narrow, necessitating nearly identical (``standard-candle'') sources, or a sharp cutoff in the rigidity distribution of the UHECR source population. In the latter case, the low-rigidity tail exacerbates the need for hard injection spectra that has been exposed by prior studies which performed a combined fit of UHECR observations. Our results place strong constraints on the most plausible astrophysical source classes of UHECRs.  

Alternatively, it is possible that exotic mechanisms limit the maximum rigidities of accelerators to the same value, e.g.~\cite{Anchordoqui:2022ejw}, or that the observed flux of UHECRs is dominated by a single local source. Such a single- or few-source scenario seems however incompatible with the observed level of anisotropy of the cosmic-ray arrival directions at UHE unless deflections of cosmic rays in the Galactic and extragalactic magnetic fields are much larger than commonly expected. An analysis of the effect of cosmic variance is beyond the scope of this paper, but we comment that the fitted scenarios result in typical maximum rigidities that correspond to cosmic-ray energies below the onset of photo-nuclear interactions with the cosmic microwave background radiation. Thus the energy-loss lengths of nuclei are large, and the volume of UHECR sources contributing to the flux at Earth can be $\mathcal{O}(\text{Gpc}^3)$.

\begin{acknowledgments}
We thank the anonymous referee for the constructive report. We would like to thank Jonathan Biteau, Damiano Caprioli, Bj\"orn
Eichmann, Glennys Farrar, Michael Kachelrie\ss{}, Ioana Mari\c{s} and Marco
Muzio for useful feedback on this study.  This work was made possible
by Institut Pascal at Université Paris-Saclay during the Paris-Saclay
Astroparticle Symposium 2021, with the support of the P2IO Laboratory
of Excellence (program “Investissements d’avenir”
ANR-11-IDEX-0003-01 Paris-Saclay and ANR-10-LABX-0038), the P2I
research departments of the Paris-Saclay university, as well as
IJCLab, CEA, IPhT, APPEC, the IN2P3 master project UCMN and EuCAPT. We
also wish to thank Sergio Petrera for providing an update of the
parameters of the $\xmax$ parameterisation \cite{PierreAuger:2013xim}
for {\scshape Sibyll2.3}c.
\end{acknowledgments}

\appendix

\section{Transformation of the Emitted Flux}\label{apx:flux_boosting}
The apparent brightness of a highly relativistic source depends on the angle $\theta$ between the observer and the direction
of motion and is affected by relativistic beaming (headlight effect)
and the relativistic Doppler effect. However, for charged particles,
we can assume that their direction of motion is isotropised after
emission from the source, rendering geometrical
beaming irrelevant. Only the Lorentz boost from
the rest frame of the acceleration region to the observer frame must be considered. For sources with a flux cutoff
$f_\text{cut}$, the differential flux within the jet frame is
given by
\begin{equation}
    \phi'(R')=\frac{\dif^{\,2} N'}{\dif R'\dif t'} = \phi_0'\,R'^{-\gs}\,f_\text{cut}(R'/R_0).
\end{equation}
To transform this flux into the observer frame
the following transformations need to be taken into
account
\begin{align}\label{eq:lorentz_transformation}
    R = R'\,\Gamma^\alpha~ \text{and}~ t = t'\,\Gamma^\xi.
\end{align}
Conservation of particle number implies $N = N'$.
The first transformation is the boost in energy for particles; with $\alpha=1$ for particles accelerated in the jet frame and emitted isotropically\,\footnote{For a general Lorentz boost of $\Gamma\,(1-\beta\cos\theta)$ with isotropic emission angle $\theta$ the high-energy tail of the rigidity spectrum retains its spectral shape, and we concentrate on the simpler case of a boost by $\Gamma$.}, and $\alpha=2$ for the espresso mechanism. The second transformation is due to time dilation, which also depends on the acceleration process. If production of UHECRs within the jet is considered, then the relativistic motion of the source region will stretch the observed time by a factor of $\Gamma$ ($\xi=1$). On the other hand, if an espresso-like mechanism is assumed, where the jet merely re-accelerates a pre-existing flux of cosmic rays $\dif N/\dif t$, then no dilation is expected, assuming that the rate of particles entering and exiting the jet is the same, i.e.\ $\dif N/\dif t|_\text{out}=\dif N/\dif t|_\text{in}$, and thus $\xi=0$.
The observed flux can then be written as
\begin{align}
    \phi(R) &= \frac{\dif^{\,2} N}{\dif R\dif t} = \frac{\dif^{\,2} N(R'(R))}{\dif R'\dif t'}\,\left|\frac{\dif R'}{\dif R}\right|\,\left|\frac{\dif t'}{\dif t}\right| \nonumber\\
    &= \phi_0'\,R^{-\gs} \,\Gamma^{\alpha(\gs-1)-\dila}\,f_\text{cut}{\textstyle\left(-\frac{R}{\Gamma^\alpha R_0}\right)} \nonumber\\
    &= \phi_0'\, R^{-\gs}\,\left(\textstyle\frac{\rmax}{R_0}\right)^{\gs-1-\dila/\alpha}\,  f_\text{cut}{\textstyle\left(-\frac{R}{R_\text{max}}\right)},
\end{align}
where in the last step \cref{eq:rmax_lorentz} was used. To evaluate the convolution of source spectra and \rmax distributions, the product $\phi(R, \rmax) \times \prmax$ needs to be evaluated using the \rmax distribution from \cref{eq:listerrmax}. The resulting product can be re-written in the ``usual'' form used in \cref{sec:pop}, \begin{align}
    \phi(R, \rmax)& \times \prmax =\nonumber\\
    & \phi_0 \, R^{-\gs} \,
    f_\text{cut}{\textstyle\left(-\frac{R}{R_\text{max}}\right)} \,
    \textstyle\frac{\bp-1}{R_0}\,\left( \textstyle
      \frac{\rmax}{R_0}\right)^{-\bp},
\end{align}
with definitions
\begin{equation}
    \phi_0 = \phi_0'\, \frac{\eta-1}{\eta + \alpha\,(1-\gs) -\dila - 1}
\end{equation}
and
\begin{equation}
    \bp = \frac{\eta-1}{\alpha} + 2 - \gs + \dila/\alpha.
\end{equation}
Therefore, the same analytical forms derived in \cref{sec:pop} can be used in this case, but the interpretation of $\bp$ is more complex as it depends on several source properties.
\section{Broken Power-Law Maximum Rigidity Distribution}\label{apx:broken_power-law_distribution}
    Source properties, e.g.\ luminosity (\cref{sec:intro_luminosity}), often follow broken power-law (BPL) distributions for some likely UHECR source classes rather than single power laws (PL). A general broken power-law distribution of maximum rigidities can be written as
    \begin{equation}\label{eq:dpdrmax_bpl}
        \prmax = \frac{ R_0^{-1} }{ C }
        \cdot\begin{cases}
            \left(\frac{\rmax}{R_0}\right)^{-\rOne}& \rmax \leq R_0    \\
            \left(\frac{\rmax}{R_0}\right)^{-\rTwo}&\rmax > R_0,
        \end{cases}
    \end{equation}
    with a break at $R_0$ and slope $\rmax^{-\rOne}$ ($\rmax^{-\rTwo}$) below (above). Normalisability imposes $\rOne<1$ and $\rTwo>1$. Under a physically more plausible scenario, with some minimum and maximum $\rmax$ ($\rmax^\rll/\rmax^\rul$) for the population of sources, these conditions can be relaxed and the normalisation constant can be expressed as
    \begin{equation}
        C = \frac{ 1 - \left(\frac{\rmax^\rll}{R_0}\right)^{-\beta_1 + 1} }{1 - \beta_1} + \frac{ 1 - \left(\frac{\rmax^\rul}{R_0}\right)^{-\beta_2 + 1} }{\beta_2 - 1}\,.
    \end{equation}
    If $\beta_1 < 1$ and $\beta_2 > 1$, for $\rmax^\rll\to0$ and $\rmax^\rul\to\infty$ this simplifies to
    \begin{equation}
        C = \frac{1}{1-\beta_1} + \frac{1}{\beta_2-1}.
    \end{equation}
    
    Assuming source spectra with super-exponential cutoff (\cref{eq:source_super_expon}), the associated population spectrum reads
    \begin{align}\label{eq:pp_BPL_sexp}
        \phi_\pp &= \frac{ \phi_0 }{ C\cdot\lc }\,R^{-\gs}\cdot\Big[L + H\Big]    \\
                &L = \left(\frac{R}{R_0}\right)^{-\rOne+1}\cdot\Gamma\left(\frac{\rOne-1}{\lc},\left(\frac{R}{R_0}\right)^{\lc}\right) \nonumber    \\
                &H = \left(\frac{R}{R_0}\right)^{-\rTwo+1}\cdot\gamma\left(\frac{\rTwo-1}{\lc},\left(\frac{R}{R_0}\right)^{\lc}\right), \nonumber
    \end{align}
    with ``$L$'' the contribution from sources with $\rmax$ below the break at $R_0$ and ``$H$'' for  sources above. For $\rOne\to-\infty$ a single power law is recovered. Because of the convergence properties of the incomplete gamma functions ($\Gamma,\gamma$),
    \begin{equation}
        \lim\limits_{R/R_0\to\infty} L = 0 \quad\text{and}\quad \lim\limits_{R/R_0\to0} H = 0\,;   \\
    \end{equation}
    i.e. in the limits of very small and very large rigidities, one of the terms dominates. Only in the vicinity of the break are their contributions of similar magnitude.
    
\begin{figure}
    	\centering
    	\includegraphics[width=\linewidth]{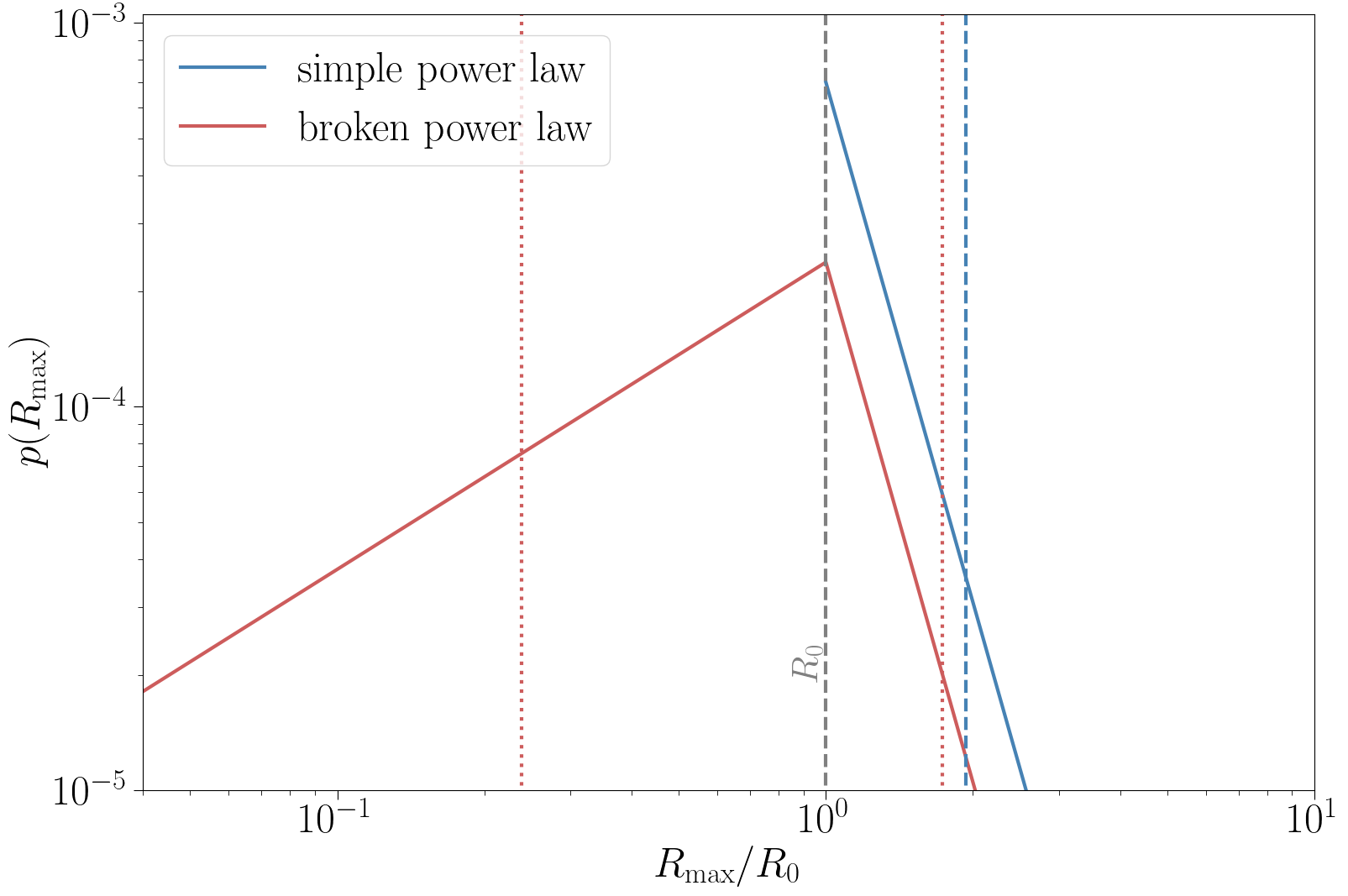}
    	\caption{Power-law (solid blue) and broken power-law (solid red) distribution of maximum rigidities $\prmax$ with break/starting-point at $R_0$, and $\gs=2,\,\rOne=-0.8,\,\rTwo=4.5$. The one-sided $90\%$ quantile of the PL (blue dashed line), and two-sided $90\%$ quantile of the BPL (red dotted lines) are also displayed. In this scenario, the PL approach underestimates the population variance by a factor of $\sim3.8$.}\label{fig:dpdrmax_SPL_vs_BPL}
	\end{figure}

    \cref{eq:pp_BPL_sexp} suggests that the slope of the population spectrum at any rigidity is completely specified by a set of three parameters, with the most obvious choice ($\gamma_\src,\beta_1,\beta_2$). However, an approximate description with only two parameters is possible, similar to the power-law scenario. The simplification is exact in the limit of very small and very large rigidities but not in the transition region where $L$ and $H$ have competitive levels. Two different cases can be identified depending on the value of $\beta_1$. The behaviour in the high-rigidity limit is unaffected by the distribution below the break and is always $\phi_\pp\propto R^{-\gs-\beta_2+1}$.
    \begin{itemize}
        \item $\beta_1 \leq 1$ : Because the probability $\prmax$ increases only slowly for $\rmax\to0$ it can be shown that $\lim\limits_{R\to0} \phi_\pp \propto R^{-\gs}$, suggesting that after interpretation $\bp=\beta_2$ we approximately retrieve the same population spectrum as for the single power law where $\beta_1 = -\infty$. The slope of the population spectrum can then be adequately described by the set of $(\gs,\bp=\beta_2)$.
        \item $\beta_1 > 1$ : The contribution from sources below the break does not vanish and we obtain $\lim\limits_{R\to0} \phi_\pp \propto R^{-\gs-\beta_1+1}$. Initially, this does not appear compatible with a PL-like description, however, re-interpretation in terms of effective population spectral index $\gp=\gs+\beta_1-1$ and effective $\rmax$-distribution $\bp=\beta_2-\beta_1+1$ leads to the usual behaviour
        \begin{align}
        \lim\limits_{R\to0} &\phi_\pp \propto R^{-\gp}   \\
        \lim\limits_{R\to\infty} &\phi_\pp \propto R^{-\gp-\beta_\pp+1}\,,
        \end{align}
        with the difference that the distribution is now characterised by the re-defined parameter set $(\gp,\bp)$.
    \end{itemize}
    The above prescriptions can be used to interpret the fit results for a PL distribution given in \cref{sec:results} for a population of sources with BPL($\rmax$). Yet, if not $\beta_1\to-\infty$, the population variance predicted from \cref{eq:Rq90_def} will underestimate the true variance since sources below the break are neglected (\cref{fig:dpdrmax_SPL_vs_BPL}). All source classes proposed in \cref{tab:beta_key} III.2 belong to the $\beta_1 < 1$ category, enabling the simplified treatment.
    
    Although we do not predict steep sub-break distributions for any evaluated astrophysical source classes, it is still useful to study the scenario where $\beta_1 > 1$. The relevant parameters are now $(\gp,\bp)$, defined as
    \begin{align}
        (a)~~ &\gp = \gs + \beta_1 - 1 \label{eq:index_redefinition_ys}\\
        (b)~~ &\bp = \beta_2 - \beta_1 + 1\,. \label{eq:index_redefinition_bp}
    \end{align}
    Solutions of this system are degenerate and different choices of $(\gs,\beta_1,\beta_2)$ can lead to the same population spectrum\,\footnote{Except for the cutoff shape at $R_0$. Combinations of steep source distributions and hard source spectra will lead to sharper cutoffs than more identical sources with softer spectra.}. One of the parameters must be fixed with additional information to break the degeneracy. This is a remarkable finding -- a similar population spectrum can be achieved for highly identical sources with softer spectral index and for diverse sources with harder spectral index. Therefore, the distribution below the cutoff can be steep and sources very diverse, provided individual sources have ultra-hard spectra so that \cref{eq:index_redefinition_ys} remains fulfilled. If this condition is satisfied and the break at $R_0$ is sufficiently strong, then the UHECR flux is dominated by sources very close to the break at $R_0$, explaining the apparently low population variance. Sources further below the break still contribute to the flux at lower rigidities by modifying the population spectrum as $\phi_\pp\propto R^{-\gs}\to R^{-\gs-\beta_1+1}$ if $\beta_1 > 1$. For flatter sub-break distributions, their contribution is approximately negligible except close to the break. This suggests that the sources of UHECRs can appear very similar even if the underlying population is diverse, provided the UHE termination of the $\rmax$-distribution is sufficiently sharp, and the distribution before the cutoff is not too steep after taking into account the hardness of individual source-spectra.

\bibliography{bibliography}

\begin{thebibliography}{101}%
\makeatletter
\providecommand \@ifxundefined [1]{%
 \@ifx{#1\undefined}
}%
\providecommand \@ifnum [1]{%
 \ifnum #1\expandafter \@firstoftwo
 \else \expandafter \@secondoftwo
 \fi
}%
\providecommand \@ifx [1]{%
 \ifx #1\expandafter \@firstoftwo
 \else \expandafter \@secondoftwo
 \fi
}%
\providecommand \natexlab [1]{#1}%
\providecommand \enquote  [1]{``#1''}%
\providecommand \bibnamefont  [1]{#1}%
\providecommand \bibfnamefont [1]{#1}%
\providecommand \citenamefont [1]{#1}%
\providecommand \href@noop [0]{\@secondoftwo}%
\providecommand \href [0]{\begingroup \@sanitize@url \@href}%
\providecommand \@href[1]{\@@startlink{#1}\@@href}%
\providecommand \@@href[1]{\endgroup#1\@@endlink}%
\providecommand \@sanitize@url [0]{\catcode `\\12\catcode `\$12\catcode
  `\&12\catcode `\#12\catcode `\^12\catcode `\_12\catcode `\%12\relax}%
\providecommand \@@startlink[1]{}%
\providecommand \@@endlink[0]{}%
\providecommand \url  [0]{\begingroup\@sanitize@url \@url }%
\providecommand \@url [1]{\endgroup\@href {#1}{\urlprefix }}%
\providecommand \urlprefix  [0]{URL }%
\providecommand \Eprint [0]{\href }%
\providecommand \doibase [0]{https://doi.org/}%
\providecommand \selectlanguage [0]{\@gobble}%
\providecommand \bibinfo  [0]{\@secondoftwo}%
\providecommand \bibfield  [0]{\@secondoftwo}%
\providecommand \translation [1]{[#1]}%
\providecommand \BibitemOpen [0]{}%
\providecommand \bibitemStop [0]{}%
\providecommand \bibitemNoStop [0]{.\EOS\space}%
\providecommand \EOS [0]{\spacefactor3000\relax}%
\providecommand \BibitemShut  [1]{\csname bibitem#1\endcsname}%
\let\auto@bib@innerbib\@empty
\bibitem [{\citenamefont {Anchordoqui}(2019)}]{Anchordoqui:2018qom}%
  \BibitemOpen
  \bibfield  {author} {\bibinfo {author} {\bibfnamefont {L.~A.}\ \bibnamefont
  {Anchordoqui}},\ }\href {https://doi.org/10.1016/j.physrep.2019.01.002}
  {\bibfield  {journal} {\bibinfo  {journal} {Phys. Rept.}\ }\textbf {\bibinfo
  {volume} {801}},\ \bibinfo {pages} {1} (\bibinfo {year} {2019})},\ \Eprint
  {https://arxiv.org/abs/1807.09645} {arXiv:1807.09645 [astro-ph.HE]}
  \BibitemShut {NoStop}%
\bibitem [{\citenamefont {Alves~Batista}\ \emph
  {et~al.}(2019{\natexlab{a}})\citenamefont {Alves~Batista} \emph
  {et~al.}}]{AlvesBatista:2019tlv}%
  \BibitemOpen
  \bibfield  {author} {\bibinfo {author} {\bibfnamefont {R.}~\bibnamefont
  {Alves~Batista}} \emph {et~al.},\ }\href
  {https://doi.org/10.3389/fspas.2019.00023} {\bibfield  {journal} {\bibinfo
  {journal} {Front. Astron. Space Sci.}\ }\textbf {\bibinfo {volume} {6}},\
  \bibinfo {pages} {23} (\bibinfo {year} {2019}{\natexlab{a}})},\ \Eprint
  {https://arxiv.org/abs/1903.06714} {arXiv:1903.06714 [astro-ph.HE]}
  \BibitemShut {NoStop}%
\bibitem [{\citenamefont {di~Matteo}\ \emph {et~al.}(2021)\citenamefont
  {di~Matteo} \emph {et~al.}}]{TelescopeArray:2021gxg}%
  \BibitemOpen
  \bibfield  {author} {\bibinfo {author} {\bibfnamefont {A.}~\bibnamefont
  {di~Matteo}} \emph {et~al.} (\bibinfo {collaboration} {Telescope Array,
  Pierre Auger}),\ }\href {https://doi.org/10.22323/1.395.0308} {\bibfield
  {journal} {\bibinfo  {journal} {PoS}\ }\textbf {\bibinfo {volume}
  {ICRC2021}},\ \bibinfo {pages} {308} (\bibinfo {year} {2021})},\ \Eprint
  {https://arxiv.org/abs/2111.12366} {arXiv:2111.12366 [astro-ph.HE]}
  \BibitemShut {NoStop}%
\bibitem [{\citenamefont {Allard}\ \emph
  {et~al.}(2007{\natexlab{a}})\citenamefont {Allard}, \citenamefont {Parizot},\
  and\ \citenamefont {Olinto}}]{Allard:2005cx}%
  \BibitemOpen
  \bibfield  {author} {\bibinfo {author} {\bibfnamefont {D.}~\bibnamefont
  {Allard}}, \bibinfo {author} {\bibfnamefont {E.}~\bibnamefont {Parizot}},\
  and\ \bibinfo {author} {\bibfnamefont {A.~V.}\ \bibnamefont {Olinto}},\
  }\href {https://doi.org/10.1016/j.astropartphys.2006.09.006} {\bibfield
  {journal} {\bibinfo  {journal} {Astropart. Phys.}\ }\textbf {\bibinfo
  {volume} {27}},\ \bibinfo {pages} {61} (\bibinfo {year}
  {2007}{\natexlab{a}})},\ \Eprint {https://arxiv.org/abs/astro-ph/0512345}
  {arXiv:astro-ph/0512345 [astro-ph]} \BibitemShut {NoStop}%
\bibitem [{\citenamefont {Hooper}\ \emph {et~al.}(2007)\citenamefont {Hooper},
  \citenamefont {Sarkar},\ and\ \citenamefont {Taylor}}]{Hooper:2006tn}%
  \BibitemOpen
  \bibfield  {author} {\bibinfo {author} {\bibfnamefont {D.}~\bibnamefont
  {Hooper}}, \bibinfo {author} {\bibfnamefont {S.}~\bibnamefont {Sarkar}},\
  and\ \bibinfo {author} {\bibfnamefont {A.~M.}\ \bibnamefont {Taylor}},\
  }\href {https://doi.org/10.1016/j.astropartphys.2006.10.008} {\bibfield
  {journal} {\bibinfo  {journal} {Astropart. Phys.}\ }\textbf {\bibinfo
  {volume} {27}},\ \bibinfo {pages} {199} (\bibinfo {year} {2007})},\ \Eprint
  {https://arxiv.org/abs/astro-ph/0608085} {arXiv:astro-ph/0608085}
  \BibitemShut {NoStop}%
\bibitem [{\citenamefont {Allard}\ \emph
  {et~al.}(2007{\natexlab{b}})\citenamefont {Allard}, \citenamefont {Olinto},\
  and\ \citenamefont {Parizot}}]{Allard:2007gx}%
  \BibitemOpen
  \bibfield  {author} {\bibinfo {author} {\bibfnamefont {D.}~\bibnamefont
  {Allard}}, \bibinfo {author} {\bibfnamefont {A.~V.}\ \bibnamefont {Olinto}},\
  and\ \bibinfo {author} {\bibfnamefont {E.}~\bibnamefont {Parizot}},\ }\href
  {https://doi.org/10.1051/0004-6361:20077478} {\bibfield  {journal} {\bibinfo
  {journal} {Astron. Astrophys.}\ }\textbf {\bibinfo {volume} {473}},\ \bibinfo
  {pages} {59} (\bibinfo {year} {2007}{\natexlab{b}})},\ \Eprint
  {https://arxiv.org/abs/astro-ph/0703633} {arXiv:astro-ph/0703633 [ASTRO-PH]}
  \BibitemShut {NoStop}%
\bibitem [{\citenamefont {Allard}\ \emph {et~al.}(2008)\citenamefont {Allard},
  \citenamefont {Busca}, \citenamefont {Decerprit}, \citenamefont {Olinto},\
  and\ \citenamefont {Parizot}}]{Allard:2008gj}%
  \BibitemOpen
  \bibfield  {author} {\bibinfo {author} {\bibfnamefont {D.}~\bibnamefont
  {Allard}}, \bibinfo {author} {\bibfnamefont {N.~G.}\ \bibnamefont {Busca}},
  \bibinfo {author} {\bibfnamefont {G.}~\bibnamefont {Decerprit}}, \bibinfo
  {author} {\bibfnamefont {A.~V.}\ \bibnamefont {Olinto}},\ and\ \bibinfo
  {author} {\bibfnamefont {E.}~\bibnamefont {Parizot}},\ }\href
  {https://doi.org/10.1088/1475-7516/2008/10/033} {\bibfield  {journal}
  {\bibinfo  {journal} {JCAP}\ }\textbf {\bibinfo {volume} {0810}},\ \bibinfo
  {pages} {033}},\ \Eprint {https://arxiv.org/abs/0805.4779} {arXiv:0805.4779
  [astro-ph]} \BibitemShut {NoStop}%
\bibitem [{\citenamefont {Globus}\ \emph {et~al.}(2015)\citenamefont {Globus},
  \citenamefont {Allard}, \citenamefont {Mochkovitch},\ and\ \citenamefont
  {Parizot}}]{Globus:2014fka}%
  \BibitemOpen
  \bibfield  {author} {\bibinfo {author} {\bibfnamefont {N.}~\bibnamefont
  {Globus}}, \bibinfo {author} {\bibfnamefont {D.}~\bibnamefont {Allard}},
  \bibinfo {author} {\bibfnamefont {R.}~\bibnamefont {Mochkovitch}},\ and\
  \bibinfo {author} {\bibfnamefont {E.}~\bibnamefont {Parizot}},\ }\href
  {https://doi.org/10.1093/mnras/stv893} {\bibfield  {journal} {\bibinfo
  {journal} {Mon. Not. Roy. Astron. Soc.}\ }\textbf {\bibinfo {volume} {451}},\
  \bibinfo {pages} {751} (\bibinfo {year} {2015})},\ \Eprint
  {https://arxiv.org/abs/1409.1271} {arXiv:1409.1271 [astro-ph.HE]}
  \BibitemShut {NoStop}%
\bibitem [{\citenamefont {{N. Globus, D. Allard, and E.
  Parizot}}(2015)}]{Globus+15}%
  \BibitemOpen
  \bibfield  {author} {\bibinfo {author} {\bibnamefont {{N. Globus, D. Allard,
  and E. Parizot}}},\ }\href {https://doi.org/10.1103/PhysRevD.92.021302}
  {\bibfield  {journal} {\bibinfo  {journal} {Phys. Rev. D}\ }\textbf {\bibinfo
  {volume} {92}},\ \bibinfo {pages} {021302} (\bibinfo {year} {2015})},\
  \Eprint {https://arxiv.org/abs/1505.01377} {arXiv:1505.01377 [astro-ph.HE]}
  \BibitemShut {NoStop}%
\bibitem [{\citenamefont {Unger}\ \emph {et~al.}(2015)\citenamefont {Unger},
  \citenamefont {Farrar},\ and\ \citenamefont {Anchordoqui}}]{Unger:2015laa}%
  \BibitemOpen
  \bibfield  {author} {\bibinfo {author} {\bibfnamefont {M.}~\bibnamefont
  {Unger}}, \bibinfo {author} {\bibfnamefont {G.~R.}\ \bibnamefont {Farrar}},\
  and\ \bibinfo {author} {\bibfnamefont {L.~A.}\ \bibnamefont {Anchordoqui}},\
  }\href {https://doi.org/10.1103/PhysRevD.92.123001} {\bibfield  {journal}
  {\bibinfo  {journal} {Phys. Rev. D}\ }\textbf {\bibinfo {volume} {92}},\
  \bibinfo {pages} {123001} (\bibinfo {year} {2015})},\ \Eprint
  {https://arxiv.org/abs/1505.02153} {arXiv:1505.02153 [astro-ph.HE]}
  \BibitemShut {NoStop}%
\bibitem [{\citenamefont {Aab}\ \emph {et~al.}(2017)\citenamefont {Aab} \emph
  {et~al.}}]{PierreAuger:2016use}%
  \BibitemOpen
  \bibfield  {author} {\bibinfo {author} {\bibfnamefont {A.}~\bibnamefont
  {Aab}} \emph {et~al.} (\bibinfo {collaboration} {Pierre Auger}),\ }\href
  {https://doi.org/10.1088/1475-7516/2017/04/038} {\bibfield  {journal}
  {\bibinfo  {journal} {JCAP}\ }\textbf {\bibinfo {volume} {04}},\ \bibinfo
  {pages} {038}},\ \bibinfo {note} {[Erratum: JCAP 03, E02 (2018)]},\ \Eprint
  {https://arxiv.org/abs/1612.07155} {arXiv:1612.07155 [astro-ph.HE]}
  \BibitemShut {NoStop}%
\bibitem [{\citenamefont {{K. Fang and K. Murase}}(2018)}]{Fang+17}%
  \BibitemOpen
  \bibfield  {author} {\bibinfo {author} {\bibnamefont {{K. Fang and K.
  Murase}}},\ }\href {https://doi.org/10.1038/s41567-017-0025-4} {\bibfield
  {journal} {\bibinfo  {journal} {Phys. Lett.}\ }\textbf {\bibinfo {volume}
  {14}},\ \bibinfo {pages} {396} (\bibinfo {year} {2018})},\ \bibinfo {note}
  {[Nature Phys.14,no.4,396(2018)]},\ \Eprint
  {https://arxiv.org/abs/1704.00015} {arXiv:1704.00015 [astro-ph.HE]}
  \BibitemShut {NoStop}%
\bibitem [{\citenamefont {Kachelrie\ss{}}\ \emph {et~al.}(2017)\citenamefont
  {Kachelrie\ss{}}, \citenamefont {Kalashev}, \citenamefont {Ostapchenko},\
  and\ \citenamefont {Semikoz}}]{Kachelriess:2017tvs}%
  \BibitemOpen
  \bibfield  {author} {\bibinfo {author} {\bibfnamefont {M.}~\bibnamefont
  {Kachelrie\ss{}}}, \bibinfo {author} {\bibfnamefont {O.}~\bibnamefont
  {Kalashev}}, \bibinfo {author} {\bibfnamefont {S.}~\bibnamefont
  {Ostapchenko}},\ and\ \bibinfo {author} {\bibfnamefont {D.~V.}\ \bibnamefont
  {Semikoz}},\ }\href {https://doi.org/10.1103/PhysRevD.96.083006} {\bibfield
  {journal} {\bibinfo  {journal} {Phys. Rev. D}\ }\textbf {\bibinfo {volume}
  {96}},\ \bibinfo {pages} {083006} (\bibinfo {year} {2017})},\ \Eprint
  {https://arxiv.org/abs/1704.06893} {arXiv:1704.06893 [astro-ph.HE]}
  \BibitemShut {NoStop}%
\bibitem [{\citenamefont {Boncioli}\ \emph {et~al.}(2019)\citenamefont
  {Boncioli}, \citenamefont {Biehl},\ and\ \citenamefont
  {Winter}}]{Boncioli:2018lrv}%
  \BibitemOpen
  \bibfield  {author} {\bibinfo {author} {\bibfnamefont {D.}~\bibnamefont
  {Boncioli}}, \bibinfo {author} {\bibfnamefont {D.}~\bibnamefont {Biehl}},\
  and\ \bibinfo {author} {\bibfnamefont {W.}~\bibnamefont {Winter}},\ }\href
  {https://doi.org/10.3847/1538-4357/aafda7} {\bibfield  {journal} {\bibinfo
  {journal} {Astrophys. J.}\ }\textbf {\bibinfo {volume} {872}},\ \bibinfo
  {pages} {110} (\bibinfo {year} {2019})},\ \Eprint
  {https://arxiv.org/abs/1808.07481} {arXiv:1808.07481 [astro-ph.HE]}
  \BibitemShut {NoStop}%
\bibitem [{\citenamefont {Muzio}\ \emph {et~al.}(2019)\citenamefont {Muzio},
  \citenamefont {Unger},\ and\ \citenamefont {Farrar}}]{Muzio:2019leu}%
  \BibitemOpen
  \bibfield  {author} {\bibinfo {author} {\bibfnamefont {M.~S.}\ \bibnamefont
  {Muzio}}, \bibinfo {author} {\bibfnamefont {M.}~\bibnamefont {Unger}},\ and\
  \bibinfo {author} {\bibfnamefont {G.~R.}\ \bibnamefont {Farrar}},\ }\href
  {https://doi.org/10.1103/PhysRevD.100.103008} {\bibfield  {journal} {\bibinfo
   {journal} {Phys. Rev. D}\ }\textbf {\bibinfo {volume} {100}},\ \bibinfo
  {pages} {103008} (\bibinfo {year} {2019})},\ \Eprint
  {https://arxiv.org/abs/1906.06233} {arXiv:1906.06233 [astro-ph.HE]}
  \BibitemShut {NoStop}%
\bibitem [{\citenamefont {Heinze}\ \emph {et~al.}(2020)\citenamefont {Heinze},
  \citenamefont {Biehl}, \citenamefont {Fedynitch}, \citenamefont {Boncioli},
  \citenamefont {Rudolph},\ and\ \citenamefont {Winter}}]{Heinze:2020zqb}%
  \BibitemOpen
  \bibfield  {author} {\bibinfo {author} {\bibfnamefont {J.}~\bibnamefont
  {Heinze}}, \bibinfo {author} {\bibfnamefont {D.}~\bibnamefont {Biehl}},
  \bibinfo {author} {\bibfnamefont {A.}~\bibnamefont {Fedynitch}}, \bibinfo
  {author} {\bibfnamefont {D.}~\bibnamefont {Boncioli}}, \bibinfo {author}
  {\bibfnamefont {A.}~\bibnamefont {Rudolph}},\ and\ \bibinfo {author}
  {\bibfnamefont {W.}~\bibnamefont {Winter}},\ }\href
  {https://doi.org/10.1093/mnras/staa2751} {\bibfield  {journal} {\bibinfo
  {journal} {Mon. Not. Roy. Astron. Soc.}\ }\textbf {\bibinfo {volume} {498}},\
  \bibinfo {pages} {5990} (\bibinfo {year} {2020})},\ \Eprint
  {https://arxiv.org/abs/2006.14301} {arXiv:2006.14301 [astro-ph.HE]}
  \BibitemShut {NoStop}%
\bibitem [{\citenamefont {Muzio}\ \emph {et~al.}(2022)\citenamefont {Muzio},
  \citenamefont {Farrar},\ and\ \citenamefont {Unger}}]{Muzio:2021zud}%
  \BibitemOpen
  \bibfield  {author} {\bibinfo {author} {\bibfnamefont {M.~S.}\ \bibnamefont
  {Muzio}}, \bibinfo {author} {\bibfnamefont {G.~R.}\ \bibnamefont {Farrar}},\
  and\ \bibinfo {author} {\bibfnamefont {M.}~\bibnamefont {Unger}},\ }\href
  {https://doi.org/10.1103/PhysRevD.105.023022} {\bibfield  {journal} {\bibinfo
   {journal} {Phys. Rev. D}\ }\textbf {\bibinfo {volume} {105}},\ \bibinfo
  {pages} {023022} (\bibinfo {year} {2022})},\ \Eprint
  {https://arxiv.org/abs/2108.05512} {arXiv:2108.05512 [astro-ph.HE]}
  \BibitemShut {NoStop}%
\bibitem [{\citenamefont {Abreu}\ \emph {et~al.}(2021)\citenamefont {Abreu}
  \emph {et~al.}}]{PierreAuger:2021mmt}%
  \BibitemOpen
  \bibfield  {author} {\bibinfo {author} {\bibfnamefont {P.}~\bibnamefont
  {Abreu}} \emph {et~al.} (\bibinfo {collaboration} {Pierre Auger}),\ }\href
  {https://doi.org/10.22323/1.395.0311} {\bibfield  {journal} {\bibinfo
  {journal} {PoS}\ }\textbf {\bibinfo {volume} {ICRC2021}},\ \bibinfo {pages}
  {311} (\bibinfo {year} {2021})}\BibitemShut {NoStop}%
\bibitem [{\citenamefont {Bergman}(2021)}]{Bergman:2021djm}%
  \BibitemOpen
  \bibfield  {author} {\bibinfo {author} {\bibfnamefont {D.}~\bibnamefont
  {Bergman}} (\bibinfo {collaboration} {Telescope Array}),\ }\href
  {https://doi.org/10.22323/1.395.0338} {\bibfield  {journal} {\bibinfo
  {journal} {PoS}\ }\textbf {\bibinfo {volume} {ICRC2021}},\ \bibinfo {pages}
  {338} (\bibinfo {year} {2021})}\BibitemShut {NoStop}%
\bibitem [{\citenamefont {Halim}\ \emph {et~al.}(2022)\citenamefont {Halim}
  \emph {et~al.}}]{PierreAuger:2022atd}%
  \BibitemOpen
  \bibfield  {author} {\bibinfo {author} {\bibfnamefont {A.~A.}\ \bibnamefont
  {Halim}} \emph {et~al.} (\bibinfo {collaboration} {Pierre Auger}),\
  }\href@noop {} {\  (\bibinfo {year} {2022})},\ \Eprint
  {https://arxiv.org/abs/2211.02857} {arXiv:2211.02857 [astro-ph.HE]}
  \BibitemShut {NoStop}%
\bibitem [{\citenamefont {Peters}(1961)}]{Peters:1961}%
  \BibitemOpen
  \bibfield  {author} {\bibinfo {author} {\bibfnamefont {B.}~\bibnamefont
  {Peters}},\ }\href {https://doi.org/10.1007/BF02783106} {\bibfield  {journal}
  {\bibinfo  {journal} {Il Nuovo Cimento}\ }\textbf {\bibinfo {volume} {22}},\
  \bibinfo {pages} {800} (\bibinfo {year} {1961})}\BibitemShut {NoStop}%
\bibitem [{\citenamefont {Gaisser}\ \emph {et~al.}(2016)\citenamefont
  {Gaisser}, \citenamefont {Engel},\ and\ \citenamefont
  {Resconi}}]{Gaisser:2016uoy}%
  \BibitemOpen
  \bibfield  {author} {\bibinfo {author} {\bibfnamefont {T.~K.}\ \bibnamefont
  {Gaisser}}, \bibinfo {author} {\bibfnamefont {R.}~\bibnamefont {Engel}},\
  and\ \bibinfo {author} {\bibfnamefont {E.}~\bibnamefont {Resconi}},\
  }\href@noop {} {\emph {\bibinfo {title} {{Cosmic Rays and Particle Physics}:
  {2nd Edition}}}}\ (\bibinfo  {publisher} {Cambridge University Press},\
  \bibinfo {year} {2016})\BibitemShut {NoStop}%
\bibitem [{\citenamefont {Ptitsyna}\ and\ \citenamefont
  {Troitsky}(2010)}]{Ptitsyna:2008zs}%
  \BibitemOpen
  \bibfield  {author} {\bibinfo {author} {\bibfnamefont {K.~V.}\ \bibnamefont
  {Ptitsyna}}\ and\ \bibinfo {author} {\bibfnamefont {S.~V.}\ \bibnamefont
  {Troitsky}},\ }\href {https://doi.org/10.3367/UFNe.0180.201007c.0723}
  {\bibfield  {journal} {\bibinfo  {journal} {Phys. Usp.}\ }\textbf {\bibinfo
  {volume} {53}},\ \bibinfo {pages} {691} (\bibinfo {year} {2010})},\ \Eprint
  {https://arxiv.org/abs/0808.0367} {arXiv:0808.0367 [astro-ph]} \BibitemShut
  {NoStop}%
\bibitem [{\citenamefont {Ahlers}\ \emph {et~al.}(2013)\citenamefont {Ahlers},
  \citenamefont {Anchordoqui},\ and\ \citenamefont {Taylor}}]{Ahlers:2012az}%
  \BibitemOpen
  \bibfield  {author} {\bibinfo {author} {\bibfnamefont {M.}~\bibnamefont
  {Ahlers}}, \bibinfo {author} {\bibfnamefont {L.~A.}\ \bibnamefont
  {Anchordoqui}},\ and\ \bibinfo {author} {\bibfnamefont {A.~M.}\ \bibnamefont
  {Taylor}},\ }\href {https://doi.org/10.1103/PhysRevD.87.023004} {\bibfield
  {journal} {\bibinfo  {journal} {Phys. Rev. D}\ }\textbf {\bibinfo {volume}
  {87}},\ \bibinfo {pages} {023004} (\bibinfo {year} {2013})},\ \Eprint
  {https://arxiv.org/abs/1209.5427} {arXiv:1209.5427 [astro-ph.HE]}
  \BibitemShut {NoStop}%
\bibitem [{\citenamefont {Eichmann}\ \emph {et~al.}(2018)\citenamefont
  {Eichmann}, \citenamefont {Rachen}, \citenamefont {Merten}, \citenamefont
  {van Vliet},\ and\ \citenamefont {Becker~Tjus}}]{Eichmann:2017iyr}%
  \BibitemOpen
  \bibfield  {author} {\bibinfo {author} {\bibfnamefont {B.}~\bibnamefont
  {Eichmann}}, \bibinfo {author} {\bibfnamefont {J.~P.}\ \bibnamefont
  {Rachen}}, \bibinfo {author} {\bibfnamefont {L.}~\bibnamefont {Merten}},
  \bibinfo {author} {\bibfnamefont {A.}~\bibnamefont {van Vliet}},\ and\
  \bibinfo {author} {\bibfnamefont {J.}~\bibnamefont {Becker~Tjus}},\ }\href
  {https://doi.org/10.1088/1475-7516/2018/02/036} {\bibfield  {journal}
  {\bibinfo  {journal} {JCAP}\ }\textbf {\bibinfo {volume} {02}},\ \bibinfo
  {pages} {036}},\ \Eprint {https://arxiv.org/abs/1701.06792} {arXiv:1701.06792
  [astro-ph.HE]} \BibitemShut {NoStop}%
\bibitem [{\citenamefont {Eichmann}\ \emph {et~al.}(2022)\citenamefont
  {Eichmann}, \citenamefont {Kachelrie\ss{}},\ and\ \citenamefont
  {Oikonomou}}]{Eichmann:2022ias}%
  \BibitemOpen
  \bibfield  {author} {\bibinfo {author} {\bibfnamefont {B.}~\bibnamefont
  {Eichmann}}, \bibinfo {author} {\bibfnamefont {M.}~\bibnamefont
  {Kachelrie\ss{}}},\ and\ \bibinfo {author} {\bibfnamefont {F.}~\bibnamefont
  {Oikonomou}},\ }\href {https://doi.org/10.1088/1475-7516/2022/07/006}
  {\bibfield  {journal} {\bibinfo  {journal} {JCAP}\ }\textbf {\bibinfo
  {volume} {07}}\bibfield  {number} {\bibinfo  {number} { (07)},\ \bibinfo
  {pages} {006}},\ }\Eprint {https://arxiv.org/abs/2202.11942}
  {arXiv:2202.11942 [astro-ph.HE]} \BibitemShut {NoStop}%
\bibitem [{\citenamefont {Rodrigues}\ \emph {et~al.}(2021)\citenamefont
  {Rodrigues}, \citenamefont {Heinze}, \citenamefont {Palladino}, \citenamefont
  {van Vliet},\ and\ \citenamefont {Winter}}]{Rodrigues:2020pli}%
  \BibitemOpen
  \bibfield  {author} {\bibinfo {author} {\bibfnamefont {X.}~\bibnamefont
  {Rodrigues}}, \bibinfo {author} {\bibfnamefont {J.}~\bibnamefont {Heinze}},
  \bibinfo {author} {\bibfnamefont {A.}~\bibnamefont {Palladino}}, \bibinfo
  {author} {\bibfnamefont {A.}~\bibnamefont {van Vliet}},\ and\ \bibinfo
  {author} {\bibfnamefont {W.}~\bibnamefont {Winter}},\ }\href
  {https://doi.org/10.1103/PhysRevLett.126.191101} {\bibfield  {journal}
  {\bibinfo  {journal} {Phys. Rev. Lett.}\ }\textbf {\bibinfo {volume} {126}},\
  \bibinfo {pages} {191101} (\bibinfo {year} {2021})},\ \Eprint
  {https://arxiv.org/abs/2003.08392} {arXiv:2003.08392 [astro-ph.HE]}
  \BibitemShut {NoStop}%
\bibitem [{\citenamefont {Das}\ \emph {et~al.}(2021)\citenamefont {Das},
  \citenamefont {Razzaque},\ and\ \citenamefont {Gupta}}]{Das:2020nvx}%
  \BibitemOpen
  \bibfield  {author} {\bibinfo {author} {\bibfnamefont {S.}~\bibnamefont
  {Das}}, \bibinfo {author} {\bibfnamefont {S.}~\bibnamefont {Razzaque}},\ and\
  \bibinfo {author} {\bibfnamefont {N.}~\bibnamefont {Gupta}},\ }\href
  {https://doi.org/10.1140/epjc/s10052-021-08885-4} {\bibfield  {journal}
  {\bibinfo  {journal} {Eur. Phys. J. C}\ }\textbf {\bibinfo {volume} {81}},\
  \bibinfo {pages} {59} (\bibinfo {year} {2021})},\ \Eprint
  {https://arxiv.org/abs/2004.07621} {arXiv:2004.07621 [astro-ph.HE]}
  \BibitemShut {NoStop}%
\bibitem [{\citenamefont {Mollerach}\ and\ \citenamefont
  {Roulet}(2020)}]{Mollerach:2020mhr}%
  \BibitemOpen
  \bibfield  {author} {\bibinfo {author} {\bibfnamefont {S.}~\bibnamefont
  {Mollerach}}\ and\ \bibinfo {author} {\bibfnamefont {E.}~\bibnamefont
  {Roulet}},\ }\href {https://doi.org/10.1103/PhysRevD.101.103024} {\bibfield
  {journal} {\bibinfo  {journal} {Phys. Rev. D}\ }\textbf {\bibinfo {volume}
  {101}},\ \bibinfo {pages} {103024} (\bibinfo {year} {2020})},\ \Eprint
  {https://arxiv.org/abs/2004.04253} {arXiv:2004.04253 [astro-ph.HE]}
  \BibitemShut {NoStop}%
\bibitem [{\citenamefont {Matthews}\ and\ \citenamefont
  {Taylor}(2021)}]{Matthews:2021nik}%
  \BibitemOpen
  \bibfield  {author} {\bibinfo {author} {\bibfnamefont {J.~H.}\ \bibnamefont
  {Matthews}}\ and\ \bibinfo {author} {\bibfnamefont {A.~M.}\ \bibnamefont
  {Taylor}},\ }\href {https://doi.org/10.1093/mnras/stab758} {\bibfield
  {journal} {\bibinfo  {journal} {Mon. Not. Roy. Astron. Soc.}\ }\textbf
  {\bibinfo {volume} {503}},\ \bibinfo {pages} {5948} (\bibinfo {year}
  {2021})},\ \Eprint {https://arxiv.org/abs/2103.06900} {arXiv:2103.06900
  [astro-ph.HE]} \BibitemShut {NoStop}%
\bibitem [{\citenamefont {Lipari}(2021)}]{Lipari:2020szc}%
  \BibitemOpen
  \bibfield  {author} {\bibinfo {author} {\bibfnamefont {P.}~\bibnamefont
  {Lipari}},\ }\href {https://doi.org/10.1016/j.astropartphys.2020.102507}
  {\bibfield  {journal} {\bibinfo  {journal} {Astropart. Phys.}\ }\textbf
  {\bibinfo {volume} {125}},\ \bibinfo {pages} {102507} (\bibinfo {year}
  {2021})},\ \Eprint {https://arxiv.org/abs/2001.00982} {arXiv:2001.00982
  [astro-ph.HE]} \BibitemShut {NoStop}%
\bibitem [{\citenamefont {Yuan}\ \emph {et~al.}(2011)\citenamefont {Yuan},
  \citenamefont {Zhang},\ and\ \citenamefont {Bi}}]{Yuan:2011ys}%
  \BibitemOpen
  \bibfield  {author} {\bibinfo {author} {\bibfnamefont {Q.}~\bibnamefont
  {Yuan}}, \bibinfo {author} {\bibfnamefont {B.}~\bibnamefont {Zhang}},\ and\
  \bibinfo {author} {\bibfnamefont {X.-J.}\ \bibnamefont {Bi}},\ }\href
  {https://doi.org/10.1103/PhysRevD.84.043002} {\bibfield  {journal} {\bibinfo
  {journal} {Phys. Rev. D}\ }\textbf {\bibinfo {volume} {84}},\ \bibinfo
  {pages} {043002} (\bibinfo {year} {2011})},\ \Eprint
  {https://arxiv.org/abs/1104.3357} {arXiv:1104.3357 [astro-ph.HE]}
  \BibitemShut {NoStop}%
\bibitem [{\citenamefont {Kachelriess}\ and\ \citenamefont
  {Semikoz}(2006)}]{Kachelriess:2005xh}%
  \BibitemOpen
  \bibfield  {author} {\bibinfo {author} {\bibfnamefont {M.}~\bibnamefont
  {Kachelriess}}\ and\ \bibinfo {author} {\bibfnamefont {D.~V.}\ \bibnamefont
  {Semikoz}},\ }\href {https://doi.org/10.1016/j.physletb.2006.01.009}
  {\bibfield  {journal} {\bibinfo  {journal} {Phys. Lett. B}\ }\textbf
  {\bibinfo {volume} {634}},\ \bibinfo {pages} {143} (\bibinfo {year}
  {2006})},\ \Eprint {https://arxiv.org/abs/astro-ph/0510188}
  {arXiv:astro-ph/0510188} \BibitemShut {NoStop}%
\bibitem [{\citenamefont {Shibata}\ \emph {et~al.}(2010)\citenamefont
  {Shibata}, \citenamefont {Katayose}, \citenamefont {Huang},\ and\
  \citenamefont {Chen}}]{Shibata:2010zza}%
  \BibitemOpen
  \bibfield  {author} {\bibinfo {author} {\bibfnamefont {M.}~\bibnamefont
  {Shibata}}, \bibinfo {author} {\bibfnamefont {Y.}~\bibnamefont {Katayose}},
  \bibinfo {author} {\bibfnamefont {J.}~\bibnamefont {Huang}},\ and\ \bibinfo
  {author} {\bibfnamefont {D.}~\bibnamefont {Chen}},\ }\href
  {https://doi.org/10.1088/0004-637X/716/2/1076} {\bibfield  {journal}
  {\bibinfo  {journal} {Astrophys. J.}\ }\textbf {\bibinfo {volume} {716}},\
  \bibinfo {pages} {1076} (\bibinfo {year} {2010})}\BibitemShut {NoStop}%
\bibitem [{\citenamefont {Bell}(1978{\natexlab{a}})}]{Bell:1978zc}%
  \BibitemOpen
  \bibfield  {author} {\bibinfo {author} {\bibfnamefont {A.~R.}\ \bibnamefont
  {Bell}},\ }\href {https://doi.org/10.1093/mnras/182.3.443} {\bibfield
  {journal} {\bibinfo  {journal} {Mon. Not. Roy. Astron. Soc.}\ }\textbf
  {\bibinfo {volume} {182}},\ \bibinfo {pages} {147} (\bibinfo {year}
  {1978}{\natexlab{a}})}\BibitemShut {NoStop}%
\bibitem [{\citenamefont {Bell}(1978{\natexlab{b}})}]{Bell:1978fj}%
  \BibitemOpen
  \bibfield  {author} {\bibinfo {author} {\bibfnamefont {A.~R.}\ \bibnamefont
  {Bell}},\ }\href {https://doi.org/10.1093/mnras/182.2.147} {\bibfield
  {journal} {\bibinfo  {journal} {Mon. Not. Roy. Astron. Soc.}\ }\textbf
  {\bibinfo {volume} {182}},\ \bibinfo {pages} {443} (\bibinfo {year}
  {1978}{\natexlab{b}})}\BibitemShut {NoStop}%
\bibitem [{\citenamefont {Protheroe}\ and\ \citenamefont
  {Stanev}(1999)}]{Protheroe:1998pj}%
  \BibitemOpen
  \bibfield  {author} {\bibinfo {author} {\bibfnamefont {R.~J.}\ \bibnamefont
  {Protheroe}}\ and\ \bibinfo {author} {\bibfnamefont {T.}~\bibnamefont
  {Stanev}},\ }\href {https://doi.org/10.1016/S0927-6505(98)00055-3} {\bibfield
   {journal} {\bibinfo  {journal} {Astropart. Phys.}\ }\textbf {\bibinfo
  {volume} {10}},\ \bibinfo {pages} {185} (\bibinfo {year} {1999})},\ \Eprint
  {https://arxiv.org/abs/astro-ph/9808129} {arXiv:astro-ph/9808129}
  \BibitemShut {NoStop}%
\bibitem [{\citenamefont {Protheroe}(2004)}]{Protheroe:2004rt}%
  \BibitemOpen
  \bibfield  {author} {\bibinfo {author} {\bibfnamefont {R.~J.}\ \bibnamefont
  {Protheroe}},\ }\href {https://doi.org/10.1016/j.astropartphys.2004.02.004}
  {\bibfield  {journal} {\bibinfo  {journal} {Astropart. Phys.}\ }\textbf
  {\bibinfo {volume} {21}},\ \bibinfo {pages} {415} (\bibinfo {year} {2004})},\
  \Eprint {https://arxiv.org/abs/astro-ph/0401523} {arXiv:astro-ph/0401523}
  \BibitemShut {NoStop}%
\bibitem [{\citenamefont {Zirakashvili}\ and\ \citenamefont
  {Aharonian}(2007)}]{Zirakashvili:2006pv}%
  \BibitemOpen
  \bibfield  {author} {\bibinfo {author} {\bibfnamefont {V.~N.}\ \bibnamefont
  {Zirakashvili}}\ and\ \bibinfo {author} {\bibfnamefont {F.}~\bibnamefont
  {Aharonian}},\ }\href {https://doi.org/10.1051/0004-6361:20066494} {\bibfield
   {journal} {\bibinfo  {journal} {Astron. Astrophys.}\ }\textbf {\bibinfo
  {volume} {465}},\ \bibinfo {pages} {695} (\bibinfo {year} {2007})},\ \Eprint
  {https://arxiv.org/abs/astro-ph/0612717} {arXiv:astro-ph/0612717}
  \BibitemShut {NoStop}%
\bibitem [{\citenamefont {Hillas}(1984)}]{Hillas:1984ijl}%
  \BibitemOpen
  \bibfield  {author} {\bibinfo {author} {\bibfnamefont {A.~M.}\ \bibnamefont
  {Hillas}},\ }\href {https://doi.org/10.1146/annurev.aa.22.090184.002233}
  {\bibfield  {journal} {\bibinfo  {journal} {Ann. Rev. Astron. Astrophys.}\
  }\textbf {\bibinfo {volume} {22}},\ \bibinfo {pages} {425} (\bibinfo {year}
  {1984})}\BibitemShut {NoStop}%
\bibitem [{\citenamefont {Caprioli}(2015)}]{Caprioli:2015zka}%
  \BibitemOpen
  \bibfield  {author} {\bibinfo {author} {\bibfnamefont {D.}~\bibnamefont
  {Caprioli}},\ }\href {https://doi.org/10.1088/2041-8205/811/2/L38} {\bibfield
   {journal} {\bibinfo  {journal} {Astrophys. J. Lett.}\ }\textbf {\bibinfo
  {volume} {811}},\ \bibinfo {pages} {L38} (\bibinfo {year} {2015})},\ \Eprint
  {https://arxiv.org/abs/1505.06739} {arXiv:1505.06739 [astro-ph.HE]}
  \BibitemShut {NoStop}%
\bibitem [{\citenamefont {Mbarek}\ and\ \citenamefont
  {Caprioli}(2019)}]{Mbarek:2019glq}%
  \BibitemOpen
  \bibfield  {author} {\bibinfo {author} {\bibfnamefont {R.}~\bibnamefont
  {Mbarek}}\ and\ \bibinfo {author} {\bibfnamefont {D.}~\bibnamefont
  {Caprioli}},\ }\href {https://doi.org/10.3847/1538-4357/ab4a08} {\bibfield
  {journal} {\bibinfo  {journal} {Astrophys. J.}\ }\textbf {\bibinfo {volume}
  {886}},\ \bibinfo {pages} {8} (\bibinfo {year} {2019})},\ \Eprint
  {https://arxiv.org/abs/1904.02720} {arXiv:1904.02720 [astro-ph.HE]}
  \BibitemShut {NoStop}%
\bibitem [{\citenamefont {Mbarek}\ and\ \citenamefont
  {Caprioli}(2021)}]{Mbarek:2021bay}%
  \BibitemOpen
  \bibfield  {author} {\bibinfo {author} {\bibfnamefont {R.}~\bibnamefont
  {Mbarek}}\ and\ \bibinfo {author} {\bibfnamefont {D.}~\bibnamefont
  {Caprioli}},\ }\href {https://doi.org/10.3847/1538-4357/ac1da8} {\bibfield
  {journal} {\bibinfo  {journal} {Astrophys. J.}\ }\textbf {\bibinfo {volume}
  {921}},\ \bibinfo {pages} {85} (\bibinfo {year} {2021})},\ \Eprint
  {https://arxiv.org/abs/2105.05262} {arXiv:2105.05262 [astro-ph.HE]}
  \BibitemShut {NoStop}%
\bibitem [{\citenamefont {Lister}\ and\ \citenamefont
  {Marscher}(1997)}]{Lister_1997}%
  \BibitemOpen
  \bibfield  {author} {\bibinfo {author} {\bibfnamefont {M.~L.}\ \bibnamefont
  {Lister}}\ and\ \bibinfo {author} {\bibfnamefont {A.~P.}\ \bibnamefont
  {Marscher}},\ }\href {https://doi.org/10.1086/303629} {\bibfield  {journal}
  {\bibinfo  {journal} {The Astrophysical Journal}\ }\textbf {\bibinfo {volume}
  {476}},\ \bibinfo {pages} {572} (\bibinfo {year} {1997})}\BibitemShut
  {NoStop}%
\bibitem [{\citenamefont {{Lister}}\ \emph {et~al.}(2019)\citenamefont
  {{Lister}} \emph {et~al.}}]{2019ApJ...874...43L}%
  \BibitemOpen
  \bibfield  {author} {\bibinfo {author} {\bibfnamefont {M.~L.}\ \bibnamefont
  {{Lister}}} \emph {et~al.},\ }\href
  {https://doi.org/10.3847/1538-4357/ab08ee} {\bibfield  {journal} {\bibinfo
  {journal} {\apj}\ }\textbf {\bibinfo {volume} {874}},\ \bibinfo {eid} {43}
  (\bibinfo {year} {2019})},\ \Eprint {https://arxiv.org/abs/1902.09591}
  {arXiv:1902.09591 [astro-ph.GA]} \BibitemShut {NoStop}%
\bibitem [{\citenamefont {Lovelace}(1976)}]{Lovelace:1976}%
  \BibitemOpen
  \bibfield  {author} {\bibinfo {author} {\bibfnamefont {R.~V.~E.}\
  \bibnamefont {Lovelace}},\ }\href {https://doi.org/10.1038/262649a0}
  {\bibfield  {journal} {\bibinfo  {journal} {Nature}\ }\textbf {\bibinfo
  {volume} {262}},\ \bibinfo {pages} {649–652} (\bibinfo {year}
  {1976})}\BibitemShut {NoStop}%
\bibitem [{\citenamefont {Waxman}(1995)}]{Waxman:1995vg}%
  \BibitemOpen
  \bibfield  {author} {\bibinfo {author} {\bibfnamefont {E.}~\bibnamefont
  {Waxman}},\ }\href {https://doi.org/10.1103/PhysRevLett.75.386} {\bibfield
  {journal} {\bibinfo  {journal} {Phys. Rev. Lett.}\ }\textbf {\bibinfo
  {volume} {75}},\ \bibinfo {pages} {386} (\bibinfo {year} {1995})},\ \Eprint
  {https://arxiv.org/abs/astro-ph/9505082} {arXiv:astro-ph/9505082}
  \BibitemShut {NoStop}%
\bibitem [{\citenamefont {Waxman}(2001)}]{Waxman:2001tk}%
  \BibitemOpen
  \bibfield  {author} {\bibinfo {author} {\bibfnamefont {E.}~\bibnamefont
  {Waxman}},\ }\href@noop {} {\bibfield  {journal} {\bibinfo  {journal} {ICTP
  Lect. Notes Ser.}\ }\textbf {\bibinfo {volume} {4}},\ \bibinfo {pages} {309}
  (\bibinfo {year} {2001})},\ \Eprint {https://arxiv.org/abs/astro-ph/0103186}
  {arXiv:astro-ph/0103186} \BibitemShut {NoStop}%
\bibitem [{\citenamefont {Blandford}(2000)}]{Blandford_2000}%
  \BibitemOpen
  \bibfield  {author} {\bibinfo {author} {\bibfnamefont {R.~D.}\ \bibnamefont
  {Blandford}},\ }\href {https://doi.org/10.1238/physica.topical.085a00191}
  {\bibfield  {journal} {\bibinfo  {journal} {Physica Scripta}\ }\textbf
  {\bibinfo {volume} {T85}},\ \bibinfo {pages} {191} (\bibinfo {year}
  {2000})}\BibitemShut {NoStop}%
\bibitem [{\citenamefont {Lemoine}\ and\ \citenamefont
  {Waxman}(2009)}]{Lemoine:2009pw}%
  \BibitemOpen
  \bibfield  {author} {\bibinfo {author} {\bibfnamefont {M.}~\bibnamefont
  {Lemoine}}\ and\ \bibinfo {author} {\bibfnamefont {E.}~\bibnamefont
  {Waxman}},\ }\href {https://doi.org/10.1088/1475-7516/2009/11/009} {\bibfield
   {journal} {\bibinfo  {journal} {JCAP}\ }\textbf {\bibinfo {volume} {11}},\
  \bibinfo {pages} {009}},\ \Eprint {https://arxiv.org/abs/0907.1354}
  {arXiv:0907.1354 [astro-ph.HE]} \BibitemShut {NoStop}%
\bibitem [{\citenamefont {Kachelriess}(2022)}]{Kachelriess:2022phl}%
  \BibitemOpen
  \bibfield  {author} {\bibinfo {author} {\bibfnamefont {M.}~\bibnamefont
  {Kachelriess}},\ }\href {https://doi.org/10.22323/1.395.0018} {\bibfield
  {journal} {\bibinfo  {journal} {PoS}\ }\textbf {\bibinfo {volume}
  {ICRC2021}},\ \bibinfo {pages} {018} (\bibinfo {year} {2022})},\ \Eprint
  {https://arxiv.org/abs/2201.04535} {arXiv:2201.04535 [astro-ph.HE]}
  \BibitemShut {NoStop}%
\bibitem [{\citenamefont {Rieger}(2022)}]{Rieger:2022qhs}%
  \BibitemOpen
  \bibfield  {author} {\bibinfo {author} {\bibfnamefont {F.~M.}\ \bibnamefont
  {Rieger}},\ }\href {https://doi.org/10.3390/universe8110607} {\bibfield
  {journal} {\bibinfo  {journal} {Universe}\ }\textbf {\bibinfo {volume} {8}},\
  \bibinfo {pages} {607} (\bibinfo {year} {2022})},\ \Eprint
  {https://arxiv.org/abs/2211.12202} {arXiv:2211.12202 [astro-ph.HE]}
  \BibitemShut {NoStop}%
\bibitem [{\citenamefont {Santoro}\ \emph {et~al.}(2020)\citenamefont
  {Santoro}, \citenamefont {Tadhunter}, \citenamefont {Baron}, \citenamefont
  {Morganti},\ and\ \citenamefont {Holt}}]{Santoro_2020}%
  \BibitemOpen
  \bibfield  {author} {\bibinfo {author} {\bibfnamefont {F.}~\bibnamefont
  {Santoro}}, \bibinfo {author} {\bibfnamefont {C.}~\bibnamefont {Tadhunter}},
  \bibinfo {author} {\bibfnamefont {D.}~\bibnamefont {Baron}}, \bibinfo
  {author} {\bibfnamefont {R.}~\bibnamefont {Morganti}},\ and\ \bibinfo
  {author} {\bibfnamefont {J.}~\bibnamefont {Holt}},\ }\href
  {https://doi.org/10.1051/0004-6361/202039077} {\bibfield  {journal} {\bibinfo
   {journal} {Astronomy \& Astrophysics}\ }\textbf {\bibinfo {volume} {644}},\
  \bibinfo {pages} {A54} (\bibinfo {year} {2020})}\BibitemShut {NoStop}%
\bibitem [{\citenamefont {Ajello}\ \emph {et~al.}(2009)\citenamefont {Ajello}
  \emph {et~al.}}]{Ajello:2009ip}%
  \BibitemOpen
  \bibfield  {author} {\bibinfo {author} {\bibfnamefont {M.}~\bibnamefont
  {Ajello}} \emph {et~al.},\ }\href
  {https://doi.org/10.1088/0004-637X/699/1/603} {\bibfield  {journal} {\bibinfo
   {journal} {Astrophys. J.}\ }\textbf {\bibinfo {volume} {699}},\ \bibinfo
  {pages} {603} (\bibinfo {year} {2009})},\ \Eprint
  {https://arxiv.org/abs/0905.0472} {arXiv:0905.0472 [astro-ph.CO]}
  \BibitemShut {NoStop}%
\bibitem [{\citenamefont {Marcotulli}\ \emph {et~al.}(2022)\citenamefont
  {Marcotulli} \emph {et~al.}}]{BASS:2022gdj}%
  \BibitemOpen
  \bibfield  {author} {\bibinfo {author} {\bibfnamefont {L.}~\bibnamefont
  {Marcotulli}} \emph {et~al.} (\bibinfo {collaboration} {BASS}),\ }\href
  {https://doi.org/10.3847/1538-4357/ac937f} {\bibfield  {journal} {\bibinfo
  {journal} {Astrophys. J.}\ }\textbf {\bibinfo {volume} {940}},\ \bibinfo
  {pages} {77} (\bibinfo {year} {2022})},\ \Eprint
  {https://arxiv.org/abs/2209.09929} {arXiv:2209.09929 [astro-ph.HE]}
  \BibitemShut {NoStop}%
\bibitem [{\citenamefont {van Velzen}(2018)}]{vanVelzen:2017qum}%
  \BibitemOpen
  \bibfield  {author} {\bibinfo {author} {\bibfnamefont {S.}~\bibnamefont {van
  Velzen}},\ }\href {https://doi.org/10.3847/1538-4357/aa998e} {\bibfield
  {journal} {\bibinfo  {journal} {Astrophys. J.}\ }\textbf {\bibinfo {volume}
  {852}},\ \bibinfo {pages} {72} (\bibinfo {year} {2018})},\ \Eprint
  {https://arxiv.org/abs/1707.03458} {arXiv:1707.03458 [astro-ph.HE]}
  \BibitemShut {NoStop}%
\bibitem [{\citenamefont {Lin}\ \emph {et~al.}(2022)\citenamefont {Lin},
  \citenamefont {Jiang}, \citenamefont {Kong}, \citenamefont {Huang},
  \citenamefont {Lin}, \citenamefont {Zhu},\ and\ \citenamefont
  {Wang}}]{Lin:2022jvw}%
  \BibitemOpen
  \bibfield  {author} {\bibinfo {author} {\bibfnamefont {Z.}~\bibnamefont
  {Lin}}, \bibinfo {author} {\bibfnamefont {N.}~\bibnamefont {Jiang}}, \bibinfo
  {author} {\bibfnamefont {X.}~\bibnamefont {Kong}}, \bibinfo {author}
  {\bibfnamefont {S.}~\bibnamefont {Huang}}, \bibinfo {author} {\bibfnamefont
  {Z.}~\bibnamefont {Lin}}, \bibinfo {author} {\bibfnamefont {J.}~\bibnamefont
  {Zhu}},\ and\ \bibinfo {author} {\bibfnamefont {Y.}~\bibnamefont {Wang}},\
  }\href {https://doi.org/10.3847/2041-8213/ac9c63} {\bibfield  {journal}
  {\bibinfo  {journal} {Astrophys. J. Lett.}\ }\textbf {\bibinfo {volume}
  {939}},\ \bibinfo {pages} {L33} (\bibinfo {year} {2022})},\ \Eprint
  {https://arxiv.org/abs/2210.14950} {arXiv:2210.14950 [astro-ph.HE]}
  \BibitemShut {NoStop}%
\bibitem [{\citenamefont {Wanderman}\ and\ \citenamefont
  {Piran}(2010)}]{Wanderman:2009es}%
  \BibitemOpen
  \bibfield  {author} {\bibinfo {author} {\bibfnamefont {D.}~\bibnamefont
  {Wanderman}}\ and\ \bibinfo {author} {\bibfnamefont {T.}~\bibnamefont
  {Piran}},\ }\href {https://doi.org/10.1111/j.1365-2966.2010.16787.x}
  {\bibfield  {journal} {\bibinfo  {journal} {Mon. Not. Roy. Astron. Soc.}\
  }\textbf {\bibinfo {volume} {406}},\ \bibinfo {pages} {1944} (\bibinfo {year}
  {2010})},\ \Eprint {https://arxiv.org/abs/0912.0709} {arXiv:0912.0709
  [astro-ph.HE]} \BibitemShut {NoStop}%
\bibitem [{\citenamefont {Ueda}\ \emph {et~al.}(2014)\citenamefont {Ueda},
  \citenamefont {Akiyama}, \citenamefont {Hasinger}, \citenamefont {Miyaji},\
  and\ \citenamefont {Watson}}]{Ueda:2014tma}%
  \BibitemOpen
  \bibfield  {author} {\bibinfo {author} {\bibfnamefont {Y.}~\bibnamefont
  {Ueda}}, \bibinfo {author} {\bibfnamefont {M.}~\bibnamefont {Akiyama}},
  \bibinfo {author} {\bibfnamefont {G.}~\bibnamefont {Hasinger}}, \bibinfo
  {author} {\bibfnamefont {T.}~\bibnamefont {Miyaji}},\ and\ \bibinfo {author}
  {\bibfnamefont {M.~G.}\ \bibnamefont {Watson}},\ }\href
  {https://doi.org/10.1088/0004-637X/786/2/104} {\bibfield  {journal} {\bibinfo
   {journal} {Astrophys. J.}\ }\textbf {\bibinfo {volume} {786}},\ \bibinfo
  {pages} {104} (\bibinfo {year} {2014})},\ \Eprint
  {https://arxiv.org/abs/1402.1836} {arXiv:1402.1836 [astro-ph.CO]}
  \BibitemShut {NoStop}%
\bibitem [{\citenamefont {Saikia}\ \emph {et~al.}(2016)\citenamefont {Saikia},
  \citenamefont {K\"ording},\ and\ \citenamefont {Falcke}}]{Saikia:2016blk}%
  \BibitemOpen
  \bibfield  {author} {\bibinfo {author} {\bibfnamefont {P.}~\bibnamefont
  {Saikia}}, \bibinfo {author} {\bibfnamefont {E.}~\bibnamefont {K\"ording}},\
  and\ \bibinfo {author} {\bibfnamefont {H.}~\bibnamefont {Falcke}},\ }\href
  {https://doi.org/10.1093/mnras/stw1321} {\bibfield  {journal} {\bibinfo
  {journal} {Mon. Not. Roy. Astron. Soc.}\ }\textbf {\bibinfo {volume} {461}},\
  \bibinfo {pages} {297} (\bibinfo {year} {2016})},\ \Eprint
  {https://arxiv.org/abs/1606.06147} {arXiv:1606.06147 [astro-ph.HE]}
  \BibitemShut {NoStop}%
\bibitem [{\citenamefont {Aab}\ \emph {et~al.}(2020)\citenamefont {Aab} \emph
  {et~al.}}]{PierreAuger:2020qqz}%
  \BibitemOpen
  \bibfield  {author} {\bibinfo {author} {\bibfnamefont {A.}~\bibnamefont
  {Aab}} \emph {et~al.} (\bibinfo {collaboration} {Pierre Auger}),\ }\href
  {https://doi.org/10.1103/PhysRevD.102.062005} {\bibfield  {journal} {\bibinfo
   {journal} {Phys. Rev. D}\ }\textbf {\bibinfo {volume} {102}},\ \bibinfo
  {pages} {062005} (\bibinfo {year} {2020})},\ \Eprint
  {https://arxiv.org/abs/2008.06486} {arXiv:2008.06486 [astro-ph.HE]}
  \BibitemShut {NoStop}%
\bibitem [{\citenamefont {Aab}\ \emph {et~al.}(2014)\citenamefont {Aab} \emph
  {et~al.}}]{PierreAuger:2014sui}%
  \BibitemOpen
  \bibfield  {author} {\bibinfo {author} {\bibfnamefont {A.}~\bibnamefont
  {Aab}} \emph {et~al.} (\bibinfo {collaboration} {Pierre Auger}),\ }\href
  {https://doi.org/10.1103/PhysRevD.90.122005} {\bibfield  {journal} {\bibinfo
  {journal} {Phys. Rev. D}\ }\textbf {\bibinfo {volume} {90}},\ \bibinfo
  {pages} {122005} (\bibinfo {year} {2014})},\ \Eprint
  {https://arxiv.org/abs/1409.4809} {arXiv:1409.4809 [astro-ph.HE]}
  \BibitemShut {NoStop}%
\bibitem [{\citenamefont {Yushkov}(2020)}]{Yushkov:2020nhr}%
  \BibitemOpen
  \bibfield  {author} {\bibinfo {author} {\bibfnamefont {A.}~\bibnamefont
  {Yushkov}} (\bibinfo {collaboration} {Auger}),\ }\href
  {https://doi.org/10.22323/1.358.0482} {\bibfield  {journal} {\bibinfo
  {journal} {PoS}\ }\textbf {\bibinfo {volume} {ICRC2019}},\ \bibinfo {pages}
  {482} (\bibinfo {year} {2020})}\BibitemShut {NoStop}%
\bibitem [{\citenamefont {Kampert}\ and\ \citenamefont
  {Unger}(2012)}]{Kampert:2012mx}%
  \BibitemOpen
  \bibfield  {author} {\bibinfo {author} {\bibfnamefont {K.-H.}\ \bibnamefont
  {Kampert}}\ and\ \bibinfo {author} {\bibfnamefont {M.}~\bibnamefont
  {Unger}},\ }\href {https://doi.org/10.1016/j.astropartphys.2012.02.004}
  {\bibfield  {journal} {\bibinfo  {journal} {Astropart. Phys.}\ }\textbf
  {\bibinfo {volume} {35}},\ \bibinfo {pages} {660} (\bibinfo {year} {2012})},\
  \Eprint {https://arxiv.org/abs/1201.0018} {arXiv:1201.0018 [astro-ph.HE]}
  \BibitemShut {NoStop}%
\bibitem [{\citenamefont {Alves~Batista}\ \emph {et~al.}(2016)\citenamefont
  {Alves~Batista} \emph {et~al.}}]{Batista:2016yrx}%
  \BibitemOpen
  \bibfield  {author} {\bibinfo {author} {\bibfnamefont {R.}~\bibnamefont
  {Alves~Batista}} \emph {et~al.},\ }\href
  {https://doi.org/10.1088/1475-7516/2016/05/038} {\bibfield  {journal}
  {\bibinfo  {journal} {JCAP}\ }\textbf {\bibinfo {volume} {05}},\ \bibinfo
  {pages} {038}},\ \Eprint {https://arxiv.org/abs/1603.07142} {arXiv:1603.07142
  [astro-ph.IM]} \BibitemShut {NoStop}%
\bibitem [{\citenamefont {Ackermann}\ \emph {et~al.}(2015)\citenamefont
  {Ackermann} \emph {et~al.}}]{Fermi-LAT:2014ryh}%
  \BibitemOpen
  \bibfield  {author} {\bibinfo {author} {\bibfnamefont {M.}~\bibnamefont
  {Ackermann}} \emph {et~al.} (\bibinfo {collaboration} {Fermi-LAT}),\ }\href
  {https://doi.org/10.1088/0004-637X/799/1/86} {\bibfield  {journal} {\bibinfo
  {journal} {Astrophys. J.}\ }\textbf {\bibinfo {volume} {799}},\ \bibinfo
  {pages} {86} (\bibinfo {year} {2015})},\ \Eprint
  {https://arxiv.org/abs/1410.3696} {arXiv:1410.3696 [astro-ph.HE]}
  \BibitemShut {NoStop}%
\bibitem [{\citenamefont {Kopper}(2018)}]{Kopper:2017zzm}%
  \BibitemOpen
  \bibfield  {author} {\bibinfo {author} {\bibfnamefont {C.}~\bibnamefont
  {Kopper}} (\bibinfo {collaboration} {IceCube}),\ }\href
  {https://doi.org/10.22323/1.301.0981} {\bibfield  {journal} {\bibinfo
  {journal} {PoS}\ }\textbf {\bibinfo {volume} {ICRC2017}},\ \bibinfo {pages}
  {981} (\bibinfo {year} {2018})}\BibitemShut {NoStop}%
\bibitem [{\citenamefont {Aartsen}\ \emph {et~al.}(2018)\citenamefont {Aartsen}
  \emph {et~al.}}]{IceCube:2018fhm}%
  \BibitemOpen
  \bibfield  {author} {\bibinfo {author} {\bibfnamefont {M.~G.}\ \bibnamefont
  {Aartsen}} \emph {et~al.} (\bibinfo {collaboration} {IceCube}),\ }\href
  {https://doi.org/10.1103/PhysRevD.98.062003} {\bibfield  {journal} {\bibinfo
  {journal} {Phys. Rev. D}\ }\textbf {\bibinfo {volume} {98}},\ \bibinfo
  {pages} {062003} (\bibinfo {year} {2018})},\ \Eprint
  {https://arxiv.org/abs/1807.01820} {arXiv:1807.01820 [astro-ph.HE]}
  \BibitemShut {NoStop}%
\bibitem [{\citenamefont {Alves~Batista}\ \emph
  {et~al.}(2019{\natexlab{b}})\citenamefont {Alves~Batista}, \citenamefont
  {de~Almeida}, \citenamefont {Lago},\ and\ \citenamefont
  {Kotera}}]{AlvesBatista:2018zui}%
  \BibitemOpen
  \bibfield  {author} {\bibinfo {author} {\bibfnamefont {R.}~\bibnamefont
  {Alves~Batista}}, \bibinfo {author} {\bibfnamefont {R.~M.}\ \bibnamefont
  {de~Almeida}}, \bibinfo {author} {\bibfnamefont {B.}~\bibnamefont {Lago}},\
  and\ \bibinfo {author} {\bibfnamefont {K.}~\bibnamefont {Kotera}},\ }\href
  {https://doi.org/10.1088/1475-7516/2019/01/002} {\bibfield  {journal}
  {\bibinfo  {journal} {JCAP}\ }\textbf {\bibinfo {volume} {01}},\ \bibinfo
  {pages} {002}},\ \Eprint {https://arxiv.org/abs/1806.10879} {arXiv:1806.10879
  [astro-ph.HE]} \BibitemShut {NoStop}%
\bibitem [{\citenamefont {Alves~Batista}\ \emph {et~al.}(2015)\citenamefont
  {Alves~Batista}, \citenamefont {Boncioli}, \citenamefont {di~Matteo},
  \citenamefont {van Vliet},\ and\ \citenamefont
  {Walz}}]{AlvesBatista:2015jem}%
  \BibitemOpen
  \bibfield  {author} {\bibinfo {author} {\bibfnamefont {R.}~\bibnamefont
  {Alves~Batista}}, \bibinfo {author} {\bibfnamefont {D.}~\bibnamefont
  {Boncioli}}, \bibinfo {author} {\bibfnamefont {A.}~\bibnamefont {di~Matteo}},
  \bibinfo {author} {\bibfnamefont {A.}~\bibnamefont {van Vliet}},\ and\
  \bibinfo {author} {\bibfnamefont {D.}~\bibnamefont {Walz}},\ }\href
  {https://doi.org/10.1088/1475-7516/2015/10/063} {\bibfield  {journal}
  {\bibinfo  {journal} {JCAP}\ }\textbf {\bibinfo {volume} {10}},\ \bibinfo
  {pages} {063}},\ \Eprint {https://arxiv.org/abs/1508.01824} {arXiv:1508.01824
  [astro-ph.HE]} \BibitemShut {NoStop}%
\bibitem [{\citenamefont {Gilmore}\ \emph {et~al.}(2012)\citenamefont
  {Gilmore}, \citenamefont {Somerville}, \citenamefont {Primack},\ and\
  \citenamefont {Domínguez}}]{Gilmore:2012}%
  \BibitemOpen
  \bibfield  {author} {\bibinfo {author} {\bibfnamefont {R.~C.}\ \bibnamefont
  {Gilmore}}, \bibinfo {author} {\bibfnamefont {R.~S.}\ \bibnamefont
  {Somerville}}, \bibinfo {author} {\bibfnamefont {J.~R.}\ \bibnamefont
  {Primack}},\ and\ \bibinfo {author} {\bibfnamefont {A.}~\bibnamefont
  {Domínguez}},\ }\href {https://doi.org/10.1111/j.1365-2966.2012.20841.x}
  {\bibfield  {journal} {\bibinfo  {journal} {Mon. Not. Roy. Astron. Soc.}\
  }\textbf {\bibinfo {volume} {422}},\ \bibinfo {pages} {3189} (\bibinfo {year}
  {2012})}\BibitemShut {NoStop}%
\bibitem [{\citenamefont {Aloisio}\ and\ \citenamefont
  {Berezinsky}(2004)}]{Aloisio:2004jda}%
  \BibitemOpen
  \bibfield  {author} {\bibinfo {author} {\bibfnamefont {R.}~\bibnamefont
  {Aloisio}}\ and\ \bibinfo {author} {\bibfnamefont {V.}~\bibnamefont
  {Berezinsky}},\ }\href {https://doi.org/10.1086/421869} {\bibfield  {journal}
  {\bibinfo  {journal} {Astrophys. J.}\ }\textbf {\bibinfo {volume} {612}},\
  \bibinfo {pages} {900} (\bibinfo {year} {2004})},\ \Eprint
  {https://arxiv.org/abs/astro-ph/0403095} {arXiv:astro-ph/0403095}
  \BibitemShut {NoStop}%
\bibitem [{\citenamefont {Gonz\'alez}\ \emph {et~al.}(2021)\citenamefont
  {Gonz\'alez}, \citenamefont {Mollerach},\ and\ \citenamefont
  {Roulet}}]{Gonzalez:2021ajv}%
  \BibitemOpen
  \bibfield  {author} {\bibinfo {author} {\bibfnamefont {J.~M.}\ \bibnamefont
  {Gonz\'alez}}, \bibinfo {author} {\bibfnamefont {S.}~\bibnamefont
  {Mollerach}},\ and\ \bibinfo {author} {\bibfnamefont {E.}~\bibnamefont
  {Roulet}},\ }\href {https://doi.org/10.1103/PhysRevD.104.063005} {\bibfield
  {journal} {\bibinfo  {journal} {Phys. Rev. D}\ }\textbf {\bibinfo {volume}
  {104}},\ \bibinfo {pages} {063005} (\bibinfo {year} {2021})},\ \Eprint
  {https://arxiv.org/abs/2105.08138} {arXiv:2105.08138 [astro-ph.HE]}
  \BibitemShut {NoStop}%
\bibitem [{\citenamefont {Mollerach}\ and\ \citenamefont
  {Roulet}(2013)}]{Mollerach:2013dza}%
  \BibitemOpen
  \bibfield  {author} {\bibinfo {author} {\bibfnamefont {S.}~\bibnamefont
  {Mollerach}}\ and\ \bibinfo {author} {\bibfnamefont {E.}~\bibnamefont
  {Roulet}},\ }\href {https://doi.org/10.1088/1475-7516/2013/10/013} {\bibfield
   {journal} {\bibinfo  {journal} {JCAP}\ }\textbf {\bibinfo {volume} {10}},\
  \bibinfo {pages} {013}},\ \Eprint {https://arxiv.org/abs/1305.6519}
  {arXiv:1305.6519 [astro-ph.HE]} \BibitemShut {NoStop}%
\bibitem [{\citenamefont {Globus}\ \emph {et~al.}(2008)\citenamefont {Globus},
  \citenamefont {Allard},\ and\ \citenamefont {Parizot}}]{Globus:2007bi}%
  \BibitemOpen
  \bibfield  {author} {\bibinfo {author} {\bibfnamefont {N.}~\bibnamefont
  {Globus}}, \bibinfo {author} {\bibfnamefont {D.}~\bibnamefont {Allard}},\
  and\ \bibinfo {author} {\bibfnamefont {E.}~\bibnamefont {Parizot}},\ }\href
  {https://doi.org/10.1051/0004-6361:20078653} {\bibfield  {journal} {\bibinfo
  {journal} {Astron. Astrophys.}\ }\textbf {\bibinfo {volume} {479}},\ \bibinfo
  {pages} {97} (\bibinfo {year} {2008})},\ \Eprint
  {https://arxiv.org/abs/0709.1541} {arXiv:0709.1541 [astro-ph]} \BibitemShut
  {NoStop}%
\bibitem [{\citenamefont {Wittkowski}(2017)}]{Wittkowski:2017ZK}%
  \BibitemOpen
  \bibfield  {author} {\bibinfo {author} {\bibfnamefont {D.}~\bibnamefont
  {Wittkowski}},\ }\href {https://doi.org/10.22323/1.301.0563} {\bibfield
  {journal} {\bibinfo  {journal} {PoS}\ }\textbf {\bibinfo {volume}
  {ICRC2017}},\ \bibinfo {pages} {563} (\bibinfo {year} {2017})}\BibitemShut
  {NoStop}%
\bibitem [{\citenamefont {Abreu}\ \emph {et~al.}(2013)\citenamefont {Abreu}
  \emph {et~al.}}]{PierreAuger:2013xim}%
  \BibitemOpen
  \bibfield  {author} {\bibinfo {author} {\bibfnamefont {P.}~\bibnamefont
  {Abreu}} \emph {et~al.} (\bibinfo {collaboration} {Pierre Auger}),\ }\href
  {https://doi.org/10.1088/1475-7516/2013/02/026} {\bibfield  {journal}
  {\bibinfo  {journal} {JCAP}\ }\textbf {\bibinfo {volume} {02}},\ \bibinfo
  {pages} {026}},\ \Eprint {https://arxiv.org/abs/1301.6637} {arXiv:1301.6637
  [astro-ph.HE]} \BibitemShut {NoStop}%
\bibitem [{\citenamefont {Hillas}(2005)}]{Hillas:2005cs}%
  \BibitemOpen
  \bibfield  {author} {\bibinfo {author} {\bibfnamefont {A.~M.}\ \bibnamefont
  {Hillas}},\ }\href {https://doi.org/10.1088/0954-3899/31/5/R02} {\bibfield
  {journal} {\bibinfo  {journal} {J. Phys. G}\ }\textbf {\bibinfo {volume}
  {31}},\ \bibinfo {pages} {R95} (\bibinfo {year} {2005})}\BibitemShut
  {NoStop}%
\bibitem [{\citenamefont {Baker}\ and\ \citenamefont
  {Cousins}(1984)}]{Baker:1983tu}%
  \BibitemOpen
  \bibfield  {author} {\bibinfo {author} {\bibfnamefont {S.}~\bibnamefont
  {Baker}}\ and\ \bibinfo {author} {\bibfnamefont {R.~D.}\ \bibnamefont
  {Cousins}},\ }\href {https://doi.org/10.1016/0167-5087(84)90016-4} {\bibfield
   {journal} {\bibinfo  {journal} {Nucl. Instrum. Meth.}\ }\textbf {\bibinfo
  {volume} {221}},\ \bibinfo {pages} {437} (\bibinfo {year}
  {1984})}\BibitemShut {NoStop}%
\bibitem [{\citenamefont {Dawson}(2020)}]{Dawson:2020bkp}%
  \BibitemOpen
  \bibfield  {author} {\bibinfo {author} {\bibfnamefont {B.}~\bibnamefont
  {Dawson}} (\bibinfo {collaboration} {Pierre Auger}),\ }\href
  {https://doi.org/10.22323/1.358.0231} {\bibfield  {journal} {\bibinfo
  {journal} {PoS}\ }\textbf {\bibinfo {volume} {ICRC2019}},\ \bibinfo {pages}
  {231} (\bibinfo {year} {2020})}\BibitemShut {NoStop}%
\bibitem [{\citenamefont {Riehn}\ \emph {et~al.}(2018)\citenamefont {Riehn},
  \citenamefont {Dembinski}, \citenamefont {Engel}, \citenamefont {Fedynitch},
  \citenamefont {Gaisser},\ and\ \citenamefont {Stanev}}]{Riehn:2017mfm}%
  \BibitemOpen
  \bibfield  {author} {\bibinfo {author} {\bibfnamefont {F.}~\bibnamefont
  {Riehn}}, \bibinfo {author} {\bibfnamefont {H.~P.}\ \bibnamefont
  {Dembinski}}, \bibinfo {author} {\bibfnamefont {R.}~\bibnamefont {Engel}},
  \bibinfo {author} {\bibfnamefont {A.}~\bibnamefont {Fedynitch}}, \bibinfo
  {author} {\bibfnamefont {T.~K.}\ \bibnamefont {Gaisser}},\ and\ \bibinfo
  {author} {\bibfnamefont {T.}~\bibnamefont {Stanev}},\ }\href
  {https://doi.org/10.22323/1.301.0301} {\bibfield  {journal} {\bibinfo
  {journal} {PoS}\ }\textbf {\bibinfo {volume} {ICRC2017}},\ \bibinfo {pages}
  {301} (\bibinfo {year} {2018})},\ \Eprint {https://arxiv.org/abs/1709.07227}
  {arXiv:1709.07227 [hep-ph]} \BibitemShut {NoStop}%
\bibitem [{\citenamefont {Pierog}\ \emph {et~al.}(2015)\citenamefont {Pierog},
  \citenamefont {Karpenko}, \citenamefont {Katzy}, \citenamefont {Yatsenko},\
  and\ \citenamefont {Werner}}]{Pierog:2013ria}%
  \BibitemOpen
  \bibfield  {author} {\bibinfo {author} {\bibfnamefont {T.}~\bibnamefont
  {Pierog}}, \bibinfo {author} {\bibfnamefont {I.}~\bibnamefont {Karpenko}},
  \bibinfo {author} {\bibfnamefont {J.~M.}\ \bibnamefont {Katzy}}, \bibinfo
  {author} {\bibfnamefont {E.}~\bibnamefont {Yatsenko}},\ and\ \bibinfo
  {author} {\bibfnamefont {K.}~\bibnamefont {Werner}},\ }\href
  {https://doi.org/10.1103/PhysRevC.92.034906} {\bibfield  {journal} {\bibinfo
  {journal} {Phys. Rev. C}\ }\textbf {\bibinfo {volume} {92}},\ \bibinfo
  {pages} {034906} (\bibinfo {year} {2015})},\ \Eprint
  {https://arxiv.org/abs/1306.0121} {arXiv:1306.0121 [hep-ph]} \BibitemShut
  {NoStop}%
\bibitem [{\citenamefont {Verzi}(2019)}]{Verzi:2019AO}%
  \BibitemOpen
  \bibfield  {author} {\bibinfo {author} {\bibfnamefont {V.}~\bibnamefont
  {Verzi}},\ }\href {https://doi.org/10.22323/1.358.0450} {\bibfield  {journal}
  {\bibinfo  {journal} {PoS}\ }\textbf {\bibinfo {volume} {ICRC2019}},\
  \bibinfo {pages} {450} (\bibinfo {year} {2019})}\BibitemShut {NoStop}%
\bibitem [{\citenamefont {Greisen}(1966)}]{Greisen:1966jv}%
  \BibitemOpen
  \bibfield  {author} {\bibinfo {author} {\bibfnamefont {K.}~\bibnamefont
  {Greisen}},\ }\href {https://doi.org/10.1103/PhysRevLett.16.748} {\bibfield
  {journal} {\bibinfo  {journal} {Phys. Rev. Lett.}\ }\textbf {\bibinfo
  {volume} {16}},\ \bibinfo {pages} {748} (\bibinfo {year} {1966})}\BibitemShut
  {NoStop}%
\bibitem [{\citenamefont {Zatsepin}\ and\ \citenamefont
  {Kuzmin}(1966)}]{Zatsepin:1966jv}%
  \BibitemOpen
  \bibfield  {author} {\bibinfo {author} {\bibfnamefont {G.~T.}\ \bibnamefont
  {Zatsepin}}\ and\ \bibinfo {author} {\bibfnamefont {V.~A.}\ \bibnamefont
  {Kuzmin}},\ }\href@noop {} {\bibfield  {journal} {\bibinfo  {journal} {JETP
  Lett.}\ }\textbf {\bibinfo {volume} {4}},\ \bibinfo {pages} {78} (\bibinfo
  {year} {1966})}\BibitemShut {NoStop}%
\bibitem [{\citenamefont {Rosenfeld}(1975)}]{Rosenfeld:1975fy}%
  \BibitemOpen
  \bibfield  {author} {\bibinfo {author} {\bibfnamefont {A.~H.}\ \bibnamefont
  {Rosenfeld}},\ }\href {https://doi.org/10.1146/annurev.ns.25.120175.003011}
  {\bibfield  {journal} {\bibinfo  {journal} {Ann. Rev. Nucl. Part. Sci.}\
  }\textbf {\bibinfo {volume} {25}},\ \bibinfo {pages} {555} (\bibinfo {year}
  {1975})}\BibitemShut {NoStop}%
\bibitem [{\citenamefont {Hasinger}\ \emph {et~al.}(2005)\citenamefont
  {Hasinger}, \citenamefont {Miyaji},\ and\ \citenamefont
  {Schmidt}}]{Hasinger:2005sb}%
  \BibitemOpen
  \bibfield  {author} {\bibinfo {author} {\bibfnamefont {G.}~\bibnamefont
  {Hasinger}}, \bibinfo {author} {\bibfnamefont {T.}~\bibnamefont {Miyaji}},\
  and\ \bibinfo {author} {\bibfnamefont {M.}~\bibnamefont {Schmidt}},\ }\href
  {https://doi.org/10.1051/0004-6361:20042134} {\bibfield  {journal} {\bibinfo
  {journal} {Astron. Astrophys.}\ }\textbf {\bibinfo {volume} {441}},\ \bibinfo
  {pages} {417} (\bibinfo {year} {2005})},\ \Eprint
  {https://arxiv.org/abs/astro-ph/0506118} {arXiv:astro-ph/0506118}
  \BibitemShut {NoStop}%
\bibitem [{\citenamefont {Sun}\ \emph {et~al.}(2015)\citenamefont {Sun},
  \citenamefont {Zhang},\ and\ \citenamefont {Li}}]{Sun:2015bda}%
  \BibitemOpen
  \bibfield  {author} {\bibinfo {author} {\bibfnamefont {H.}~\bibnamefont
  {Sun}}, \bibinfo {author} {\bibfnamefont {B.}~\bibnamefont {Zhang}},\ and\
  \bibinfo {author} {\bibfnamefont {Z.}~\bibnamefont {Li}},\ }\href
  {https://doi.org/10.1088/0004-637X/812/1/33} {\bibfield  {journal} {\bibinfo
  {journal} {Astrophys. J.}\ }\textbf {\bibinfo {volume} {812}},\ \bibinfo
  {pages} {33} (\bibinfo {year} {2015})},\ \Eprint
  {https://arxiv.org/abs/1509.01592} {arXiv:1509.01592 [astro-ph.HE]}
  \BibitemShut {NoStop}%
\bibitem [{\citenamefont {Ajello}\ \emph {et~al.}(2014)\citenamefont {Ajello}
  \emph {et~al.}}]{Ajello:2013lka}%
  \BibitemOpen
  \bibfield  {author} {\bibinfo {author} {\bibfnamefont {M.}~\bibnamefont
  {Ajello}} \emph {et~al.},\ }\href
  {https://doi.org/10.1088/0004-637X/780/1/73} {\bibfield  {journal} {\bibinfo
  {journal} {Astrophys. J.}\ }\textbf {\bibinfo {volume} {780}},\ \bibinfo
  {pages} {73} (\bibinfo {year} {2014})},\ \Eprint
  {https://arxiv.org/abs/1310.0006} {arXiv:1310.0006 [astro-ph.CO]}
  \BibitemShut {NoStop}%
\bibitem [{\citenamefont {Gu\'epin}\ \emph {et~al.}(2018)\citenamefont
  {Gu\'epin}, \citenamefont {Kotera}, \citenamefont {Barausse}, \citenamefont
  {Fang},\ and\ \citenamefont {Murase}}]{Guepin:2017abw}%
  \BibitemOpen
  \bibfield  {author} {\bibinfo {author} {\bibfnamefont {C.}~\bibnamefont
  {Gu\'epin}}, \bibinfo {author} {\bibfnamefont {K.}~\bibnamefont {Kotera}},
  \bibinfo {author} {\bibfnamefont {E.}~\bibnamefont {Barausse}}, \bibinfo
  {author} {\bibfnamefont {K.}~\bibnamefont {Fang}},\ and\ \bibinfo {author}
  {\bibfnamefont {K.}~\bibnamefont {Murase}},\ }\href
  {https://doi.org/10.1051/0004-6361/201732392} {\bibfield  {journal} {\bibinfo
   {journal} {Astron. Astrophys.}\ }\textbf {\bibinfo {volume} {616}},\
  \bibinfo {pages} {A179} (\bibinfo {year} {2018})},\ \bibinfo {note}
  {[Erratum: Astron.Astrophys. 636, C3 (2020)]},\ \Eprint
  {https://arxiv.org/abs/1711.11274} {arXiv:1711.11274 [astro-ph.HE]}
  \BibitemShut {NoStop}%
\bibitem [{\citenamefont {Kochanek}(2016)}]{Kochanek:2016zzg}%
  \BibitemOpen
  \bibfield  {author} {\bibinfo {author} {\bibfnamefont {C.~S.}\ \bibnamefont
  {Kochanek}},\ }\href {https://doi.org/10.1093/mnras/stw1290} {\bibfield
  {journal} {\bibinfo  {journal} {Mon. Not. Roy. Astron. Soc.}\ }\textbf
  {\bibinfo {volume} {461}},\ \bibinfo {pages} {371} (\bibinfo {year}
  {2016})},\ \Eprint {https://arxiv.org/abs/1601.06787} {arXiv:1601.06787
  [astro-ph.HE]} \BibitemShut {NoStop}%
\bibitem [{\citenamefont {Hopkins}\ and\ \citenamefont
  {Beacom}(2006)}]{Hopkins:2006bw}%
  \BibitemOpen
  \bibfield  {author} {\bibinfo {author} {\bibfnamefont {A.~M.}\ \bibnamefont
  {Hopkins}}\ and\ \bibinfo {author} {\bibfnamefont {J.~F.}\ \bibnamefont
  {Beacom}},\ }\href {https://doi.org/10.1086/506610} {\bibfield  {journal}
  {\bibinfo  {journal} {Astrophys. J.}\ }\textbf {\bibinfo {volume} {651}},\
  \bibinfo {pages} {142} (\bibinfo {year} {2006})},\ \Eprint
  {https://arxiv.org/abs/astro-ph/0601463} {arXiv:astro-ph/0601463}
  \BibitemShut {NoStop}%
\bibitem [{\citenamefont {\'Alvarez-Mu\~niz}\ \emph {et~al.}(2020)\citenamefont
  {\'Alvarez-Mu\~niz} \emph {et~al.}}]{GRAND:2018iaj}%
  \BibitemOpen
  \bibfield  {author} {\bibinfo {author} {\bibfnamefont {J.}~\bibnamefont
  {\'Alvarez-Mu\~niz}} \emph {et~al.} (\bibinfo {collaboration} {GRAND}),\
  }\href {https://doi.org/10.1007/s11433-018-9385-7} {\bibfield  {journal}
  {\bibinfo  {journal} {Sci. China Phys. Mech. Astron.}\ }\textbf {\bibinfo
  {volume} {63}},\ \bibinfo {pages} {219501} (\bibinfo {year} {2020})},\
  \Eprint {https://arxiv.org/abs/1810.09994} {arXiv:1810.09994 [astro-ph.HE]}
  \BibitemShut {NoStop}%
\bibitem [{\citenamefont {Aartsen}\ \emph {et~al.}(2019)\citenamefont {Aartsen}
  \emph {et~al.}}]{IceCube:2019pna}%
  \BibitemOpen
  \bibfield  {author} {\bibinfo {author} {\bibfnamefont {M.~G.}\ \bibnamefont
  {Aartsen}} \emph {et~al.} (\bibinfo {collaboration} {IceCube}),\ }\href@noop
  {} {\bibinfo {title} {{Neutrino astronomy with the next generation IceCube
  Neutrino Observatory}}} (\bibinfo {year} {2019}),\ \Eprint
  {https://arxiv.org/abs/1911.02561} {arXiv:1911.02561 [astro-ph.HE]}
  \BibitemShut {NoStop}%
\bibitem [{\citenamefont {Heinze}\ \emph {et~al.}(2019)\citenamefont {Heinze},
  \citenamefont {Fedynitch}, \citenamefont {Boncioli},\ and\ \citenamefont
  {Winter}}]{Heinze:2019jou}%
  \BibitemOpen
  \bibfield  {author} {\bibinfo {author} {\bibfnamefont {J.}~\bibnamefont
  {Heinze}}, \bibinfo {author} {\bibfnamefont {A.}~\bibnamefont {Fedynitch}},
  \bibinfo {author} {\bibfnamefont {D.}~\bibnamefont {Boncioli}},\ and\
  \bibinfo {author} {\bibfnamefont {W.}~\bibnamefont {Winter}},\ }\href
  {https://doi.org/10.3847/1538-4357/ab05ce} {\bibfield  {journal} {\bibinfo
  {journal} {Astrophys. J.}\ }\textbf {\bibinfo {volume} {873}},\ \bibinfo
  {pages} {88} (\bibinfo {year} {2019})},\ \Eprint
  {https://arxiv.org/abs/1901.03338} {arXiv:1901.03338 [astro-ph.HE]}
  \BibitemShut {NoStop}%
\bibitem [{\citenamefont {Pedreira}(2021)}]{Pedreira:2021gcl}%
  \BibitemOpen
  \bibfield  {author} {\bibinfo {author} {\bibfnamefont {F.}~\bibnamefont
  {Pedreira}} (\bibinfo {collaboration} {Pierre Auger}),\ }\href
  {https://doi.org/10.22323/1.358.0979} {\bibfield  {journal} {\bibinfo
  {journal} {PoS}\ }\textbf {\bibinfo {volume} {ICRC2019}},\ \bibinfo {pages}
  {979} (\bibinfo {year} {2021})}\BibitemShut {NoStop}%
\bibitem [{\citenamefont {Allison}\ \emph {et~al.}(2016)\citenamefont {Allison}
  \emph {et~al.}}]{ARA:2015wxq}%
  \BibitemOpen
  \bibfield  {author} {\bibinfo {author} {\bibfnamefont {P.}~\bibnamefont
  {Allison}} \emph {et~al.} (\bibinfo {collaboration} {ARA}),\ }\href
  {https://doi.org/10.1103/PhysRevD.93.082003} {\bibfield  {journal} {\bibinfo
  {journal} {Phys. Rev. D}\ }\textbf {\bibinfo {volume} {93}},\ \bibinfo
  {pages} {082003} (\bibinfo {year} {2016})},\ \Eprint
  {https://arxiv.org/abs/1507.08991} {arXiv:1507.08991 [astro-ph.HE]}
  \BibitemShut {NoStop}%
\bibitem [{\citenamefont {Cummings}\ \emph {et~al.}(2021)\citenamefont
  {Cummings}, \citenamefont {Aloisio},\ and\ \citenamefont
  {Krizmanic}}]{Cummings:2020ycz}%
  \BibitemOpen
  \bibfield  {author} {\bibinfo {author} {\bibfnamefont {A.~L.}\ \bibnamefont
  {Cummings}}, \bibinfo {author} {\bibfnamefont {R.}~\bibnamefont {Aloisio}},\
  and\ \bibinfo {author} {\bibfnamefont {J.~F.}\ \bibnamefont {Krizmanic}},\
  }\href {https://doi.org/10.1103/PhysRevD.103.043017} {\bibfield  {journal}
  {\bibinfo  {journal} {Phys. Rev. D}\ }\textbf {\bibinfo {volume} {103}},\
  \bibinfo {pages} {043017} (\bibinfo {year} {2021})},\ \Eprint
  {https://arxiv.org/abs/2011.09869} {arXiv:2011.09869 [astro-ph.HE]}
  \BibitemShut {NoStop}%
\bibitem [{\citenamefont {Kimura}\ \emph {et~al.}(2018)\citenamefont {Kimura},
  \citenamefont {Murase},\ and\ \citenamefont {Zhang}}]{Kimura:2017ubz}%
  \BibitemOpen
  \bibfield  {author} {\bibinfo {author} {\bibfnamefont {S.~S.}\ \bibnamefont
  {Kimura}}, \bibinfo {author} {\bibfnamefont {K.}~\bibnamefont {Murase}},\
  and\ \bibinfo {author} {\bibfnamefont {B.~T.}\ \bibnamefont {Zhang}},\ }\href
  {https://doi.org/10.1103/PhysRevD.97.023026} {\bibfield  {journal} {\bibinfo
  {journal} {Phys. Rev. D}\ }\textbf {\bibinfo {volume} {97}},\ \bibinfo
  {pages} {023026} (\bibinfo {year} {2018})},\ \Eprint
  {https://arxiv.org/abs/1705.05027} {arXiv:1705.05027 [astro-ph.HE]}
  \BibitemShut {NoStop}%
\bibitem [{\citenamefont {Zyla}\ \emph {et~al.}(2020)\citenamefont {Zyla} \emph
  {et~al.}}]{ParticleDataGroup:2020ssz}%
  \BibitemOpen
  \bibfield  {author} {\bibinfo {author} {\bibfnamefont {P.~A.}\ \bibnamefont
  {Zyla}} \emph {et~al.} (\bibinfo {collaboration} {Particle Data Group}),\
  }\href {https://doi.org/10.1093/ptep/ptaa104} {\bibfield  {journal} {\bibinfo
   {journal} {PTEP}\ }\textbf {\bibinfo {volume} {2020}},\ \bibinfo {pages}
  {083C01} (\bibinfo {year} {2020})}\BibitemShut {NoStop}%
\bibitem [{\citenamefont {Anchordoqui}(2022)}]{Anchordoqui:2022ejw}%
  \BibitemOpen
  \bibfield  {author} {\bibinfo {author} {\bibfnamefont {L.~A.}\ \bibnamefont
  {Anchordoqui}},\ }\href {https://doi.org/10.1103/PhysRevD.106.116022}
  {\bibfield  {journal} {\bibinfo  {journal} {Phys. Rev. D}\ }\textbf {\bibinfo
  {volume} {106}},\ \bibinfo {pages} {116022} (\bibinfo {year} {2022})},\
  \Eprint {https://arxiv.org/abs/2205.13931} {arXiv:2205.13931 [hep-ph]}
  \BibitemShut {NoStop}%
\end{thebibliography}%

\end{document}